\documentclass[numbook,natbib,final,runningheads]{svjour3}
\usepackage{natbib}
\usepackage{epsfig}
\usepackage{graphicx,mathptmx}

\usepackage{makeidx}
\makeindex

\newcommand{\degr}{\ifmmode^\circ\else$^\circ$\fi}

\newcommand{\lapprox} {\, \lower3pt\hbox{$\sim$}\llap{\raise2pt\hbox{$<$}}\,}
\newcommand{\gapprox} {\, \lower3pt\hbox{$\sim$}\llap{\raise2pt\hbox{$>$}}\,}

\begin{document}

\title{Recent Advances in Understanding Particle Acceleration Processes in Solar Flares}

\author{V.~V.~Zharkova$^1$, K.~Arzner$^2$, A.~O.~Benz$^2$, P.~Browning$^3$, C.~Dauphin$^4$,
A.~G.~Emslie$^{5,6}$, L.~Fletcher$^7$, E.~P.~Kontar$^7$, G.~Mann$^8$, M.~Onofri$^9$,
V.~Petrosian$^{10}$, R.~Turkmani$^{11}$, N.~Vilmer$^{4}$, and L.~Vlahos$^9$}

\institute{$^{1}$Department of Mathematics, University of
Bradford, U.K.  \email{v.v.zharkova@brad.ac.uk}\\
$^{2}$ETH Zurich, Switzerland \\
$^{3}$Physics Department, University of Manchester, U.K. \\
$^{4}$Observatoire de Paris, LESIA, France \\
$^{5}$Department of Physics, Oklahoma State University, Stillwater, OK 74078, U.S.A.\\
$^{6}$Department of Physics and Astronomy, Western Kentucky University, Bowling Green, KY 42101, U.S.A.\\
$^{7}$Department of Physics and Astronomy, University of Glasgow, Glasgow, G12 8QQ, U.K.\\
$^{8}$Astrophysikalischen Institut, Potsdam, Germany \\
$^{9}$Dept. of Physics, University of Thessaloniki, Thessaloniki 54124, Greece \\
$^{10}$Center for Astrophysics, Stanford University, U.S.A.\\
$^{11}$Imperial College, London, U.K.\\
}

\authorrunning{Zharkova et al.}
\titlerunning{Particle Acceleration}


\maketitle

\begin{abstract}

We review basic theoretical concepts in particle acceleration,
with particular emphasis on processes likely to occur in regions
of magnetic reconnection. Several new developments are discussed,
including detailed studies of reconnection in three-dimensional
magnetic field configurations (e.g., current sheets, collapsing
traps, separatrix regions) and stochastic acceleration in a
turbulent environment.  Fluid, test-particle, and particle-in-cell
approaches are used and results compared.  While these studies show
considerable promise in accounting for
the various observational manifestations of solar flares, they are
limited by a number of factors, mostly relating to available
computational power. Not the least of these issues is the need to
explicitly incorporate the electrodynamic feedback of the
accelerated particles themselves on the environment in which they
are accelerated. A brief prognosis for future advancement is
offered.

\end{abstract}

\newpage

\setcounter{tocdepth}{6}

\tableofcontents

\section{Introduction} \label{sec:8introduction}


It is by now well established that the primary source of the
energy released in a solar flare is in stressed, current-carrying
(``non-potential'') magnetic fields\index{magnetic field!non-potential}. This magnetic energy is
released through a {\it magnetic reconnection}\index{reconnection} process, in which
magnetic field lines of opposite polarity change their topological
connectivity and the field relaxes to a state of lower energy,
thereby releasing energy in the form of accelerated particles,
plasma heating and mass motion.

Magnetic reconnection can occur in several geometries.  For
example, closed magnetic loops embedded into the solar photosphere
can collide in the upper solar atmosphere, resulting in a
localized region of relatively high magnetic shear, as in the
classical Sweet-Parker\index{reconnection!Sweet-Parker} model \citep[see,
e.g.,][]{1969ARA&A...7..149S}. 
Alternatively, open magnetic field
lines can collapse onto a closed helmet-like magnetic
configuration.\index{looptop sources} 
This leads to a region of high magnetic shear near the top of the closed structure -- as evidenced by looptop sources\index{above-the-looptop sources}\index{flare (individual)!SOL1992-01-13T17:25 (M2.0)!above-the-loop-top source} of hard X-rays first observed by \textit{Yohkoh}
\citep{1994Natur.371..495M} -- and in the energization of
successive field lines at greater and greater heights, which
produces both coronal hard X-ray sources\index{coronal sources}
\citep{2008A&ARv..16..155K} and a concomitant increase in the
separation of the associated hard X-ray and H$\alpha$ footpoints\index{footpoints}
\citep{1976SoPh...50...85K}.
\index{satellites!Yohkoh@\textit{Yohkoh}}
Magnetic reconnection can also be
accompanied by dramatic large-scale changes in the magnetic field
configuration, such as in the ``breakout'' model\index{flare models!breakout}
\citep{1999ApJ...510..485A}, in which a filament, previously
restrained by the magnetic tension of a set of magnetic field
lines exterior to it, is released into interplanetary space as a
result of reconnection of the constraining field lines.\index{filaments!eruptive}
Reconnection can also result in the formation of magnetohydrodynamic (MHD) shocks, which may be responsible for significant particle acceleration exterior
to the region of magnetic reconnection itself.
\index{shocks!and reconnection}

Whatever the geometry, the fundamental process of magnetic
reconnection somehow transfers a significant fraction
\citep[e.g.,][]{1997JGR...10214631M} of the stored magnetic energy
into accelerated charged particles, both electrons and ions. Some
of the high-energy particles may be trapped in the corona by
plasma turbulence\index{turbulence}\index{trapping!by turbulence}\index{plasma turbulence} \citep[e.g.,][]{2004ApJ...610..550P} and/or by
magnetic mirroring\index{magnetic structures!mirror geometry} \citep[e.g.,][]{1979PASAu...3..369M,
1984PhDT........33L, 1997ApJ...485..859S}, to produce hard X-ray
(HXR) looptop sources\index{hard X-rays!looptop sources} \citep[e.g.,][]{1994Natur.371..495M,
2002ApJ...569..459P, 2003ApJ...596L.251S,
2006ApJ...638.1140J,2006A&A...456..751B,2006PhDT........35L}. 
Some particles may escape the acceleration region into interplanetary
space and can contribute to solar energetic particle (SEP) events
detected at 1~AU \citep[e.g.,][]{2007ApJ...663L.109K}.\index{acceleration region!escape from}
 
Understanding the conversion of magnetic energy into accelerated
particles \index{flares!and acceleration}is one of the main goals of solar flare research. 
It also has attendant implications for our understanding of particle
acceleration in other astrophysical sites and was indeed a key
driver for the design of {\em Reuven Ramaty High-Energy Solar Spectroscopic Imager (RHESSI)}, 
which measures the diagnostic high-energy hard X-ray and gamma-ray
radiation produced by the accelerated particles with unprecedented
temporal, spectral and spatial resolution.
\index{satellites!RHESSI@\textit{RHESSI}}\index{RHESSI@\textit{RHESSI}}

At a fundamental level, of course, charged particles are
accelerated by an electric field\index{electric fields!and particle acceleration}.
Differences in the large-scale
structure and temporal evolution of the accelerating electric
field allows us to discuss particle acceleration in terms of
several basic scenarios (see Section~\ref{sec:8basic_model}).
These include large-scale sub-Dreicer\index{electric fields!sub-Dreicer} acceleration \citep[see, for
example][]{1994ApJ...435..469B}; super-Dreicer\index{electric fields!super-Dreicer} acceleration in a relatively localized 2-D current sheet\index{current sheets}
\citep[e.g.,][]{1993SoPh..146..127L}; and stochastic acceleration\index{acceleration!in MHD turbulence}
by MHD turbulence\index{turbulence!MHD} generated in fast plasma reconnection outflows
\citep[e.g.,][]{1996ApJ...461..445M, 1997ApJ...482..774P}.

In Section~\ref{sec:8developments}, we report on recent work aimed
at extending these basic mechanisms to more complex models. These
include (1) first-order Fermi acceleration, either in a collapsing
magnetic trap \citep{1997ApJ...485..859S} or in a fast termination\index{shocks!termination}
shock \citep{1998ApJ...495L..67T, 2007A&A...471L..37A}, or by a
super-Dreicer electric field in a simple 3-D magnetic topology
\citep{2005SSRv..121..165Z,2009JPlPh..75..159Z}; and (2)
acceleration in more elaborate three-dimensional magnetic
topologies  derived from MHD simulations \citep
{2006ApJ...640L..99D,2008A&A...491..289D}. We also discuss some
recent suggestions of acceleration by large-scale Alfv\'en waves
\citep{2008ApJ...675.1645F} and acceleration involving multiple
reconnection regions or in a turbulent environment
\citep{2004ApJ...608..540V,2005ApJ...620L..59T,2007A&A...468..289D}.

In the latter portion of Section~\ref{sec:8developments} we also
consider both ``local'' and ``global'' aspects of the acceleration
process. At the local level, accurate simulations of particle
trajectories are needed to correctly infer the general properties
of the acceleration.  At the ``global'' level, we recognize that,
in view of the large number of accelerated particles involved, the
traditional ``test-particle'' approach\index{acceleration!test-particle approach}, in which the
electromagnetic environment in the acceleration region is {\it
prescribed}, is not applicable (Section~\ref{sec:8test_p}) -- a
self-consistent approach requires that we incorporate the electric
and magnetic fields associated with the accelerated particles
themselves in current sheets with and without tearing
instabilities (Sections~\ref{sec:8pic} and~\ref{sec:8island}).\index{acceleration region!test-particle approach}

In Sections~\ref{sec:8explain} and~\ref{sec:8summary} we relate
the results of the recent modeling efforts to flare observations,
particularly from {\em RHESSI}, and we offer a brief prognosis for
future models of the particle acceleration process.

\section{Review of Pertinent Observations} \label{sec:8obs_review}

\subsection{Hard X-ray lightcurves and spectra} \label{sec:8photons}
\index{hard X-rays!lightcurves and spectra}

\subsubsection{Lightcurves}\label{sec:zharkova_light}

Spatially-integrated hard X-ray light curves obtained by {\em
RHESSI} confirm many of the temporal features established by
previous observations: sharp increases (bursts) of hard X-ray
intensity over a relatively short ($\sim$0.5-5~s)
timescale accompanied by a more slowly-varying component with a
timescale up to a few tens of minutes \citep{Chapter3}. This
suggests that electron acceleration\index{acceleration!electrons!timescales} is characterized by two
fundamentally distinct timescales: a very short timescale
associated with the actual {\it acceleration} of ambient electrons
to high, bremsstrahlung-emitting energies and a much longer
timescale associated with the {\it duration} of the acceleration
process, a process that, as evidenced by steady hard X-ray\index{hard X-rays!gradual emission}
emission for a substantial fraction of an hour or even longer, can
continue to accelerate particles over such a sustained period.

At lower photon energies $\epsilon$, where thermal bremsstrahlung
\index{bremsstrahlung!thermal} dominates the total emission, the
emission at higher photon energies is weighted more heavily by
plasma at high temperatures $T$.

As a result, the decrease in
conductive cooling time \index{cooling!conductive} with
temperature -- $\tau \approx T^{-5/2}$ for collisionally-dominated
conductive processes \citep[see][]{1962pfig.book.....S} -- results
in the emission at higher energies peaking sooner
\citep{2007ApJ...661.1242A}, and hence the light curve peaking
progressively earlier with increase in photon energy. At higher
energies, where {\it non-thermal} bremsstrahlung dominates, the
relative timing of the emission at different energies depends on a
balance between the reduced ``time-of-flight'' 
\index{time-of-flight analysis} for higher energy electrons before they impact upon the
thick target\index{thick-target model}\index{flare models!thick-target}
 of the lower atmosphere \citep[which tends to {\it
advance} high energy emissions relative to low energy ones
--][]{1996ApJ...464..974A, 1998ApJ...509..911B} and on the
decrease in collision frequency\index{frequency!collision} with
energy \citep[which tends to {\it delay} high energy emissions
relative to low energy ones --][]{1997ApJ...487..936A}.

\subsubsection{Photon and electron energy spectra} \label{sec:zharkova_energy_sp}

Hard X-ray emission during flares typically exhibits a very steep
spectrum at lower energies $\epsilon \approx 10$~keV, indicative of a
thermal process. It must be noted, however, that the frequent
assumption \citep[e.g.,][]{2003ApJ...595L..97H} of an {\it
iso}thermal plasma\index{hard X-rays!thermal interpretation} 
is neither expected\index{isothermal spectrum!improbability of} on the basis of simple
physics nor required by the observations
\citep[e.g.,][]{1974IAUS...57..395B}. Indeed,
\citet{2007ApJ...661.1242A} has shown that the assumption of an
isothermal source is {\it inconsistent} with the observations of
temporal variations of the hard X-ray spectrum in certain flares
observed by {\em RHESSI}.  
Indeed, in most cases analysis of the variation of
residuals (observed hard X-ray flux vs. flux calculated using an
isothermal spectrum) with photon energy shows that 
that an isothermal model {\it does} provide an excellent fit to
the observed spectrum; generally a broad-band hard X-ray spectrometer shows the (unique) hottest component of a multithermal source.
Nevertheless \textit{RHESSI's} superior resolution and good photometric calibration have made it possible to derive 
a two-temperature source characterization from a single bremsstrahlung spectrum \citep[e.g.,][]{2010ApJ...725L.161C} in some cases.\index{bremsstrahlung!multithermal}\index{soft X-rays!multi-thermal bremsstrahlung modeling}

In most powerful flares the photon spectra reveal double power
laws\index{bremsstrahlung!nonthermal}\index{hard X-rays!spectrum!double power law} below 300~keV, with the
spectral index below $\sim$30-60~keV being smaller (flatter)
than that above $\sim$30-60~keV by 2-4 units
\citep{2006A&A...458..641G,2008A&ARv..16..155K}. For the most
energetic flares, there can be discernible emission up to a few
hundred~MeV \citep[for reviews,
see][]{2009RAA.....9...11C,Chapter4}. 
At photon energies $\sim$500-1000~keV, the spectrum \index{spectrum!hard X-ray} tends to
flatten again, a flattening that has been, at least partially,
attributed to the increasing contribution of electron-electron
bremsstrahlung \citep{2007ApJ...670..857K}. However, the
contribution of electron-electron bremsstrahlung to this
\index{bremsstrahlung!electron-electron}
flattening\index{bremsstrahlung!nonthermal} does not exceed $\sim$0.4, while the observed spectral hardening, as observed in a few events with \textit{GRANAT}/PHEBUS, can be as high as 1.5-2 units at
energies from 600~keV up to several tens of~MeV \citep[see,
e.g.,][]{1998A&A...334.1099T,1999A&A...342..575V}.
\index{satellites!GRANAT@\textit{GRANAT}!PHEBUS}

\begin{figure}
\centering
\includegraphics[width=0.8\textwidth]{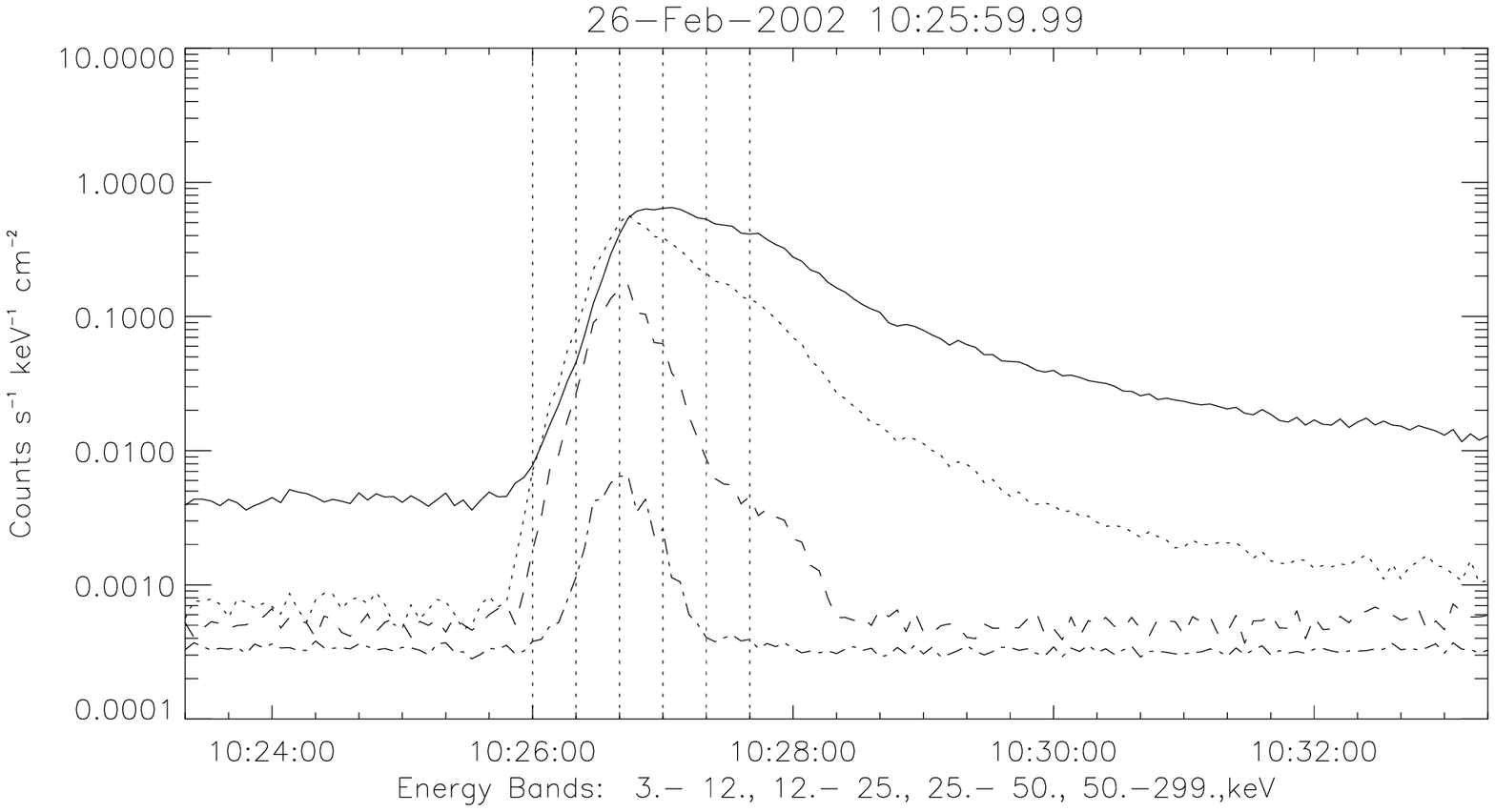}
\includegraphics[width=0.8\textwidth]{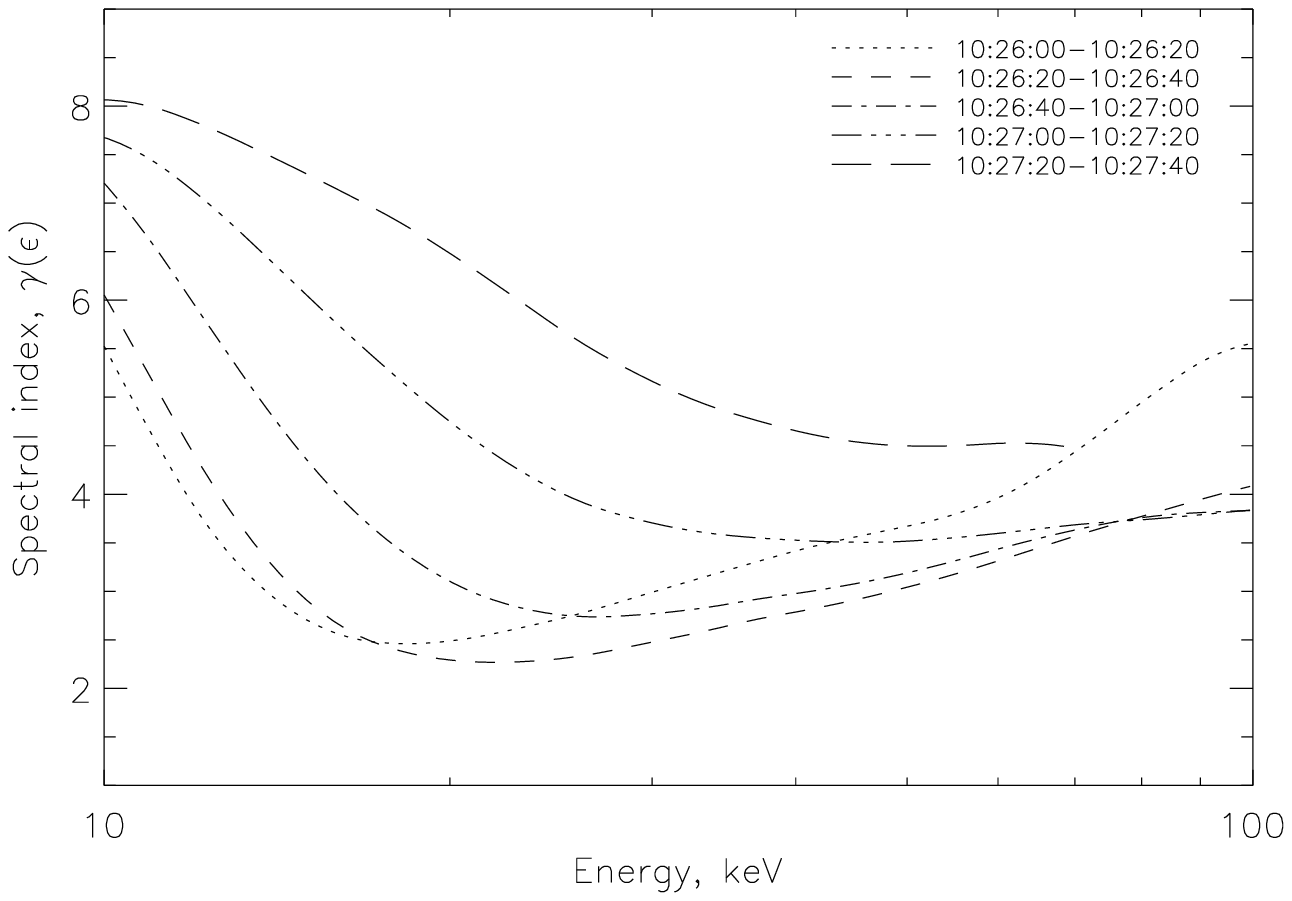}
\caption{{\it Upper Panel}: Temporal variation (4-second cadence)
of the count rates in seven front {\em RHESSI} segments for SOL2002-02-26T10:27 (C9.6)
\index{flare (individual)!SOL2002-02-26T10:27 (C9.6)!illustration} (10:27~UT).
The vertical lines show five 20-second
accumulation intervals for spectral analysis. {\it Lower Panel}:
Temporal variation of the energy-dependent photon spectral index
$\gamma(\epsilon) = - d \log I(\epsilon)/d \log \epsilon$. Each
line corresponds to one of the time intervals in the upper plot.
From \citet{2005SoPh..227..299K}.}
\label{fig:zharkova_g_time}
\end{figure}

{\em RHESSI} observations reveal a relationship between spectral
hardness and time. Values of the ``local'' spectral index\index{local@``local'' spectral index}
$\gamma(\epsilon) = - d \log I(\epsilon)/d \log \epsilon$ are
generally lowest around the peak of the event -- a so-called
``soft-hard-soft'' evolution originally noted by
\index{hard X-rays!soft-hard-soft}
\citet{1969ApJ...155L.117P}. {\em RHESSI}'s superior spectral
resolution shows that nearly every discernible intensity peak
(which can number up to a dozen or more in a single flare) has a
soft-hard-soft structure \citep{2004A&A...426.1093G}. Also the
decrease in spectral index before, and increase in spectral index
after, this time does not occur at the same rate for all photon
energies $\epsilon$ \citep{2005SoPh..227..299K}; see Figure~\ref{fig:zharkova_g_time}.
Rather, a flare typically starts with a spectral index $\gamma$ that is
strongly dependent on energy\index{accelerated particles!energy spectra!energy-dependent index}, with the minimum value of $\gamma$
showing a tendency to grow with time.

Non-thermal emission in the corona is identified in the impulsive
phase by its softer spectrum \citep{1999ApJ...514..484M,
2002ApJ...569..459P}, consistent with the small column depth of
the coronal part of the source\index{hard X-rays!coronal sources}.
The absence of a significant
amount of (energy-dependent) collisional losses\index{electrons!collision losses} in this relatively
thin target\index{hard X-rays!thin-target} 
should result in a spectrum two
powers steeper than the target-averaged spectrum
\citep{1971SoPh...18..489B,1972SoPh...24..414H,
1973SoPh...32..459D}.

It should be noted that even if the accelerated electrons have a
pure power-law energy spectrum $F_0(E_0) \approx E_0^{-\delta}$, for
which a characteristic energy does not exist, characteristic
energies associated with either electron transport and/or
radiation physics may produce deviations from the power-law
behavior in the observed spatially-integrated photon spectrum.
Further, as elaborated upon by \citet{Chapter7}, anisotropy in the
mean electron distribution\index{electrons!distribution function!mean electron distribution}, combined with the intrinsic
directivity\index{directivity!hard X-ray}\index{hard X-rays!directivity} of the bremsstrahlung
emission process, produces an anisotropic distribution of primary photons.
Compton backscattering \index{Compton scattering}of photons from the
photosphere \index{albedo} \citep[``X-ray
albedo;''][]{1977ApJ...215..666L,1978ApJ...219..705B} not only
influences the observed photon spectrum, but also has a diagnostic
potential for determining the electron angular distribution
\citep{2006ApJ...653L.149K}.
\index{hard X-rays!albedo}\index{scattering!Compton}

\subsubsection{Electron numbers} \label{sec:zharkova_e_numbers}
\index{electrons!number problem}

Analysis of {\em RHESSI} data \citep{Brown2007,Dennis2007} has
confirmed the important point that a significant fraction of the
energy released in flares is deposited into accelerated particles.
\index{flares!energy content!accelerated particles}
In addition, recent calorimetric analysis of the evolution of a
radio spectrum from a dense flare \citep{Bastian2007} reveals that
a substantial fraction of the energy in the energetic electrons
can be deposited into the coronal loop, rather than into the
chromosphere; it is estimated that the energy deposited in the
corona can approach about 30\% of the estimated magnetic energy
released in the whole flare.

\begin{figure}
\center
\includegraphics[width=1.0\textwidth]{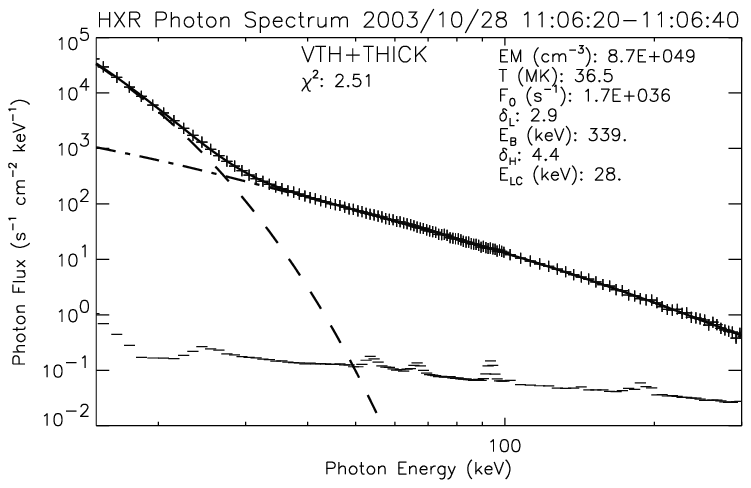}
\caption{{\em RHESSI} hard X-ray photon spectrum for SOL2003-10-28T11:10 (X17.2)\index{flare (individual)!SOL2003-10-28T11:10 (X17.2)!illustration}
(11:06:20-11:06:40 UT). The photon flux (plus signs) is
fit with a thermal bremsstrahlung contribution from an
isothermal plasma (dashed curve) plus a double power-law
non-thermal electron flux component with a low-energy cutoff
(dash-dotted curve). \index{spectrum!hard X-rays!spectrum!forward fit}
The best-fit parameters are shown in the plot.}
\label{fig:zharkova_ma_2}
\end{figure}

As has been pointed out by several authors
\citep[e.g.,][]{1971SoPh...18..489B,1988psf..book.....T}, the
total injected electron flux depends critically on the value of
the low-energy cutoff $E_c$\index{low-energy cutoff}.
A brief discussion of this issue is in
order. 
Historically, a low-energy cutoff $E_c$ was assumed simply
in order to keep the injected power $\int_{E_c}^\infty E_0
F_0(E_0) \, dE_0$ finite \citep[e.g.,
][]{1971SoPh...18..489B,2003ApJ...595L..97H}. To determine whether
or not such a cut-off is actually {\it required} by observations,
it is best to adopt a non-parametric approach to interpreting the
photon spectrum $I(\epsilon)$ -- i.e., to infer from $I(\epsilon)$
what range of mean electron source functions $\overline F(E)$
\citep{2003ApJ...595L.115B} allow a statistically acceptable fit
to $I(\epsilon)$. \citet{2005SoPh..232...63K}, in their analysis
of SOL2002-08-20T08:25 (M3.4)\index{flare (individual)!SOL2002-08-20T08:25 (M3.4)!low-energy cutoff},
have shown that a low-energy cutoff\index{low-energy cutoff!and albedo}, or even a gap, in
the mean source electron spectrum exists if the observed spectrum
is considered as primary bremsstrahlung only; however, such cutoffs
and/or gaps disappear if an albedo correction\index{albedo} (for an isotropic
primary source) is applied to the observed photon spectrum \citep[see, e.g.,][]{Chapter7}.
Further, \citet{2003ApJ...595L.119E} has shown that, when
allowance is made for warm-target effects\index{warm target}\index{low-energy cutoff!and warm target} in the electron energy-loss rate, the injected electron energy $\int_{0}^\infty E_0
F_0(E_0) \, dE_0$ corresponding to a pure power-law photon
spectrum can be finite even if no low-energy cutoff exists.

Figure~\ref{fig:zharkova_ma_2} shows the photon spectrum for the
time interval indicated during SOL2003-10-28T11:10 (X17.2)\index{flare (individual)!SOL2003-10-28T11:10 (X17.2)}. Assuming that
the non-thermal hard X-ray emission is thick-target bremsstrahlung\index{bremsstrahlung!thick-target}
\citep {1971SoPh...18..489B}, one can use a forward-fitting
method\index{inverse problem!forward fit}, including both thermal and nonthermal fit components,
\citep[e.g.,][]{2003ApJ...595L..97H} to derive the injected
nonthermal electron\index{electrons!injected distribution} acceleration rate, differential in energy
(electrons~s$^{-1}$~keV$^{-1}$).  This method
\citep[see][]{Chapter7} also determines an observational upper
limit to $E_c$ (as the intersection of the thermal and nonthermal
fit components), and hence a lower limit on the associated
electron number and energy fluxes. Integration over energy
provides the total electron acceleration rate. For this event, one
obtains a total non-thermal electron production
rate\index{electrons!accelerated!flux} $F_{e} = 1.7 \times
10^{36}$~s$^{-1}$ above $E_c = 28$~keV, corresponding to an
injected power\index{electrons!accelerated!energy content} 
$P_{e} = 1.5 \times 10^{29}$~erg~s$^{-1}$ above the same energy. For a nominal
footpoint area $S \lapprox 10^{18}$~cm$^{2}$, such a particle rate
corresponds to an electron flux $F_{e} \gapprox
10^{18}$~cm$^{-2}$~s$^{-1}$ and hence, since the average electron
velocity $v \approx (c/3) \approx 10^{10}$~cm~s$^{-1}$, to a beam
density\index{beams}\index{electrons!accelerated!density} 
$n_b \gapprox 10^8$~cm$^{-3}$. 
This is a substantial fraction of the ambient
density $n \approx 10^{10}$~cm$^{-3}$.

The total electric {\it current}
\index{electrons!accelerated!associated current} 
corresponding to such a particle acceleration rate, {\it if the acceleration is
unidirectional}, is $I \approx 3 \times 10^{17}$~A, and the
associated current density $j = I/S \approx 0.3$~A~cm$^{-2}$. Such
high current densities, especially since they are introduced over
a timescale of seconds, produce unacceptably large inductive
electric and magnetic fields\index{magnetic field!inductive}, unless (1) the acceleration is
near-isotropic, as in stochastic acceleration models
(Section~\ref{sec:8stochastic}), (2) the source has very fine
structure \citep{1985ApJ...293..584H}, or (3) the beam current is
effectively neutralized by a co-spatial return current\index{return current} formed by
both the ambient and returning beam electrons
\citep[see][]{1977ApJ...218..306K, 1980ApJ...235.1055E,
1984A&A...131L..11B,1989SoPh..120..343L,
1990A&A...234..496V,2006ApJ...651..553Z}.

Finally, the total number of electrons\index{electrons!accelerated!number} 
in the acceleration region
${\cal N} = n \, V \approx 10^{37}$, a number which corresponds to
only a few seconds of acceleration at the rates quoted above.
Hence the issue of replenishment of the acceleration region
\citep{1995ApJ...446..371E} has to be accounted for by any
acceleration and/or transport mechanism.\index{acceleration region!replenishment}

\subsection{Lightcurves and energy spectra of $\gamma$-rays} \label{sec:8ions}

\subsubsection{Gamma-ray lightcurves}
\index{gamma-rays!lightcurves}

Due to the limited sensitivity of current gamma-ray detectors,
information regarding the time scales for ion acceleration is less
stringent than that for energetic electrons
(Section~\ref{sec:8photons}).

The prompt gamma-ray line emission is known to peak
roughly simultaneously,within $\pm1$~s,
of the hard X-ray continuum emission,
based on a few events \citep[e.g.,][]{1986ApJ...300L..95K}.
\index{continuum!hard X-ray}
\index{gamma-rays!synchronism with hard X-rays} 
This near-simultaneity\index{hard X-rays!near-simultaneity with $\gamma$-rays}\index{gamma@$\gamma$-rays!near-simultaneity with HXRs} of hard
X-ray and gamma-ray emissions indicates that both ion acceleration
to a few~MeV, and interaction\index{transport!ions} of the
accelerated ions with the ambient atmosphere to produce nuclear
lines \citep{Chapter4}, can occur within a timescale $\sim$1~s.
\index{ions!transport time scales}
However, in a few flares there {\it are} time delays between
the appearance of hard X-rays and gamma-ray lines \citep[see
section 2.2.6 in][]{Chapter4}; this is often interpreted in terms
of differential transport of electrons and ions between the
acceleration and emission sites.

{\em RHESSI} observations \citep[see, e.g., Figure 3.1 in][]{Chapter4} have
provided unprecedented information on the temporal evolution of
``pure'' gamma-ray lines (i.e., lines that occur in spectral
regions that are relatively uncontaminated by electron
bremsstrahlung). 
\index{RHESSI@\textit{RHESSI}!pure@``pure'' gamma-ray lines}
Such observations indicate the continuous
production of very energetic ions ($>$~a~few hundreds of~MeV) for
hours after the impulsive phase of the flare, and so pose a
formidable constraint on the ion acceleration mechanism \citep{Chapter4}.

{\em RHESSI} observations have also provided important information
on the temporal evolution of prompt gamma-ray lines
\citep{2003ApJ...595L..69L}. Comparison with the temporal
evolution of the X-ray flux at 150~keV shows that the temporal
evolution of hard X-rays and gamma-rays are roughly similar,
indicating a common origin of accelerated electrons and ions.
However, for SOL2002-07-23T00:35 (X4.8), 
there is a small delay of around 10~s between the maxima of the X-ray and gamma-ray
fluxes. 
\index{flare (individual)!SOL2002-07-23T00:35 (X4.8)!time delays}
This has been interpreted in terms of transport of
energetic electrons and ions by \citet{2007A&A...468..289D}, and is
discussed in \citet{Chapter4}.

\subsubsection{Energy spectra and abundances of ions in flares}
\index{abundances}

As reviewed by \citet{Chapter4}, interactions of energetic ions in
the energy range from $\sim$1~Mev~nucleon$^{-1}$ to
100~Mev~nucleon$^{-1}$ produce an extensive spectrum of gamma-ray line
emission. 
Narrow gamma-ray lines\index{gamma-rays!narrow lines} result from the bombardment of
the ambient nuclei by accelerated protons and $\alpha$-particles,
while broad lines\index{gamma-rays!broad lines} occur from the inverse reactions, in which
accelerated heavy nuclei collide with ambient hydrogen or helium
nuclei. 
De-excitation gamma-ray line\index{gamma-rays!deexcitation lines} spectra provide information
on flare ions of energies above $\sim$2~Mev~nucleon$^{-1}$.
\index{ions!distribution functions}
The shape of the ion distribution below this energy is unknown, but is comparably important for the evaluation of the total ion energy\index{ions!total energy!dependence on spectral distribution}
as is the low-energy cutoff\index{low-energy cutoff} to the evaluation of the total electron energy content\index{electrons!total energy}
(Section~\ref{sec:8photons}) \citep[see][]{1997ApJ...485..430E}.

Quantitative gamma-ray line observations have now been obtained
for more than twenty events. 
The accelerated ion spectra are found
to extend as unbroken power laws down to at least
2~Mev~nucleon$^{-1}$, if a reasonable ambient Ne/O abundance ratio\index{abundances!Ne/O ratio} is
used \citep[for details, see][]{Chapter4}.
For the 19~flares observed by the \textit{Solar Maximum Mission (SMM)} prior to
the launch of {\em RHESSI}, the average spectral
index was around 4.3; however, the events observed with {\em
RHESSI} tend to have much harder slopes
\citep{2003ApJ...595L..69L,2006GMS...165..177S}.
\index{satellites!SMM@\textit{SMM}}

Even though most of the energy contained in ions resides in
protons and $\alpha$-particles, the abundances of heavier
accelerated ions provide crucial constraints for acceleration
processes. Information on the abundances of such flare-accelerated
ions was deduced for the \textit{SMM}/GRS (Gamma-Ray Spectrometer)
flares as well as for a few flares observed by {\em RHESSI}. 
\index{satellites!SMM@\textit{SMM}!GRS} 
A noticeable enhancement has been
observed in the numbers of $\alpha$-particles, as well as of
accelerated $^3$He isotopes, compared to the standard coronal
abundances. Furthermore, accelerated heavy ions, such as Ne, Mg
and Fe, are also generally found to be over-abundant\index{abundances!and acceleration of heavy ions} with respect to normal coronal compositions, as pointed out by \citet{Chapter4}.
This poses another constraint on models for ion acceleration.

\index{accelerated particles!heavy ions}

\subsubsection{Ion numbers}
\index{ions!number problem}

Since the first detection of solar flare gamma-ray lines in 1972
\citep{1973Natur.241..333C}, information on the energy content in
ions has been obtained for more than 20 events. The analysis
carried out using 19 events observed by \textit{SMM}/GRS and a few events
observed with {\em RHESSI} shows that the energy contained in
$>$1~MeV ions ranges from 10$^{29}$ to 10$^{32}$~erg, comparable
to the energy contained in sub-relativistic electrons. However,
there are still large uncertainties in the determination of these
quantities and a large dispersion of the ratio of electron ($\sim$20~keV) to ion ($\sim$1~Mev~nucleon$^{-1}$) energy contents from one
flare to another \citep[for details, see][]{Chapter4}.

The issue of relative importance of electron and ion
acceleration can also be addressed by comparing the production
ratios of ions to quasi-relativistic electrons ($E \gapprox
300$~keV). Such an analysis has been performed using both \textit{SMM}/GRS
\citep{1983AIPC..101....3F,1984AIPC..115..641C,1994ApJS...90..511C}
and {\em RHESSI} \citep{2009ApJ...698L.152S} observations. Good
correlations have been found between the total energy content\index{ions!total energy} in
protons above 30~MeV and the total energy content in electrons\index{electrons!total energy}
above 300~keV, suggesting that high-energy electrons and ions are
linked through a common acceleration processes.  It should be
noted, however, that in the statistical study by
\citet{2009AAS...21347506S}, a correlation between the number of
protons with energies above 30~MeV\index{gamma-rays!correlation with hard X-rays}, 
the electron number above
50~keV, and the soft X-ray \textit{Geostationary Operational Environmental Satellite (GOES)} class
exists, but only above a
certain threshold in the level of ion production.
\index{satellites!GOES@\textit{GOES}}
\index{GOES@\textit{GOES} classification}

\subsection{Geometry of hard X-ray and gamma-ray sources} \label{sec:8geometry}
\index{magnetic structures!hard X-ray and $\gamma$-ray sources}

\begin{figure}
\includegraphics[width=1.0\textwidth]{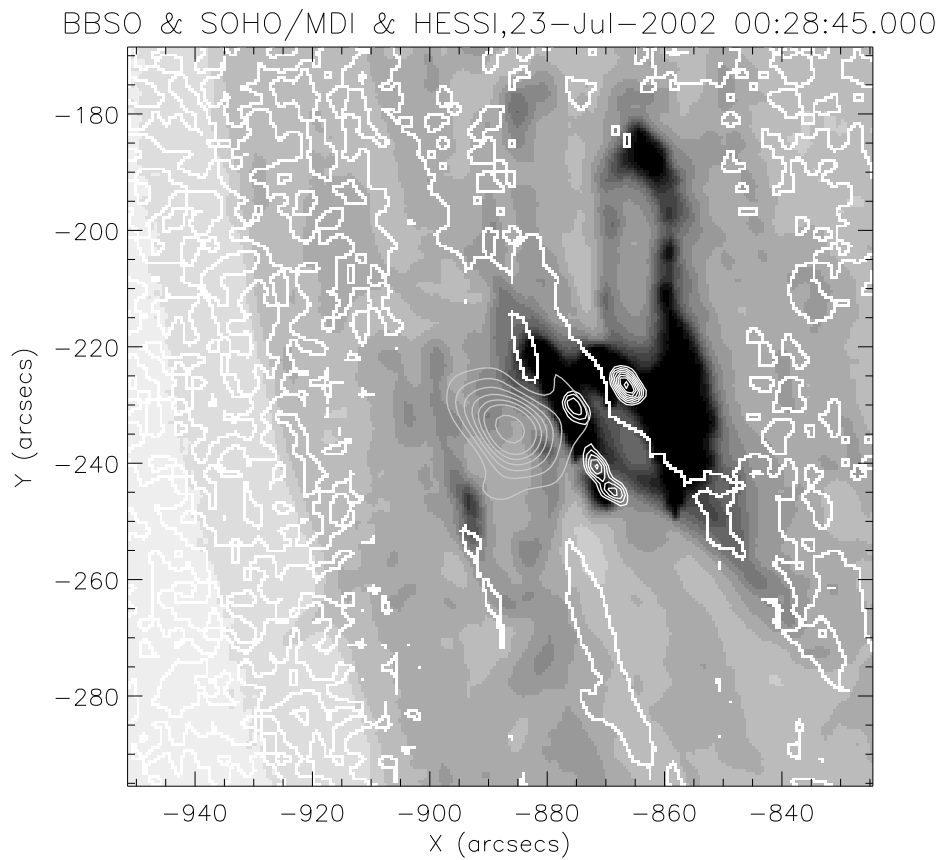}
\caption{{\em RHESSI} hard X-ray images of SOL2002-07-23T00:35 (X4.8)
(white thin contours) 
taken at 00:28~UT, overlaid on Michelson Doppler Imager (MDI)
neutral-line magnetograms (white thick contours) and (negative)
H$\alpha$ taken at 00:28:45~UT. 
One extended (coronal) and three compact (footpoint) hard X-ray sources are
evident. From \citet{2005JGRA..11008104Z}.}
\index{flare (individual)!SOL2002-07-23T00:35 (X4.8)!hard X-rays} 
\index{flare (individual)!SOL2002-07-23T00:35 (X4.8)!H$\alpha$!illustration} 
\index{flare (individual)!SOL2002-07-23T00:35 (X4.8)!illustration} 
\index{SOHO@\textit{SOHO}!MDI}
\index{satellites!SOHO@\textit{SOHO}}
\label{fig:zharkova_hxr_source}
\end{figure}

The imaging capabilities of {\em RHESSI}, combined with its high
spectral resolution, have allowed us to resolve in detail coronal
sources \citep[e.g.,][]{2008A&ARv..16..155K} and footpoints\index{footpoints!and simultaneous coronal sources}
occurring in the same flare (see
Figure~\ref{fig:zharkova_hxr_source}) and hence to study the
acceleration processes that give rise to such sources.

The coronal source\index{coronal sources} 
\index{hard X-rays!coronal sources!correlation with footpoint sources} 
often appears before the main flare hard
X-ray increase and the appearance
\index{hard X-rays!footpoint sources} of footpoints. 
In the impulsive phase, the coronal hard
X-ray emission is, generally, well correlated in both time and
spectrum with the footpoints \citep{2003ApJ...595L.107E,
2006A&A...456..751B}. These observations suggest strong coupling
between the corona and chromosphere during flares, a coupling that
is presumably related to transport of accelerated particles from
one region to the other.

\subsubsection{Differences in footpoint spectral indices} \label{sec:zharkova_sp_index}

An interesting {\em RHESSI} observation is the approximate
equality of the spectral indices 
\index{footpoints!spectral equality}
\index{hard X-rays!spectra!footpoints} 
in different footpoints of the same
loop. \cite{2003ApJ...595L.107E} reported differences $\Delta
\gamma \approx 0.3 - 0.4$ between the spectral indices of the two
dominant footpoints in SOL2002-07-23T00:35 (X4.8). 
\index{flare (individual)!SOL2002-07-23T00:35 (X4.8)!footpoint differences} 
In a few smaller
events analyzed by \cite{2006A&A...456..751B}, $\Delta \gamma$ is
even smaller and indeed is significant only in one out of five
cases. However, other observations of X-class flares reveal a much
stronger difference (up to 5) between the spectral indices of
footpoints of the same loop
\citep{1995PASJ...47..355T,2006A&A...456..751B,
2007ApJ...669L..49K}.

\begin{figure}
\includegraphics[width=1.0\textwidth]{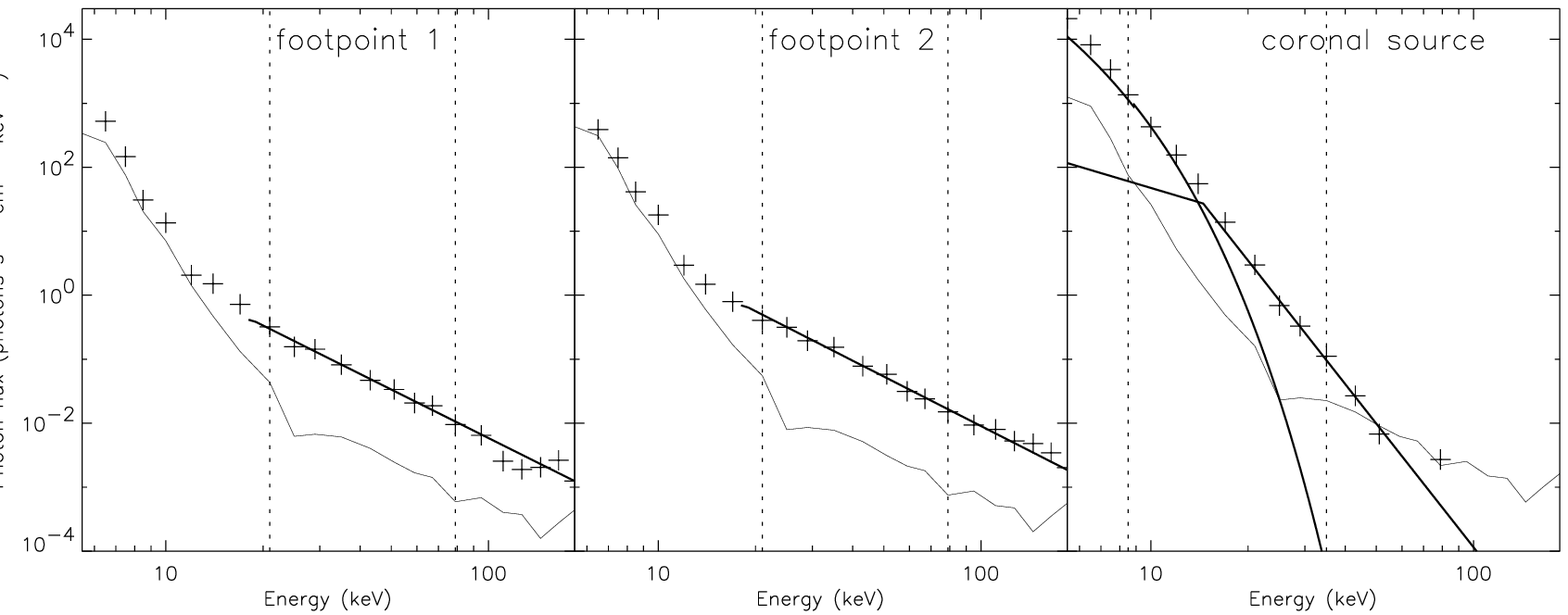}
\caption{Comparison of the spectra of the footpoint and coronal
sources of SOL2005-07-13T14:49 (M5.0). {\it Left and middle
panels:} Footpoint spectra.  A power-law has been fit between
the dotted vertical lines. {\it Right panel:} Spectrum of the
coronal source. A power-law plus an isothermal spectrum has been
fit between the dotted vertical lines. The bold line is the
power-law fit in the dashed interval and a Maxwellian fit for the
coronal source at low-energies. The thin curve is the background
(mostly instrumental) determined at a location outside the
enhanced flare emission. From \citet{2006A&A...456..751B}.}
\label{fig:zharkova_three_sources}
\end{figure}
\index{flare (individual)!SOL2005-07-13T14:49 (M5.0)!illustration}

More surprising than the occasional difference between the
spectral indices of footpoints are the differences (or lack
thereof) between the spectra of the coronal source and the
footpoints\index{hard X-rays!spectra!footpoints vs. coronal
sources}. The spectral index difference $\Delta \gamma = 2.0$
predicted \citep{1971SoPh...18..489B} between thin\index{flare models!thin-target}- and
(collision-dominated) thick-target \index{flare models!thick-target}sources is {\it not} observed
in most cases \citep{Chapter3}.
Differences $\Delta \gamma$
smaller than~2 could in principle be interpreted by invoking a
coronal target that is intermediate between thin- and thick-target
conditions; such spectral differences could also be partially
accounted for by the transport effects of precipitating electrons,
e.g., by the effect of a self-induced electric field
\citep{2006ApJ...651..553Z}. 
\index{electric fields!induced by particles}
However, $\Delta \gamma$ ranges from
0.59 to 3.68 \citep{2006A&A...458..641G, 2007ApJ...669L..49K},
which includes values beyond the capability of simple models to
explain. For example, Figure~\ref{fig:zharkova_three_sources}
\citep[after][]{2006A&A...456..751B} displays the photon energy
spectra of two footpoints and a coronal source, obtained for the
same time interval during the flare\index{flare
(individual)!SOL2005-07-13T14:49 (M5.0)} SOL2005-07-13T14:49 (M5.0). The footpoint
non-thermal power-law indices $\gamma$ are near-identical ($2.7
\pm 0.1$), while the spectral index of the non-thermal emission
for the coronal source is substantially steeper ($\gamma = 5.6 \pm
0.1$).

\subsubsection{Hard X-ray and gamma-ray sources}

Gamma-ray images\index{gamma-rays!imaging} were first obtained 
by {\em RHESSI} for the flare SOL2002-07-23T00:35 (X4.8) \citep{2003ApJ...595L..77H}.
\index{flare (individual)!SOL2002-07-23T00:35 (X4.8)!gamma@$\gamma$-rays} 
They revealed the striking result that the hard X-ray sources are separated by a
considerable distance ($\gapprox 20{^\prime}{^\prime}$) from the
2.223~MeV neutron-capture gamma-ray source. This presumably
reflects differences in the acceleration and/or transport
processes for ions and electrons, respectively \citep{Chapter4}. A
similar, but smaller, separation was also detected for the
``Halloween'' flare SOL2003-10-28T11:10 (X17.2)\index{flare
(individual)!SOL2003-10-28T11:10 (X17.2)!gamma-rays}
\citep{2006ApJ...644L..93H}.

\subsection{Hard X-ray and radio emission} \label{sec:8pre_post_flare}
\index{hard X-rays!and radio emission}

The information obtained through study of radio emission from
solar flares complements that obtained from analysis of hard X-ray
emission in several ways \citep{Chapter5}. Radiation at centimeter
and millimeter wavelengths is gyrosynchrotron
emission\index{radio emission!gyrosynchrotron} originating from
mildly relativistic electrons\index{electrons!mildly relativistic} with energies above about $\sim100$~keV.  
Analysis of this emission hence permits an estimate of
the ambient magnetic field\index{magnetic field!observation} strength, reported to be $\sim$200~G
in the corona \citep{Chapter5}. Coronal sources\index{coronal sources} have been observed
to emit in centimeter radio emission even before X-ray footpoints
appear \citep{2006PASJ...58L...1A}, suggesting that the pre-flare
heating may not be unrelated to acceleration in the main phase.
Comparison of radio and hard X-ray emissions by
\citet{2006AdSpR..38..951B} also revealed, in some events,
coherent radio emission\index{radio emission!coherent} {\it after the hard X-ray emission had
stopped} (this phenomenon was, however, reported in only 5$\%$ of
the flares studied).  
Gyrosynchrotron radio emission in flares in
general correlates well with hard X-rays in {\it time} but not in
{\it space} \citep{2008ApJ...681..644K, Chapter5}; in addition,
the electron spectra deduced from the radio emission are typically
harder than the electron spectra deduced from the hard X-ray
emission. 
Flare emission at decimeter wavelengths concurrent with
hard X-rays has also been found to originate from different
locations: for example, \citet{2009A&A...499L..33B} report the
origin of associated decimetric spike sources\index{radio emission!association with hard X-rays!decimetric spikes}\index{radio emission!decimetric spikes} separated by some
100{$^\prime$}{$^\prime$} from the coronal X-ray source\index{coronal sources!HXR and decimetric}.

Emission at meter and decimeter wavelengths originates not from
individual particles, but from coherent radiation processes
\index{radio emission!coherent}
associated with various plasma waves. \citet{2005SoPh..226..121B}
have studied the association of {\em RHESSI} hard X-rays with
coherent radio emission in the meter and decimeter band in 201
flares with flare classes larger than C5.
\index{radio emission!association with hard X-rays}
They found that coherent
radio emission occurred before the onset of hard X-ray emission in~9\% 
of the flares studied. 
Most of the radio emission was type~III
\index{radio emission!type III burst!harmonic}
\index{frequency!plasma fundamental and harmonic}
radiation at the plasma frequency and its first harmonic, emitted
by a beam of electrons moving upward in the solar atmosphere into
regions of progressively lower density\index{electron beams!and type III bursts}.
In a few cases, they also found a pulsating continuum, possibly caused by cyclotron maser emission.
\index{plasma instabilities!electron-cyclotron maser}
\index{continuum!cyclotron maser}
Both types of emission indicate the presence of electron
acceleration in the pre-flare phase, which is often accompanied by
X-ray emission from a thermal source in the corona \citep{2008A&ARv..16..155K,Chapter2}.
\index{coronal sources}
\index{acceleration!pre-flare phase}

Emission at higher frequencies, such as decimetric narrow-band
spikes, pulsations, and stationary Type~IV events, correlate more
frequently with the hard X-ray flux and thus appear to be more
directly related to the acceleration process
\citep{2005SoPh..231..117A}.
\index{radio emission!pulsations}
\index{radio emission!narrow-band spikes}
\index{radio emission!type IV burst!stationary}
While there is a good {\it association} between coherent radio emission and hard X-rays, a
strong {\it correlation} in the details of the time profile is
less frequent, so that coherent emission is not a reliable proxy
for the main flare energy release\index{radio emission!coherent!and flare energy release}.
This can be accounted for by
invoking multiple reconnection\index{reconnection!multiple sites} sites connected by common field
lines, along which accelerated particles propagate and serve as a
trigger for acceleration in other regions at some distance
\citep{2005SoPh..226..121B}.

\subsection{Magnetic field changes associated with the flare} \label{sec:8bfield}

Sharp temporal increases of hard X-ray emission are often closely
correlated in time with variations of the magnetic field
\index{magnetic field!flare-related changes} measured on the
photosphere \citep[see, e.g.,][]{2001ApJ...550L.105K,
2005ApJ...635..647S,2005ApJ...627.1031W,2005JGRA..11008104Z}. For
example, in the flare SOL2002-07-23T00:35 (X4.8)\index{flare (individual)!SOL2002-07-23T00:35 (X4.8)!flare-related changes}, the
magnetic flux change over the flare duration was about $1.2 \times
10^{21}$~Mx, most of which was concentrated in the area spanned by
the magnetic inversion line\index{magnetic structures!neutral line}.
Such magnetic field changes are
irreversible: the magnetic field reaches a new steady state and
does not return to its pre-flare value \citep{2001ApJ...550L.105K,
2005ApJ...635..647S,2005JGRA..11008104Z}.  
Cross-correlation
analysis with a time lag between the temporal variations of
magnetic field covering the neutral line and the hard X-ray light
curve reveals a significant positive correlation with a time
difference less than $\sim$100~s. 
This observation strengthens
the theoretical notion that irreversible changes in the magnetic
field are responsible for the initiation and development of the
flare phenomenon.

\section{Models of Particle Acceleration} \label{sec:8basic_model}
\index{acceleration!models}
\index{acceleration!list of mechanisms}

In this Section, we first review some basic concepts, such as
acceleration by a direct electric field in a plasma environment
(Section~\ref{sec:8dreicer}) and magnetic reconnection
(Section~\ref{sec:8reconnection}), in which magnetic energy is
released into various forms, including the acceleration of charged
particles, both electrons and ions.  Recognizing the strong
observational evidence for the formation of current sheets
(localized regions in which a strong magnetic shear is present) in
solar flares, our initial discussion relates to traditional
two-dimensional (2-D) current sheets, with subsequent elaboration
to the three-dimensional (3-D) case (Section~\ref{sec:8basic}).
Two likely products of the fundamental reconnection process are
magnetohydrodynamic shocks and stochastic MHD turbulence, and
particle acceleration by these agents is discussed in
Section~\ref{sec:8stochastic}.

\subsection{Basic physics} \label{sec:8dreicer}

As remarked in Section~\ref{sec:8introduction}, the essential
physics of particle acceleration is in principle rather
straightforward -- charged particles gain energy when subjected to
an electric field in their rest frame.  Such an electric field
${\bf E}$ may be a large-scale externally-imposed field, or a
${\bf v} \times {\bf B}$ field associated with particles crossing
magnetic field lines, or a collective field associated with the
environment in which the particle finds itself (e.g., a
collisional Coulomb field or a field associated with a level of
plasma wave energy).  The richness of these various means\index{acceleration!sources of electric field} of
creating local electric fields is reflected in the richness of
particle acceleration models in magnetized plasmas.

One of the most basic concepts\index{Dreicer field}\index{electric fields!Dreicer} is that of the
{\it Dreicer field} \citep{1974itpp.book.....C}\index{acceleration!in Dreicer field}.
Consider an
electron subject to an externally-imposed large-scale electric
field\index{electric fields!large-scale} $E$, plus the frictional force due to collisions \index{frequency!collision!and runaway} with stationary ambient particles.  
The equation of motion for such a particle is (in one dimension)
\begin{equation}\label{eqn:zharkova_dreicereq1}
{dv \over dt} = e \, E/m - f_c \, v,
\end{equation}
where $m$ and $e$ are the electron mass and (absolute value of)
charge, $f_c$ is the collision frequency, and $v$ is the velocity. 
Since for Coulomb collisions $f_c \sim v^{-3}$ \citep[e.g.,][]{1974itpp.book.....C},
we may write this as
\begin{equation}\label{eqn:zharkova_dreicereq2}
m {dv \over dt} = e \left ( E - E_D \left [{v_{\rm th} \over v}
\right ]^2 \right ),
\end{equation}
where $v_{\rm th}$ is the electron thermal speed and we have
defined the {\it Dreicer field} by
\index{Dreicer field}
\begin{equation}\label{eqn:zharkova_dreicereq3}
E_D = {f_c(v_{\rm th}) \, v_{\rm th} \over e};
\end{equation}
numerically, $E_D \simeq
10^{-8}$~$n$(cm$^{-3}$)/$T$(K)~V~cm$^{-1}$.

If $E < E_D$, then only high-velocity particles ($v > v_{\rm crit}
= v_{\rm th} [E_D/E]^{1/2}$) see a net positive force and gain
velocity. This increased velocity reduces the drag force, leading
to an even greater net acceleration and eventual runaway
acceleration\index{acceleration!conditions for runaway}.
In contrast, particles with $v < v_{\rm crit}$ suffer a net {\it
retardation} force and are not accelerated; they remain part of
the collisionally-redistributed Maxwellian core\index{electrons!distribution function!Maxwellian}. 
For a Maxwellian
distribution of velocities, the number density of accelerated
particles $\int_{v_{\rm crit}}^\infty f(v) \, dv$ (where $f(v)$ is
the phase-space distribution function) is a (rapidly) increasing
function of $E$. If $E > E_D$, all particles with velocity in
excess of the thermal speed $v_{\rm th}$ are accelerated, and the
efficient collisional re-population of the accelerated electrons
thus leads effectively to acceleration of {\it all} the electrons
in the distribution.

\subsection{Magnetic reconnection models associated with flares} \label{sec:8reconnection}
\index{reconnection}
\index{flare models!reconnection}

It is generally accepted that the energy release in solar flares
occurs through a reconstruction of the magnetic field, caused by
the change of connectivity of magnetic field lines during a
magnetic {\it reconnection}.  The electric field associated with
this changing magnetic field and/or with the associated driven
currents leads to particle acceleration\index{reconnection!and particle acceleration}\index{acceleration!and magnetic reconnection}.

Reconnection is a fundamental process defined by the
magnetohydrodynamics of a magnetized plasma
\citep[see][]{2000mare.book.....P}. The\index{magnetic diffusion
equation} magnetic diffusion equation \citep[see, e.g., equation
(3.91) of][]{1988psf..book.....T} is
\begin{equation}
{\partial {\bf B} \over \partial t} = \nabla \times ({\bf v}
\times {\bf B}) + {\eta c^2 \over 4 \pi} \, \nabla^2{\bf
B},\label{eqn:zharkova_diffusion}
\end{equation}
where $c$ is the speed of light, ${\bf B}$ is the magnetic field,
${\bf v}$ the fluid velocity and $\eta$ the resistivity.  The
order-of-magnitude ratio \index{magnetic Reynolds number} of the
terms on the right side of this equation defines the {\it magnetic
Reynolds number}
\begin{equation}
R_m = {4 \pi \ell V \over \eta c^2};\label{eqn:zharkova_rm}
\end{equation}
it measures the ratio of the advective to diffusive contributions
to the change in the magnetic field.  For typical solar coronal
values of $\eta$ and with $\ell \approx 10^9$~cm, $R_m \approx 10^{14}$;
for such high magnetic Reynolds numbers, the plasma is effectively
``frozen-in'' to the field 
\index{frozen-in field} 
and negligible change in the field topology, with its concomitant release of
magnetic energy, can occur. A topological change in the magnetic
field requires a breakdown in this ideal ``frozen-in'' flux
condition; by Equation~\ref{eqn:zharkova_rm}, this can occur
either in small-scale regions (high values of $\nabla^2 {\bf B}$)
and/or in regions in which the resistivity $\eta$ is anomalously
enhanced. As a result, magnetic reconnection fundamentally occurs
in narrow boundary layers called {\it diffusion regions}\index{diffusion region}.

\subsubsection{Basic 2-D MHD theory of magnetic reconnection}
\index{reconnection!2-D theory}

\begin{figure}
\centering
\includegraphics[width=0.8\textwidth]{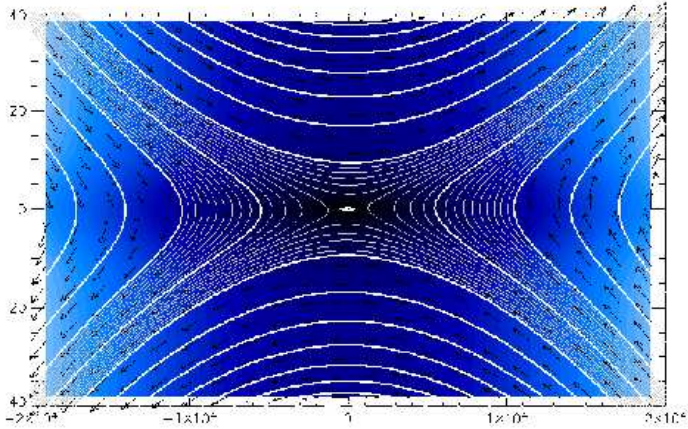}
\caption{Topology of the reconnecting magnetic field lines in the
vicinity of an X-null point $B_y = \pm \alpha B_x$.}
\label{fig:zharkova_dif_reg}
\end{figure}

In the simple neutral sheet geometry originally proposed by Sweet\index{Sweet, P. A.}
\index{reconnection!Sweet-Parker} 
and Parker\index{Parker, E. N.} \citep{1969ARA&A...7..149S},
oppositely-directed magnetic field lines ${\bf B}$ in close
proximity follow a plasma inflow ${\bf V}_i$ oriented in the
$x$-direction perpendicular to the field lines -- see Figure~\ref{fig:zharkova_dif_reg},
where the $x$-direction is vertical and the $y$-direction horizontal
\citep[note that there does
exist evidence for such reconnection geometries in flares
--][]{2000ApJ...540.1126A, 2001ApJ...554..451F,
2003ApJ...596L.251S,2009ApJ...693.1628D}. 
The increasing magnetic
pressure in the localized region of field reversal is alleviated
by reconnection in a small region of high $\nabla^2 {\bf B}$ near
the origin, which allows plasma outflows with velocity
$V_o$ toward the sides of the diffusion region\index{reconnection!2-D theory}.

\index{magnetic structures!null!illustration}

The original analysis by \citet{1969ARA&A...7..149S} showed that the reconnection
rate $V_i$ is given by
\begin{equation}
{V_i \over V_A} = {2 \over \sqrt{R_m}} \, ,
\end{equation}
where $V_A$ is the Alfv\'en velocity.  Because of the $R_m^{-1/2}$ dependence on the inflow velocity, and
hence on the rate of magnetic energy dissipation, such a neutral
sheet scenario describes a ``slow'' magnetic reconnection
regime. 
\index{reconnection!slow} 
\index{reconnection!Petschek}
In an effort to increase the
reconnection rate, \citet{1964NASSP..50..425P} proposed a model
involving the formation of shocks propagating outward from the
diffusion region; in this scenario, 
the reconnection process occurs at a significantly faster rate, with
$V_i \approx V_A$.  Both these models were originally proposed with a
classical value for the resistivity $\eta$ and both resulted in
reconnection rates too slow to explain the rapid energy release
observed in flares. 
However, later realization that the
resistivity in the reconnection region may be driven to
anomalously high values \index{resistivity!anomalous} through the
presence of plasma instabilities in the current sheet
\citep[see, for example,][and references therein] {Dahlburg1992, Bret2009}.
These include the lower hybrid drift instability\index{plasma instabilities!lower hybrid drift} leading to the Kelvin-Helmholtz
instability\index{plasma instabilities!Kelvin-Helmholtz} \citep{1960ecm..book.....L,Chen1997, Lapenta2003}, the kink instability\index{plasma instabilities!kink}
\citep{Pritchett1996, Zhu1996, Daughton1999}, the sausage instability\index{plasma instabilities!sausage} \citep{Buchner1999}
and magneto-acoustic waves\index{waves!magneto-acoustic} \citep{2007mhet.book...85S}.  
This realization revived hopes that reconnection mechanisms could indeed be a viable candidate for the energy release
in flares \citep{1981str..book...21S, 1982QB539.M23P74..., 2000ASSL..251.....S}.
Questions remaining about the stability of current sheet geometries such as those envisioned
by Sweet and Parker have been extensively
investigated using a full kinetic approach \citep [see][and
references therein]{Daughton2009}, and are discussed further in
Section~\ref{sec:8pic}.

Another simple 2-D diffusion region geometry involves magnetic
field lines of the form $B_y^2 = \alpha ^2 B_x^2$. Such a form is
associated with two types of neutral point:\index{magnetic structures!islands!neutral points} an O-type, elliptic
topology ($\alpha^2 < 0$; see Section~\ref{sec:8island}), and an
X-type, hyperbolic topology 
\index{magnetic structures!null!O-type (elliptic)} 
\index{magnetic structures!null!X-type (hyperbolic)} 
($\alpha^2 > 0$; see Section~\ref{sec:8basic};
Figure~\ref{fig:zharkova_dif_reg}).

\subsubsection{Effect of the Hall current on reconnection rates in 2-D models} \label{hall}
\index{reconnection!Hall}

In the conductive plasma within the diffusive region where
reconnection occurs, the current sheet tends to diffuse outwards
at a slow rate with a timescale of $\tau_{\rm dif}=a^2 \varpi$
where $a$ is the current sheet half-width and $\varpi$ is the magnetic diffusivity
\citep{2000mare.book.....P}. This diffusion process ohmically
converts magnetic energy into heat. As discussed in Section~\ref{sec:8basic}, 
this timescale is too slow to account for the
fast reconnection rates observed in the impulsive phase of flares
unless some other processes, such as resistive instabilities\index{plasma instabilities!resistive} or
Hall current effects, are invoked to enhance the reconnection
process\index{reconnection!Hall}.
The effects of the Hall current in MHD models of the
reconnection process are now discussed. Its effects in
particle-in-cell (PIC) models\index{simulations!PIC} will be discussed in Section
\ref{sec:8pic}, while the effect of resistive instabilities will
be discussed in Section \ref{sec:8island}.

The basic Sweet-Parker 2-D reconnection model discussed above can
produce more realistic reconnection rates if, instead of a
resistive MHD model, one uses a two-fluid MHD model, which
includes in the equation of motion a Hall current
\citep{Sonnerup1979}.  
\index{reconnection!Hall}
\index{reconnection!two-fluid}
Use of a two-fluid model means that the
velocity vector {\bf v} in Equation~\ref{eqn:zharkova_diffusion}
now represents the velocity ${\bf V}$ of the plasma as a whole, with an
additional vector ${\bf V}_H$ representing the relative velocity
between the electrons and the ions. The {\it Hall current} results
from the fact that, inside the current sheet, the electrons and ions
are magnetized\index{magnetization!differential} to significantly different degrees -- the spatial
scales for effective magnetization is $c/\omega_{pe}$ for
electrons and $c/\omega_{pi}$ for protons. (Here $\omega_{pe}$ and
$\omega_{pi}$ are the respective plasma frequencies, which are in
the ratio of the square root of the proton-to-electron mass ratio
$\sqrt{m_p/m_e} \simeq 43$.)

At length scales between $c/\omega_{pe}$ and $c/\omega_{pi}$,
the motions of the electrons and ions are decoupled, and the current
generated from the relative motion of electrons and ions can lead
to the formation of whistler waves
\citep{Mandt1994,Ma1996,Drake1997,Biskamp1997}. 
\index{waves!whistlers}
Inside such a region, the reconnection rate becomes insensitive to electron
inertia, allowing the reconnection rates to be strongly increased
to levels large enough to account for those measured in solar
flares \citep{2001JGR...106.3715B,Huba2004}. These results
indicate a compelling need to consider a full kinetic approach\index{simulations!kinetic} for
modeling magnetic reconnection processes in the diffusion region
\citep[for details, see][and references
therein]{2001JGR...106.3715B}; this will be discussed in more
detail in Sections~\ref{sec:8pic} and~\ref{sec:8island}.

\subsubsection{Further improvement of reconnection models}

\begin{figure}
\includegraphics[width=0.8\textwidth]{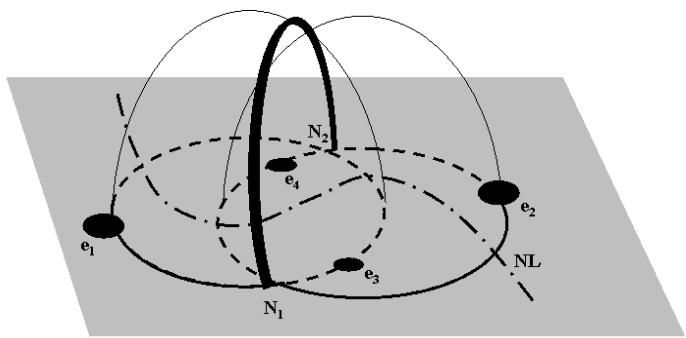}
\caption{The magnetic field geometry formed by four point-like
sources of different magnetic polarity.  The dot-dashed line
labelled `NL' is the neutral line. The {\it separatrix} surfaces
are shown as thin dome-like lines connecting the photospheric
sources. The intersection of these surfaces is the {\it
separator}, shown as a solid line. From
\citet{2002AdSpR..29.1087T}.} \label{fig:zharkova_3d_rec}
\end{figure}

It must be recognized that magnetic reconnection is inherently a
3-D process, and that the addition of the third dimension
inherently introduces fundamentally new physics. (As an elementary
example, 2-D models\index{reconnection!2-D vs. 3-D} may predict
unrealistic energy gains since the induced electric field ${\bf E}
= ({\bf V}/c) \times {\bf B}$ is infinite in extent in the
invariant direction). In three dimensions, the magnetic field has
a considerably more complicated structure
(Figure~\ref{fig:zharkova_3d_rec}), involving {\it separatrices}\index{separatrix}\index{magnetic structures!separatrix}
and {\it separators}\index{magnetic structures!separator} \citep[surfaces separating different domains
of magnetic connectivity, and lines of intersection of separatrix
surfaces, respectively;
see][]{1995JGR...10023443P,1996JGR...101.7631D,1996A&A...308..643D}.
As shown by \citet{1995JGR...10023443P}, magnetic null points are
no longer required for reconnection to take place, and, even when
they do exist, magnetic nulls\index{magnetic structures!null} have a more complicated structure,
typically involving\index{magnetic structures!fan}\index{reconnection!spine}\index{magnetic structures!spine}\index{reconnection!fan}
a ``fan'' structure oriented around an axial ``spine'' \citep{1997GApFD..84..127P}.

\index{magnetic structures!neutral line!illustration}

In summary, reconnection can proceed in a variety of different
magnetic field geometries and topologies
\citep{2002A&ARv..10..313P}.  Furthermore, depending on the
magnetic field configuration, reconnection can be accompanied by a
variety of other physical processes such as generation of plasma
waves, turbulence and shocks, each occurring on different temporal
and spatial scales. These additional processes not only directly
accelerate particles (e.g., in regions of enhanced magnetic field,
which act as moving ``mirrors''), but can also feed back on the
magnetic field topology and hence the ongoing reconnection rate.\index{magnetic structures!mirror geometry}

\subsection{Particle acceleration in a reconnecting current sheet} \label{sec:8basic}
\index{current sheets!and particle acceleration}

\begin{figure}
\begin{center}
\includegraphics[width=0.8\textwidth]{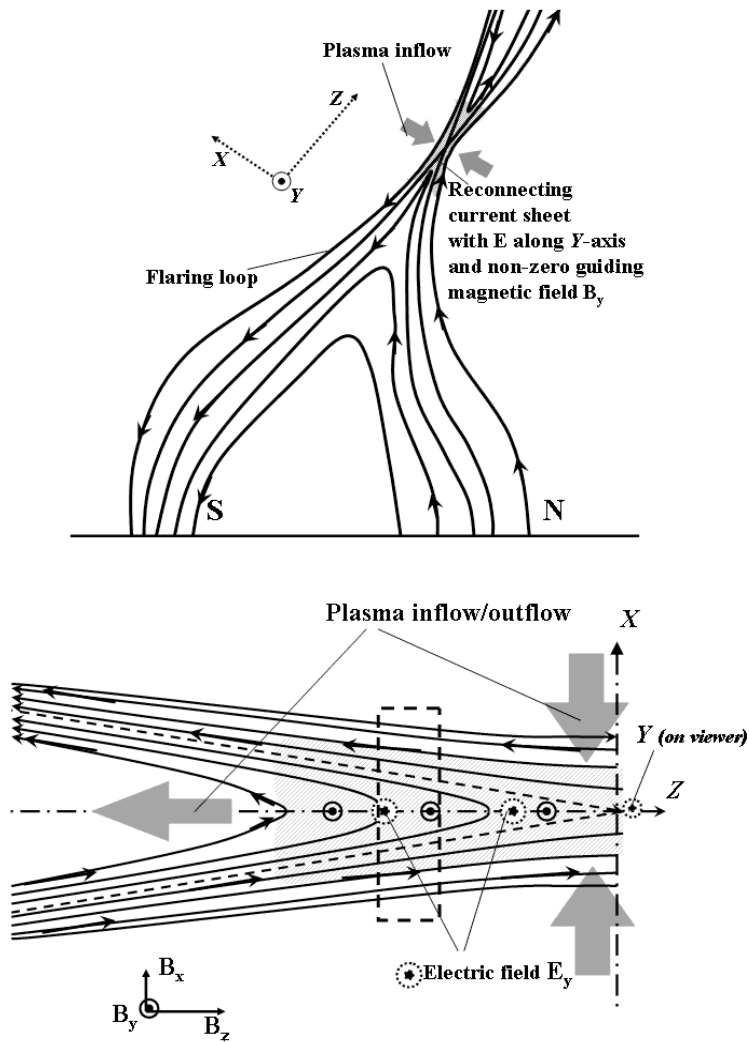}
\end{center}
\caption{\textit{Top:} current-sheet location and \textit{bottom} magnetic-field topology
in a single helmet-type reconnecting non-neutral current
sheet.
\index{current sheets!non-neutral} 
The large arrows depict plasma inflows from the top and the
bottom and outflows to the side. The $x$- and $z$-components of
the magnetic field lie in the plane of the Figure; a guiding
magnetic-field component $B_y$ and a drift electric field $E_y$
are perpendicular to the plane.} \label{fig:zharkova_helmet}
\end{figure}
\index{current sheets!non-neutral!illustration}

\begin{figure}
\begin{center}
\includegraphics[width=0.8\textwidth]{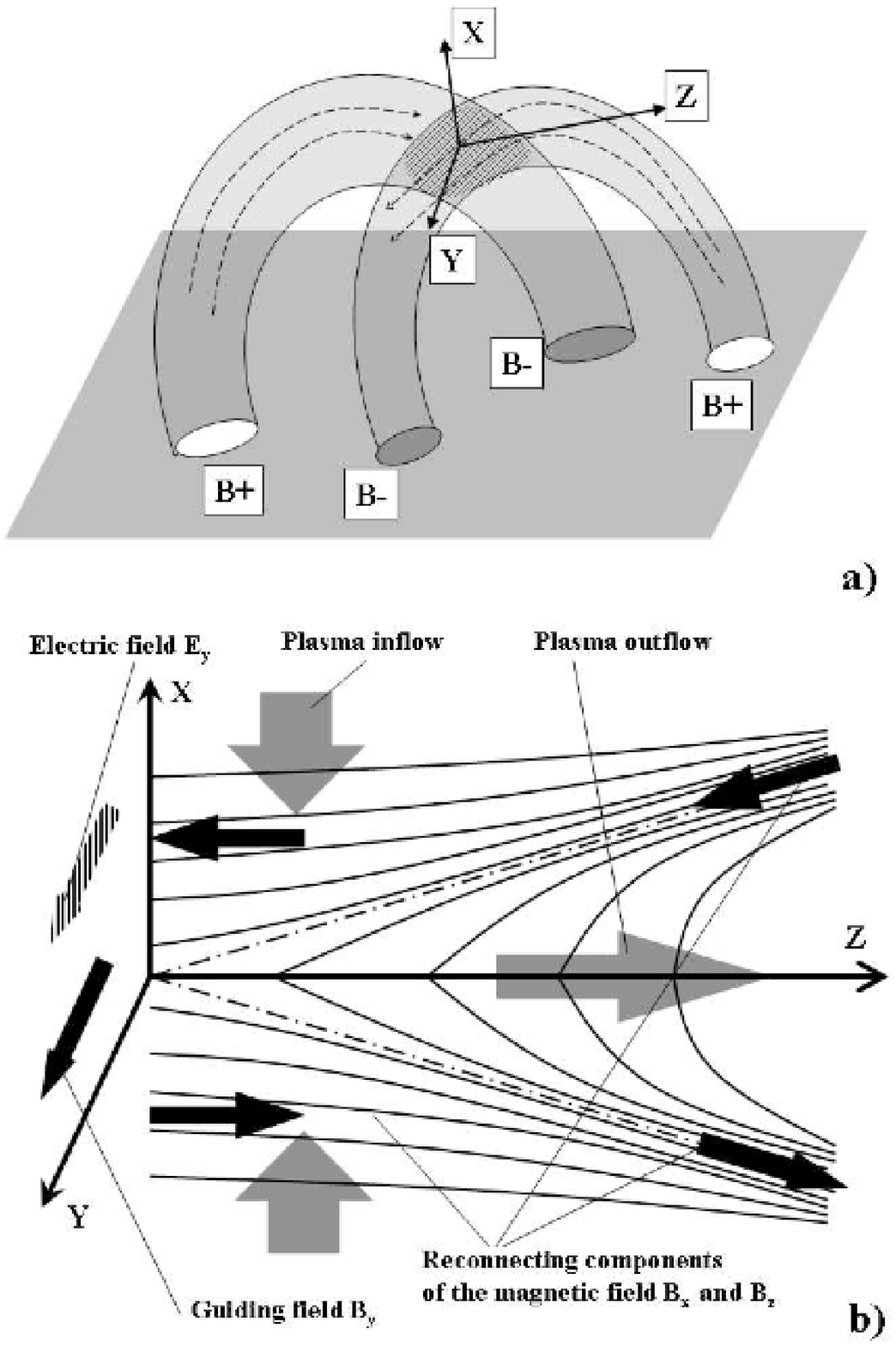}
\end{center} 
\caption{\textit{Top:} current sheet location, and \textit{bottom:}
magnetic-field topology in a reconnecting non-neutral
current sheet\index{current sheets!non-neutral} formed by two interacting loops.
\index{magnetic structures!interacting loops}
The large arrows depict the plasma inflows. As for
Figure~\ref{fig:zharkova_helmet}, the $x$- and $z$-components of
the magnetic field lie in the plane of the Figure; a guiding
magnetic field component $B_y$ and a drift electric field $E_y$
are perpendicular to the plane.} \label{fig:zharkova_twoloops}
\end{figure}
\index{current sheets!non-neutral!illustration}

Two of the most popular magnetic field topologies associated with
solar flares involve the formation of a current sheet
\citep{Gorbachev1989,2003ApJ...596L.251S}, located either at the
top of single helmet-like 
\index{magnetic structures!helmet} 
loop (Figure~\ref{fig:zharkova_helmet}) or at the intersection of two
interacting loops 
\index{magnetic structures!interacting loops}
(Figure~\ref{fig:zharkova_twoloops}). 
Other more complicated topologies are essentially combinations of these two basic ones,
and they are discussed in Section~\ref{sec:8developments}. In this
Section we therefore briefly overview the basic physics of
particle acceleration in reconnecting current sheets.

Let us, then, consider the magnetic field topology as presented in
the reconnecting current sheet\index{current sheets} 
(RCS) geometries illustrated in bottom plots of
Figures~\ref{fig:zharkova_helmet} and~\ref{fig:zharkova_twoloops}.
The tangential magnetic field component $B_z(x)$ is approximated
by the linear function
\begin{equation}\label{eqn:zharkova_bz_x}
B_z(x) = B_{0} \, \left ( - \frac x d
\right ),
\end{equation}
where $d$ is the RCS thickness \citep{1990ApJS...73..333M,
Litvinenko93}. The width $a$ of an RCS in the z-direction approaches
$10^2$ - $10^3$ m \citep{Somov1992} and magnitudes of the other two magnetic
field components $B_x$ and $B_y$ are assumed to be constant:
\begin{equation}\label{eqn:zharkova_bx_y}
\vert B_x \vert = B_{x0}, \qquad \vert B_y \vert = B_{y0}.
\end{equation}
Particles, effectively frozen into the
field lines, are advected toward the RCS plane (the $x$-axis) by
the flow field ${\bf V_i}$, where they are accelerated by the
$({\bf V_i}/c) \times {\bf B}$ electric field \citep{Bulanov1976,Litvinenko93,
Litvinenko96}\index{current sheets!electric fields}\index{electric fields!in current sheets}.\index{frozen-in field}
As the reconnection process continues, ambient
particles drift into the reconnecting current sheet; the rate of
acceleration of particles is limited by the size of the thin
diffusion region.

The induced electric field $E_y$\index{electric fields!reconnection!convective} is given by
\citep{Bulanov1976,1990ApJS...73..333M,Litvinenko93,2004ApJ...604..884Z}:
\begin{equation}
E_y = {V_{i}B_{z0} \over c} - {\eta c \over 4 \pi} \,
\frac{\partial B_z}{\partial x} \, ;
\label{eqn:zharkova_edrift}
\end{equation}
the first and second terms on the right hand side
of Equation~\ref{eqn:zharkova_edrift} dominate outside and inside the
diffusion region, respectively \citep{2000ASSL..251.....S}.
\cite{1982QB539.M23P74...} has shown that the plasma inflow
velocity $V_i$ is smaller than the local Alfv\'{e}n speed by about
two orders of magnitude, i.e., $V_i \approx 0.01 V_A$.  
\index{Alfv{\' e}n speed}
Hence the electric field is
\begin{equation}
E_{y} \approx  0.01 \, {B_{z0}^2 \over c \, \sqrt{4 \pi n \, m_p}} \,
, \label{eqn:zharkova_edriftin}
\end{equation}
where $m_p$ is the proton mass. Let us assume the following
physical conditions in the current sheet\index{current sheets!physical conditions}: temperature $T =
10^6$~K, density $n = 10^{10}$~cm$^{-3}$, and $B_{z0} \approx
100$~G. Then $V_A \approx 2 \times 10^6$~m~s$^{-1}$ and the thermal
velocities are about $10^5$~m~s$^{-1}$ for protons and
$10^7$~m~s$^{-1}$ for electrons. Taking the initial inflow
velocity $V_i$ to be the proton thermal speed, we obtain $E_y
\approx 2$~V~cm$^{-1}$, which is roughly four orders of magnitude
higher than the corresponding Dreicer field \index{Dreicer field}
$\sim$10$^{-4}$~V~cm$^{-1}$ \citep[][see
Section~\ref{sec:8dreicer}]{1990ApJS...73..333M,Litvinenko96}.
This electric field is directed along the longitudinal magnetic
field component $B_y$, allowing efficient particle acceleration in
the $y$-direction.

In these simulations the drifted, or reconnection, electric field is assumed
to exist throughout the entire diffusion region, which in the MHD approach extends
over some $10^2$~m to $10^6$~m \citep{Bulanov1976,Somov1992,Oreshina2000}.
\index{electric fields!MHD current-sheet assumption}
\index{simulations!PIC}
This assumption has been challenged by full kinetic  PIC simulations
\citep{1963PhFl....6..459F,Drake1977}, which show the reconnection field
to extend over a relatively small domain corresponding to the size of
the electron diffusion region\index{diffusion region}. 
Given the current limitations of PIC simulations (see Section~\ref{sec:8pic_limit}),
there is at present no definitive indication of which of the approaches (MHD or PIC) is more valid.
In the remainder of this Section we adopt the value provided by the MHD approach, recognizing that appropriate caveats must therefore be made\index{caveats!limits of MHD approach}.

The magnitudes of the magnetic field components $B_x$ and $B_y$
are chosen such that electrons remain magnetized in the vicinity
of the midplane while protons are unmagnetized during the
acceleration phase\index{magnetization!differential}. 
In this case, the energy of the unmagnetized
protons upon ejection can be estimated
\citep[see][]{Litvinenko96} as:
\begin{equation} \label{eqn:zharkova_enrg1}
\varepsilon = 2 m_{\alpha}c^2 \left(\frac{E_{y0}}{B_{x0}}\right)^2,
\end{equation}
where $m_{\alpha}$ is the particle mass.  However, since the
electrons are magnetized, they follow magnetic field lines and
gain an energy \citep{Litvinenko96}:
\begin{equation} \label{eqn:zharkova_enrg2}
\varepsilon = d \left| \, e E_{y0} \frac{B_{y0}}{B_{x0}} \,
\right|.
\end{equation}
where $d$ is the current sheet thickness along the $x$-direction
(see the bottom plot in Figure \ref{fig:zharkova_helmet}). 
\index{current sheets!and proton Larmor radius}
For RCSs
with thicknesses from 1~m to 100~m, corresponding to
proton Larmor radii in magnetic fields from $10^{-3}$~T to $2\times10^{-2}$~T,
electrons gain energies up to $\sim$100~keV, comparable with those of
bremsstrahlung-emitting electrons.

Although such a model does meet the challenge of (1) a
sufficiently high reconnection rate and (2) sufficient energy gain
by the accelerated electrons, the number of accelerated particles
is fundamentally limited by the small width $2a$ of the current
sheet\index{reconnection!current sheet!number problem}. 
These deficiencies are addressed in the further development
of this basic model to include three-dimensional considerations
(Section \ref{sec:8single3D}), the use of multiple current sheets\index{reconnection!multiple current sheets}
formed by the filamentation\index{magnetic structures!filamentation} of the reconnection region (Section
\ref{sec:8complex}), and a full kinetic  PIC
approach to model portions of a 3-D current sheet (Sections
\ref{sec:8pic} and \ref{sec:8island}).
\index{simulations!PIC}

\subsection{Particle acceleration by shocks and turbulence}  \label{sec:8stochastic}

In this Section, we review the basic physics involved in
acceleration models that do not rely on a large-scale coherent
electric field.

\subsubsection{Shock acceleration} \label{sec:zharkova_shock}
\index{shocks!particle acceleration}\index{acceleration!shock}

Ever since the demonstration by several authors
\citep{Blandford1978,Bell1978,1991ApJ...376..771C,1996JGR...10111095G}
that a very simple model of shock wave acceleration
\index{acceleration!shock} leads to a power-law spectrum of the
accelerated particles (in good agreement with observations of
galactic cosmic rays)\index{cosmic rays!shock acceleration}, 
shock acceleration has been frequently
invoked in both space and astrophysical plasma contexts. However,
this simple model, though very elegant, has a significant number
of shortcomings. Inclusion of losses (Coulomb at low energies and
synchrotron at high energies) or the influence of accelerated
particles on the shock structure \citep[see,
e.g.,][]{2005MNRAS.364L..76A,2005ICRC....3..261E} can cause
significant deviations from a simple power law.

Moreover, shocks are unable to accelerate low energy background
particles efficiently; shock acceleration requires the injection
of (fairly high energy) ``seed'' particles\index{acceleration!shock!seed@``seed'' particles}.  
Also, shock acceleration requires some scattering agents (most likely plasma
waves or turbulence) to cause repeated passage of the particles
through the shock front. In the case of quasi-parallel
shocks\index{acceleration!shock!quasi-parallel}, in which the
magnetic field is nearly parallel to the normal of the shock
front, the rate of energy gain is governed by the pitch-angle
diffusion rate.
\index{pitch-angle diffusion}\index{scattering!pitch-angle}

As pointed out by \cite{1983GeoRL..10..545W} and
\cite{1984AnGeo...2..449L}, in the Earth's magnetosphere, many
shocks can be quasi-perpendicular (i.e., the vector defining the
normal to the shock is almost perpendicular to the shock velocity
vector).\index{acceleration!shock!quasi-perpendicular}\index{magnetosphere!shocks}
In such a
case the particles drift primarily along the surface of the shock
and get accelerated more efficiently. However, acceleration in
quasi-perpendicular shocks suffers from the fundamental limitation
that injection of seed particles with very large velocities $v >
u_{\rm sh}\xi$ (where $u_{\rm sh}$ is the shock speed and $\xi \gg
1$ is the ratio of parallel to perpendicular diffusion
coefficients) is required \citep{2010PhRvL.104r1102A}. Moreover,
although there may be some {\it indirect} morphological evidence
for the existence of such shocks in flares \citep[see,
e.g.,][]{2003ApJ...596L.251S}, there is little {\it direct}
evidence for the occurrence of these kinds of shocks near the top
of flaring loop, where the acceleration seems to be taking place
during the impulsive phase.  
Furthermore, such shocks typically do
not appear in 3-D MHD simulations\index{simulations!MHD} of the reconnection process.

In summary, many of the features that make acceleration of cosmic
rays by shocks attractive are not directly applicable to the solar
flare environment. As a result, this mechanism has, with some
justification, received only limited attention in the literature.
However, in Section~\ref{sec:8termin_shock} we
do discuss the possibility of acceleration at fast magnetosonic 
shocks resulting from reconnection.

\subsubsection {Stochastic Acceleration} \label{sec:zharkova_stochastic}
\index{acceleration!stochastic}

A main contender for the particle acceleration mechanism in flares
is stochastic acceleration by the magnetic field\index{magnetic field!wave} component of
low-frequency magneto-acoustic waves\index{waves!magneto-acoustic!and particle acceleration} \citep{1997JGR...10214631M,
1998ApJ...492..352S,2006ApJ...644..603P}.

In the originally proposed Fermi acceleration
\index{acceleration!Fermi!first order} process
\citep{1949PhRv...75.1169F, 1956PhRv..101..351D}, particles of
velocity $v$ scatter coherently off agents moving toward each
other with a velocity $\pm u$. Since the velocity gain in a single
collision is $2u$, the rate of velocity increase is $\Delta
v/\Delta t \approx 2u/(L/v)$, where $L$ is the separation of the
scattering centers.\index{scattering!in Fermi acceleration}
Solving this, for the case in which a scatter
occurs on a timescale significantly less than that required for
$L$ to decrease, shows that the energy grows exponentially with
time: $E \propto e^{4ut/L}$. For particles scattering off {\it
randomly} moving agents, both energy gains and losses are
possible. However, the increased likelihood of a head-on
(energy-increasing) collision leads to a general (slower than
exponential) increase in energy, at a rate proportional to
$(u/v)^2 \, D_{\mu\mu}$, where $D_{\mu\mu}$ is the angular
diffusion coefficient.  This process is known as {\it
second-order} \index{acceleration!Fermi!second order} Fermi
acceleration.

In general, in a magnetized plasma characterized by a
gyrofrequency\index{frequency!Larmor} $\Omega$, a particle of velocity ${\bf v}$ will
resonate with a wave of frequency $\omega$ and wavenumber ${\bf
k}$ when
\begin{equation} \omega = {\bf k} \cdot {\bf v} - \ell
\Omega \, ,
\end{equation}
where $\ell$ is an integer.  Particles with velocities near the
Cherenkov resonance \index{Cherenkov resonance} ($v \approx
\omega/k_\parallel$) resonate with $\ell = 0$ waves. This process
is known as {\it transit-time damping}
\index{acceleration!transit-time damping} \citep[][ and references
therein]{1974JGR....79.1357H}. It acts to diffuse the particle
distribution $f({\bf p})$ in momentum space, described by the
Fokker-Planck equation \index{Fokker-Planck equation}
\begin{equation}
{\partial f({\bf p})\over \partial t} = \left( {1\over
2}\sum_{i,j} {\partial \over \partial p_i \partial p_j} D_{ij} -
\sum_i{\partial \over \partial p_i} F_{i} \right) f({\bf p}) +
Q({\bf p}) - S({\bf p}) f({\bf p}) \, .
\label{eqn:zharkova_diff_equ}
\end{equation}
The diffusion coefficient $D_{ij}$ \index{diffusion
coefficient!energy} \index{diffusion coefficient!angular} and the
secular coefficient $ F_{i}$ ($i = 1,2$ correspond to energy and
angular terms, respectively) contain the essential physics: the
action of both the accelerating waves and decelerating effects
(e.g., Coulomb collisions); the quantity $Q({\bf p})$ corresponds
to the injection of particle energy and the term $S({\bf p})f({\bf
p})$ to escape.

For solar flares, the most likely sources for the energization term
$Q({\bf p})$ are plasma wave turbulence\index{turbulence!plasma
wave} \citep{1992ApJ...398..350H, 1997ApJ...482..774P} and
cascading MHD turbulence\index{turbulence!MHD} \citep{1996ApJ...461..445M}. 
\index{acceleration!in MHD turbulence}
In the first
scenario, a presupposed level of plasma wave turbulence in the
ambient plasma accelerates ambient particles of energy $E$
stochastically with the rate $D_{EE}/E^2$. It has been shown
\citep[e.g.,][] {1992ApJ...398..350H,1996ApJ...461..445M} that for
flare conditions, plasma-wave turbulence can accelerate the
background particles to the required high energies within the very
short timescale derived from hard X-ray observations. More
importantly, at low energies, and especially in strongly
magnetized plasmas, the acceleration rate $D_{EE}/E^2$ of the
ambient electrons may significantly exceed the scattering rate
$D_{\mu\mu}$ for particular pitch angles and directions of
propagation of this turbulence \citep{1997ApJ...482..774P,
1998ApJ...495..377P}, leading to an efficient acceleration.\index{scattering!pitch-angle}
However, high-frequency waves near the plasma frequency\index{frequency!plasma} are
problematic as drivers for\index{acceleration!stochastic}
stochastic acceleration in flares, because they would couple into
decimeter radio waves and would be present at high levels in every
flare, contrary to observations \citep{2005SoPh..226..121B}.

In the second scenario, involving cascading MHD
turbulence\index{turbulence!MHD}, moving magnetic compressions
associated with fast-mode MHD waves propagating at an angle to the
magnetic field constitute the scattering centers. The wavelength
of the induced MHD turbulence is assumed to cascade through an
inertial range from large to very small, stopping at the point
where wave dissipation becomes significant due to transit-time
damping.
\index{transit-time damping} 
Resonance with one wave will result in an energy change
leading to resonance with a neighboring wave, and so on
\citep{1996ApJ...461..445M}. 
Electrons can thus be accelerated all
the way from thermal to relativistic energies through a series of
overlapping resonances 
\index{acceleration!stochastic!resonant cascade} 
with low-amplitude fast-mode waves\index{waves!fast-mode!broad-band}\index{continuum!fast-mode waves} in a continuum broad-band spectrum.

In order to be efficiently accelerated, the test particle must
have an initial speed that is greater than the speed of the
magnetic compression, i.e., of order the Alfv\'en speed. 
\index{Alfv{\' e}n speed!and injection}
For the pre-flare solar corona, {\it electrons} have speeds comparable to,
or greater than, the Alfv\'en speed, and so are efficiently
accelerated immediately. 
On the other hand, in a low-$\beta$
plasma, the protons necessarily have an initial speed that is less
than the Alfv\'en speed and so are not efficiently accelerated:
the scattering centers converge before the test particle has
suffered a sufficient number of energy-enhancing collisions.
Efficient acceleration of ions therefore requires pre-acceleration
to Alfv\'en speeds\index{acceleration!pre-acceleration of ions}. 
In the model of \citet{1995ApJ...452..912M},
this pre-acceleration is accomplished through
(parallel-propagating) Alfv\'en waves, and it is essential to
realize that {\it this process takes some time to accomplish}.
During this pre-acceleration period, most of the energy is
deposited in accelerated electrons; only after the protons are
pre-accelerated \index{acceleration!proton!need for
pre-acceleration} to the Alfv\'en speed are they effectively
accelerated -- they then gain the bulk of the released energy by
virtue of their greater mass \citep{1995ApJ...452..912M}.  As
explained by \citet{2004ApJ...602L..69E}, this may be the
explanation for the different locations of hard X-ray and
gamma-ray sources \index{hard X-rays!imaging}
\index{gamma-rays!imaging} in flares \citep[see Section
\ref{sec:8geometry} and
][]{2003ApJ...595L..77H,2006ApJ...644L..93H}; efficient proton
acceleration requires relatively long trapping times and so a
larger magnetic structure, whereas electrons can be efficiently
accelerated even in relatively small structures.

In practice, since there is a little difference {\it
mathematically} between the plasma wave turbulence and shock
mechanisms \citep{1994ApJS...90..969J}, acceleration by both
turbulence and shocks can be effectively combined\index{acceleration!similarities of shock and turbulent mechanisms}. It should be
also be noted that stochastic acceleration generally predicts an
accelerated spectrum that is not a strict power-law\index{acceleration!stochastic! spectrum not a power law}, but rather steepens with increasing energy
\citep[e.g.,][]{1996ApJ...461..445M}.  
However, a recent
development in transit-time damping theory\index{transit-time damping} involves adding both
the loss term $S({\bf p})$ in Equation~\ref{eqn:zharkova_diff_equ}
that allows electron escape, and a source term $Q({\bf p})$ due to
the return current replenishing the electrons in the acceleration
region. 
 \index{return current}
These additions cause the solution to become stationary
within less than one second under coronal conditions
\citep{2006A&A...458..641G}; furthermore, in the hard X-ray
spectral region of observed non-thermal energies, the photon
distribution is close to a power law, in agreement with
observation. The predicted spectral index $\gamma$ depends on the
wave energy density and on the interaction time between the
electrons and the waves.
\index{acceleration!stochastic!parameter dependence}
Additional trapping (e.g., by an electric
potential) tends to harden the electron spectrum.

The main limitation of stochastic acceleration models at this
juncture is the need to presuppose an \textit{ad hoc} injection of the
necessary waves (plasma or MHD).
\index{acceleration!stochastic!\textit{ad hoc} waves}
Furthermore, in the absence of
pitch-angle scattering\index{pitch-angle scattering}, the stochastic acceleration process
generally leads to a decrease in particle pitch angles.\index{scattering!pitch-angle}
This, in turn, reduces the efficiency of the acceleration process, since now
only waves with very high parallel phase speeds, or those with
pitch angles near $90^\circ$, can resonate with the particles
being accelerated. Therefore, in order for significant stochastic
acceleration to occur, the effective particle scattering rate
(which is proportional to the collision frequency\index{frequency!collision} between
particles and waves) must be greater than the frequency of Coulomb
collisions between the charged particles. The latter condition
imposes rather strict limitations on the growth rates of plasma
waves\index{waves!growth rate} or MHD turbulence, as appropriate.

\section{Recent Theoretical Developments} \label{sec:8developments}

In this Section, we highlight some recent promising developments in
our understanding of the physics of particle acceleration in the
magnetized plasma environment associated with solar flares.

In Section~\ref{sec:8stoch_new}, we turn our attention to some recent advances
in stochastic acceleration models, namely the inclusion of (1)
small-scale electric fields and (2) feedback of the accelerated
particles on the turbulence level in the acceleration region. Some
ideas on possible particle acceleration by large-scale Alfv\'en
waves in a large-scale closed magnetic configuration are discussed
in Section~\ref{sec:8waves}.\index{acceleration region!turbulence}

In Section~\ref{sec:8traps}, the evolution of the ubiquitous
``helmet''-type magnetic field configuration\index{magnetic structures!cusp} observed in flaring
regions is presented.  
\index{magnetic structures!collapsing trap}
\index{acceleration!Fermi!first order} 
\index{acceleration!betatron}
We point out two likely acceleration
regions in such a geometry: the collapse of previously-stretched
field lines onto a closed magnetic loop (which leads to
acceleration by both first-order Fermi and betatron processes,
associated with the shrinking and compression of field line
bundles, respectively), and the formation of ``termination''
shocks associated with high-speed jets, which accelerate particles through shock-drift acceleration.\index{acceleration!shock-drift}
\index{shocks!termination}
\index{jets!reconnection outflow}

In Section~\ref{sec:8single3D}, we consider particle trajectories
in a realistic 3-D current sheet geometry\index{current sheets!3-D}, considering, in
particular, the asymmetric acceleration of electrons and ions and
the resulting particle energy spectra, which turn out to be power
laws. In Section~\ref{sec:8fan3d}, we discuss particle acceleration
in 3-D magnetic field configurations, in which magnetic nulls\index{magnetic structures!null} have
a fundamentally different topological structure than they do in a
simple 2-D reconnection geometry. 
Both ``fan'' and ``spine''
reconnection modes are considered, and the resulting (rather
hard power-law) spectra of the accelerated particles are
presented.

In Section~\ref{sec:8complex}, we turn our attention to
acceleration in {\it multiple} acceleration sites, considering
cellular automaton (CA) models\index{flare models!cellular automaton}\index{cellular automata}, MHD turbulent environments, and
stochastic acceleration in a distribution of elementary current
sheet sites.

In Section~\ref{sec:8test_p}, we turn our attention to a critical
aspect of all acceleration models -- the need to recognize that,
since a substantial fraction of the magnetic energy released in
flares appears in the form of accelerated particles, these
particles simply cannot be considered as an ensemble of test
particles interacting with a {\it prescribed} electrodynamic
environment\index{acceleration!test-particle approach}\index{caveats!limitation of test-particle approach}.
Rather, the electric and magnetic fields\index{magnetic field!induced by particles} generated by
the accelerated particles themselves are an essential element of
this environment; the self-consistent inclusion of these fields
represent a paramount consideration in acceleration models,
without which unphysical and even paradoxical conclusions can
result. We review the limitations of existing acceleration models
in this context.  In Section~\ref{sec:8pic}, we discuss recent
attempts to include this electrodynamic feedback through the use
of PIC models, for which the self-consistent
electric and magnetic fields associated with the accelerated
particles {\it do} form an essential element of the modeling.
\index{electrodynamic feedback}
\index{simulations!PIC}
Finally, in Section~\ref{sec:8island}, we briefly discuss a
mechanism for particle acceleration that involves multiple
small-scale closed magnetic ``islands'' formed by the tearing-mode
instability in a reconnecting current sheet.
\index{plasma instabilities!tearing mode}
\index{magnetic structures!islands!and tearing mode}

\subsection{Stochastic acceleration} \label{sec:8stoch_new}

As discussed in Section~\ref{sec:8photons}, the low value of the
low-energy cutoff of the electron energy distribution suggested
by {\em RHESSI} observations
\citep{2005SoPh..226..317K,2005A&A...435..743S} has significant
implications for the number of accelerated flare electrons.
\index{low-energy cutoff}
 \index{accelerated particles!!energy spectra!low-energy cutoff}
 Indeed, for sufficiently low cutoff energies, the number of electrons
required in a collisional thick-target model
\citep{1971SoPh...18..489B} becomes so great that models in which
the accelerated particles\index{electrons!dominant tail population} form a subset of the ambient
distribution become untenable\index{flare models!thick-target}.  
The only viable mechanism of particle acceleration is then acceleration of the entire plasma
within the acceleration region through a stochastic process that
results in little or no net electrical current\index{acceleration!stochastic!net electrical current}, to avoid the issues presented in Section~\ref{sec:8photons}.\index{acceleration region!bulk acceleration}

As discussed in Section~\ref{sec:8stochastic}, high-frequency
waves near the plasma frequency\index{frequency!plasma} are problematic as drivers
for
\index{acceleration!stochastic} 
stochastic acceleration,
because they would couple into decimeter radio waves and would be
present at high levels in every flare, contrary to observations
\citep{2005SoPh..226..121B}. On the other hand, acceleration by
\index{acceleration!transit-time damping} 
transit-time damping do
not suffer from this drawback -- the driving waves have a
frequency that is far below the plasma frequency.

\subsubsection {Improvements to the MHD stochastic acceleration model} \label{sec:zharkova_stoch_new}

The basic transit-time damping stochastic acceleration mechanism
has been improved by adding low-frequency fluctuating electric
fields parallel to the magnetic field
\citep{2004A&A...426.1093G,2005A&A...434.1173G}. Such fields may
originate from low-frequency and high-amplitude turbulence, such
as kinetic Alfv\'en waves, as discussed in
Section~\ref{sec:8transport} \citep[see
also][]{1996ApJ...461..445M,1997ApJ...482..774P}. They can
accelerate and decelerate electrons (and protons), leading to a
net diffusion in energy space \citep{2004ApJ...605L..69A}. As
for other models for stochastic acceleration, the authors assume
an {\it ad hoc} distribution of turbulence.
\index{turbulence!ad hoc}\index{acceleration!stochastic!limitations of models}
In this model, the non-thermal electron distributions
\index{accelerated particles!energy spectra} 
in coronal sources ``grow'' out of the thermal population: the
coronal source is initially purely thermal, then a soft
non-thermal population develops, getting harder at the hard X-ray
peak and softening toward the end of the emission.

As discussed in Section~\ref{sec:8photons}, such a
``soft-hard-soft''
\index{hard X-rays!coronal sources!soft-hard-soft}\index{soft-hard-soft!in coronal sources}
behavior has been observed in a coronal source
\citep{2007A&A...466..713B}.
Indeed, \citet{2004A&A...426.1093G}
report {\em RHESSI} observations suggesting the existence of a point in the
spectrum at which the
hard X-ray flux does not change with time during an event.
\index{hard X-rays!spectra!pivot point}
Above this {\it pivot point} the spectral
variation is consistent with the
characteristic ``soft-hard-soft'' behavior of the photon spectrum.
Note, however, that \citet{2006ApJ...651..553Z} have pointed out
that such a ``pivot-point'' behavior in the variation of the hard
X-ray spectrum with time could also be associated with an electron beam accelerated by the super-Dricer
electric field in a current sheet and precipitating into a loop leg with the
self-induced (return-current) electric field
\citep[e.g.,][]{1977ApJ...218..306K,1981ApJ...245..711E}
associated with an electron beam accelerated by a directed electric field.
\index{beams!electric field}\index{electron beams!and return current} 

\index{electric fields!induced by particles}

\subsubsection{Stochastic acceleration with particle feedback}
\label{sec:zharkova_stochastic_self}

A more self-consistent approach to stochastic acceleration was
recently reported by \citet{Bykov2009}, who simulated electron
acceleration by MHD waves\index{acceleration!electrons!by Alfv{\' e}n waves}, 
but included the effect on the MHD
turbulence\index{turbulence!MHD} (spectrum and intensity) caused by the injection of
high energy particle beams into the acceleration region.\index{acceleration region!turbulence}
The
initial MHD turbulence is prescribed, and is assumed to be
produced in the form of transverse motions with a Gaussian
spectrum on the scale $2\pi/k_0$. Through mode coupling, this
turbulence produces a corresponding level of longitudinal
turbulence in a system of a finite size $R$ (where $Rk_0 > 1$).
The model includes only large scale energy-conserving motions,
with the kinetics of particles on smaller scale being determined
by turbulent advection, valid only for electron energies up to
$\sim$1~MeV \citep{Bykov2009}; thus, the model can account for
the feedback of the accelerated particles on the acceleration
mechanism on large scales only.  The phase-space diffusion
coefficients $D_{ij}$ in Equation~\ref{eqn:zharkova_diff_equ}
are expressed in terms of the spectral functions that describe
correlations between large scale turbulent motions \citep[for
details, see][]{Bykov1993}.

The model assumes a continuous injection of monoenergetic
particles (both electrons and protons) into the acceleration
region.  The initial phase of the acceleration is characterized by
a linear growth regime which results in effective particle
acceleration by the longitudinal large scale turbulent motions and
thus leads to a spectral hardening. However, because the
accelerated particles eventually accumulate a considerable
fraction of the turbulent energy, the efficiency of the
acceleration decreases; this predominantly affects the
higher-energy particles, leading to a spectral softening.

\subsection{Particle acceleration by large-scale Alfv\'en waves} \label{sec:8waves}
\index{flare models!wave energy transport}

\begin{figure}
\begin{center}
\includegraphics[width=0.7\textwidth]{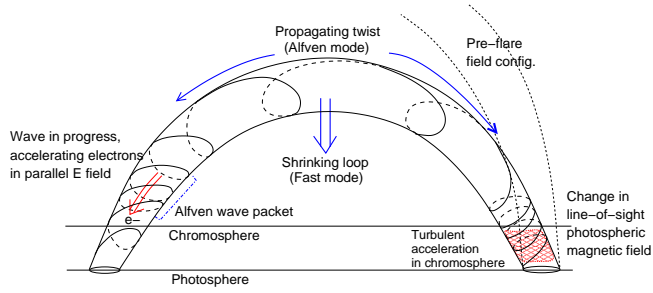}
\end{center}
\caption{The reconfiguring coronal field launches a torsional
Alfv\'en wave pulse through the corona and into the chromosphere,
as well as a fast-mode wave pulse. The Alfv\'en wave, which
propagates in the inertial regime, can lead to electron
acceleration in the corona. Any Alfv\'en wave energy entering the
chromosphere can also lead to electron acceleration there. The
wave will be partially reflected from the steep gradients
in the chromosphere and then re-enter
the corona \citep[from][]{2008ApJ...675.1645F}.}
\label{fig:zharkova_alfvencartoon}
\end{figure}
\index{waves!Alfv{\' e}n!illustration}
\index{flare models!Poynting flux!illustration}
\index{cartoons!Poynting-flux transport!illustration}

In this model, reconnection is still at the heart of the magnetic
energy release process, in that it allows the pre-flare field to
reconfigure, releasing the energy needed to power the flare.
However, motivated by {\it in situ} observations and theories for
energy transport and electron acceleration in the Earth's
magnetosphere\index{magnetosphere} during sub-storms \citep{2000SSRv...92..423S}, it is
suggested \citep{2008ApJ...675.1645F} that the energy is
transported through the corona not by particle beams but by Alfv\'en-wave
\index{acceleration!Alfv{\' e}n waves} 
pulses, launched in the primary reconnection site and carrying a large Poynting flux\index{Poynting flux} 
towards the chromosphere. 
The presence of such pulses has recently been verified in MHD simulations by
\citet{birnetal09}.  They may play a direct role in accelerating
the electrons either in the field-aligned electric fields that
they generate in the corona, or through the magnetic energy that
they transport to the chromosphere, leading to acceleration of
electrons locally in the chromosphere by mechanisms as yet
\index{acceleration!in chromosphere}
unspecified. 
A cartoon realization of this model is shown in
Figure~\ref{fig:zharkova_alfvencartoon}.

To explain the rapid time variability observed in the flare time
profiles, and also the simultaneity\index{footpoint simultaneity} of hard X-ray footpoint
sources within around 0.1~s, the wave pulses must move quickly --
on the order of $0.1 c$, which is significantly larger than the
values usually quoted for the coronal Alfv{\' e}n speed\index{Alfv{\' e}n speed!in the impulsive phase}\index{footpoints!Alfv{\' e}n-wave transport}.
However, such speeds {\it are} consistent with the properties of the core
of an active region before a flare, where the density may be
$10^9$~cm$^{-3}$ or less, and the magnetic field\index{magnetic field!in core of active region} on the order of a
few hundred Gauss \citep[e.g.,][]{2006ApJ...641L..69B}. Linear
wave pulses traveling at such speeds do not damp significantly
\textit{en route} through the corona.\index{active regions!high Alfv{\' e}n speed}

\citet{2008ApJ...675.1645F} calculate that for a coronal field of
500~G and an ambient density of $10^9$~cm$^{-3}$, a magnetic field
perturbation of 50~G is more than enough to deliver sufficient
Poynting flux to the chromosphere to produce the flare energies
observed. 
\index{Poynting flux}
Of course, such a perturbation, on the order of a few to
10\%, does stretch the linear approximation for Alfv\'en waves.
\index{waves!Alfv{\' e}n mode}
An ideal, incompressible, non-linear Alfv{\' e}n mode traveling in a completely uniform plasma
does not steepen into a shock, or dissipate due to Landau damping
\citep{1974JGR....79.2302B}, but the presence of density
fluctuations, including those generated by the wave itself, can
lead it to decay via a parametric instability\index{plasma instabilities!parametric} into a
forward-propagating ion-acoustic wave and a backward-propagating
Alfv\'en wave \citep[e.g.][]{1978ApJ...219..700G}.\index{waves!ion-acoustic}

Analytical work by \cite{1993JGR....9819049J} suggests that, in the
low-$\beta$ coronal regime, the maximum growth rate of this
instability for a wave with an amplitude of a few percent is a few
times the wave frequency, so that this decay occurs for all but
long-wavelength perturbations. However, in the low-$\beta$ case,
the Alfv\'en speed exceeds the sound speed, so the front of the
Alfv\'en wave train (and presumably also the front of a simple
Alfv\'en wave pulse) propagates ahead of the ion-acoustic wave\index{waves!ion-acoustic}
produced, and is, therefore, not significantly affected by the
instability. This has been demonstrated by
\cite{2003A&A...409..813T} in numerical simulations for roughly
the parameter regime of interest. The process has not yet been
fully explored in an environment where collisionless damping is
important, though this may be the case for  coronal depths in
flares.

Because the Alfv\'en speed in such an environment is so high
(higher than the electron thermal speed, such that an average
electron lags behind the wave), the details of the wave
propagation require two-fluid\index{simulations!two-fluid} 
or even full kinetic simulations\index{Alfv{\' e}n speed!and electron thermal speed}.
Such simulations performed in the case of the magnetosphere
\citep{1994JGR....9911095K} indicate that a parallel electric
field is generated by the wave pulse, in which ambient electrons
can -- if at the appropriate ``injection'' energy -- be
accelerated to twice the Alfv\'en speed. 
Similar considerations for solar coronal conditions indicated that modest non-thermal
energies of a few tens of~keV could be achieved in this manner.
\citet{2009ApJ...693.1494M} have also recently confirmed this for
solar plasma in the test-particle limit. This process is found
to damp the wave energy, though it remains to be seen by how much.

On arriving at the steep gradient in Alfv\'en speed represented by
the chromosphere, it is to be expected that this wave pulse would
be at least partially reflected \citep{1982SoPh...80...99E}.
\index{Alfv{\' e}n speed!gradient in chromosphere}
\index{waves!Alfv{\' e}n!reflection at chromosphere}
However, any component of the wave energy entering the chromosphere could be
locally damped -- for example by ion-neutral coupling -- leading to
chromospheric heating. 
\index{ion-neutral coupling} 
If one considers the wave pulse as
imparting stress to the chromospheric magnetic field\index{magnetic field!chromospheric}, then the
multiple current sheets that this might generate could provide an
environment for particle acceleration, as suggested by
\citet{2005ApJ...620L..59T}. The timescales are also such that a
perpendicular MHD cascade\index{turbulence!MHD!perpendicular cascade} could be
established in the reflected wave field, providing another
opportunity for stochastic acceleration of the chromospheric
electrons, if the acceleration timescale is less than the
collisional loss timescale as discussed in Section~\ref{sec:8stochastic}.

Significant theoretical work is yet required to demonstrate
theoretically whether any of these options are viable, e.g.,
whether any chromospheric electron acceleration mechanism can
reproduce the timing delays observed between high and low energy
hard X-rays \citep[e.g.,][]{1996ApJ...468..398A}, which are
normally interpreted as strong evidence for a coronal electron
accelerator. It should also be noted that in the proposed model of
energy {\it transport}, magnetic reconnection 
still plays a central role as the mechanism of energy {\it release}, and as the driver of
the Alfv{\' e}n waves.\index{waves!Alfv{\' e}n!and magnetic reconnection}
\index{reconnection!and Poynting-flux transport}
This means that concurrent electron
acceleration in the corona can also take place by ``traditional''
means (e.g., by electric fields in current sheets or by turbulent
acceleration by various MHD wave modes).

\subsection{Electron acceleration in collapsing current sheets} \label{sec:8traps}
\index{magnetic structures!collapsing trap}

\begin{figure}
\center
\includegraphics[width=.4\hsize]{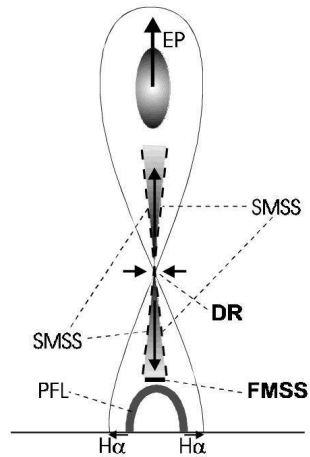}
\caption{Sketch of the flare scenario.  EP: erupting prominence;
DR: diffusion region; SMSS: slow-mode standing shock; FMSS:
fast-mode standing (also called termination) shock; PFL: post
flare loop. The grey shaded areas with the arrows show the outflow
jets. 
From \citet{2006A&A...454..969M}.} \label{fig:zharkova_ma_1}
\end{figure}

Figure~\ref{fig:zharkova_ma_1} depicts a helmet-type magnetic
structure \index{magnetic structures!helmet} often observed in
solar flares 
\citep[see, for example,][]{2003ApJ...596L.251S},
in which a prominence is destabilized due to photospheric footpoint
motions and rises upwards as an 
\index{prominence!eruptive}  {\it eruptive prominence}. 
As the prominence rises, the underlying
field lines are stretched, leading to the formation of a current
sheet. 
\index{footpoints!motions}
\index{resistivity!anomalous} 
\index{plasma instabilities!various}
We suppose that when the current exceeds a critical value,
the resistivity is suddenly enhanced
(say by plasma waves excited by various instabilities) and rapid
magnetic reconnection begins in the diffusion region (DR), as
shown earlier in Figure~\ref{fig:zharkova_dif_reg}.

\index{flare models!standard!illustration}
\index{jets!reconnection outflow!illustration}

\begin{figure}
\begin{center}
\includegraphics[scale=0.25]{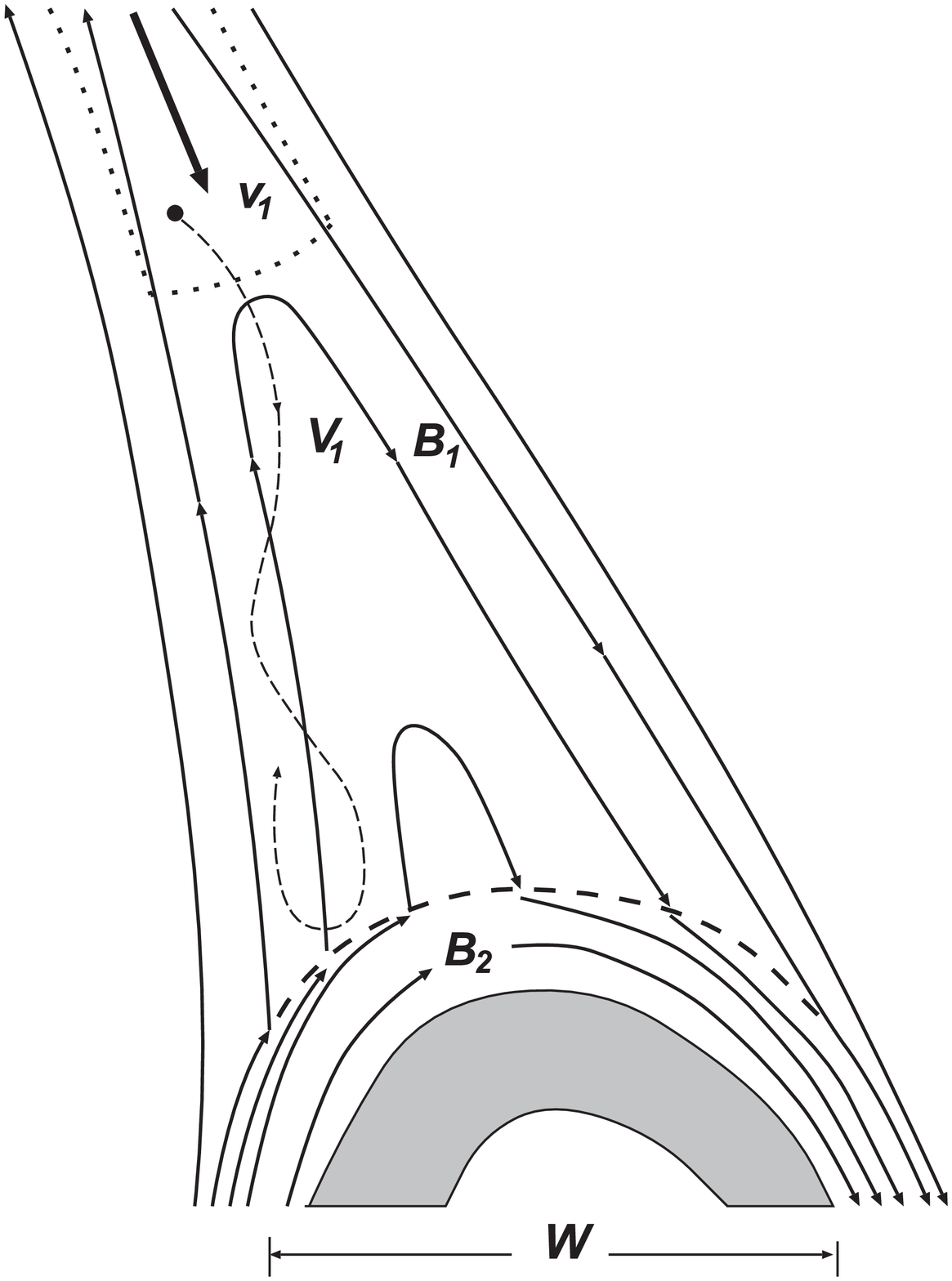}
\includegraphics[scale=0.25]{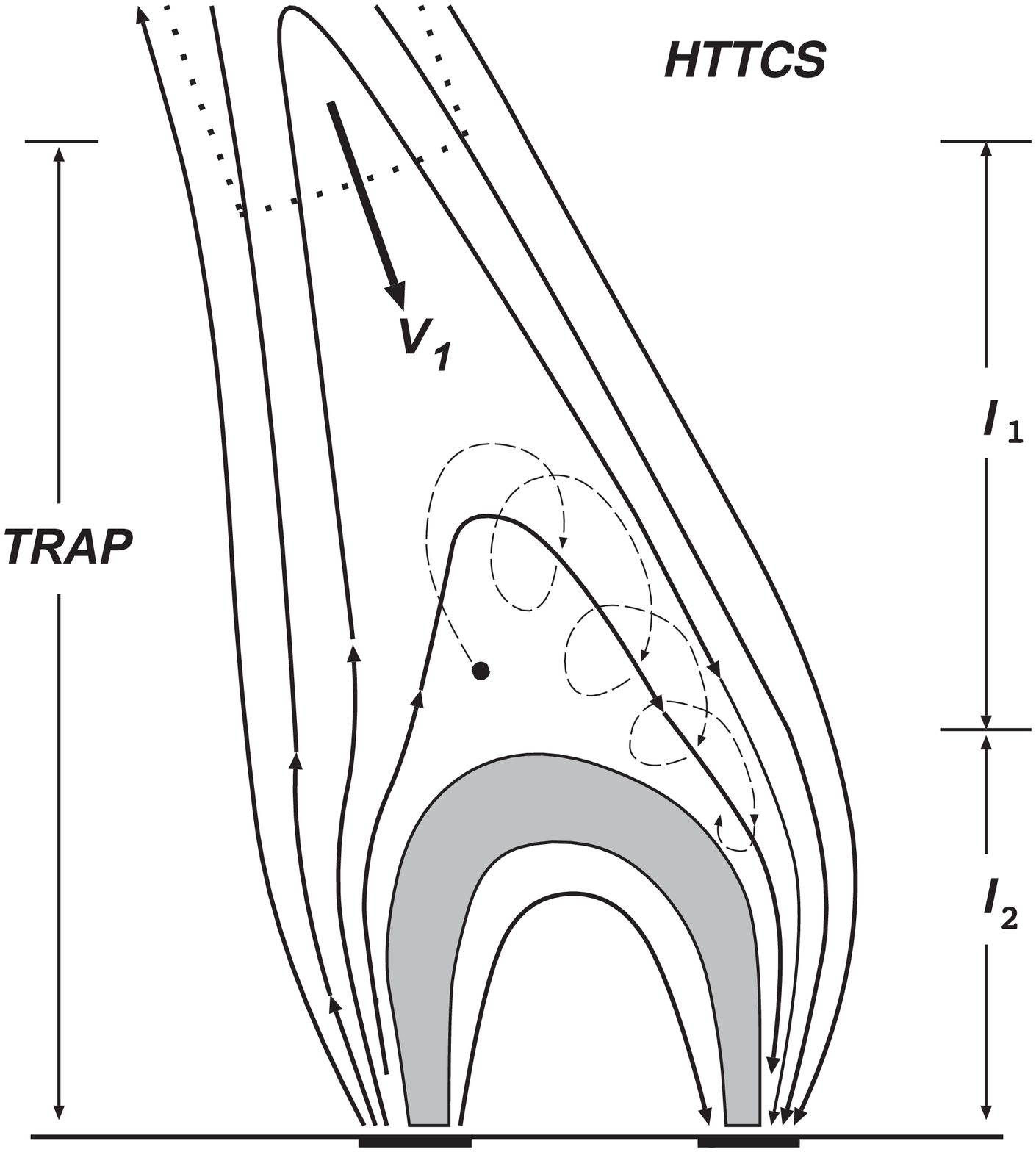}
\end{center}
\caption{Schematic models of collapsing magnetic field lines
occurring in MHD models of magnetic reconnection with (left
panel, dashed line) and without (right panel) a termination shock.
From \citet{1997ApJ...485..859S}.
HTTCS denotes ``high temperature turbulent current sheet.'' } \label{fig:zharkova_coll_traps}
\end{figure}

MHD solutions\index{simulations!MHD!standard flare model} 
for the large-scale magnetic configuration plotted
in Figure \ref{fig:zharkova_ma_1} allow us to investigate (see
Section~\ref{sec:8reconnection}) the reconnection process inside
the diffusion region, characterized by plasma inflows with
velocity $V_0$ in the $x$-direction (from the sides in
Figure~\ref{fig:zharkova_ma_1}; from the top and bottom in Figure
\ref{fig:zharkova_dif_reg}) and outflowing with velocities $V_1$
in the $z$-direction (to the top and bottom of
Figure~\ref{fig:zharkova_ma_1}; to the left or right sides of
Figure \ref{fig:zharkova_dif_reg}). The outflows are generally
very fast and create a rather complicated structure in the outer
region at the RCS edge\index{current sheets!outflows}.
The plasma confined on the newly
reconnected field lines will start moving with the velocity $V_1$,
both upwards (not shown in the cartoon), and downwards, where it
will encounter a magnetic obstacle in the form of the underlying
loop.
\index{cartoons!standard}
If the distance to this obstacle is large, a collapsing
magnetic trap (see the bottom plot in
Figure~\ref{fig:zharkova_coll_traps}) is formed. Moreover, the
downward outflow is moving into a region of generally higher
density and hence lower Alfv\'en speed\index{Alfv{\' e}n speed!and termination shock}.\index{termination shock}
If a point is reached at which the outflow velocity $V_1$ exceeds the local
magneto-acoustic wave velocity, then a fast MHD shock, termed a
``termination shock,''\index{shocks!termination} can appear ahead of this magnetic obstacle \citep{1997ApJ...485..859S},
as shown by the dashed line in the upper panel of
Figure~\ref{fig:zharkova_coll_traps}.

\index{cartoons!collapsing trap!illustration}

\subsubsection{Acceleration in collapsing magnetic traps} \label{sec:8collapsing_trap}

\citet{1997ApJ...485..859S}, \citet{2004A&A...419.1159K} and
\citet{2005AstL...31..537B, 2007AstL...33...54B} have carried out
simulations of the acceleration of electrons in collapsing
magnetic traps. 
\index{acceleration!collapsing magnetic trap}
\index{magnetic structures!collapsing trap}
Particles with an initial Maxwellian distribution are injected at
random points of a fragmented current sheet geometry obtained
using a three-dimensional MHD simulation.

As a result of its motion from the reconnection region towards the
chromosphere, the length $L$ of the magnetic trap decreases,
leading to an increase in energy of the trapped particles via a
first-order Fermi \index{acceleration!Fermi!first order} process
associated with conservation of the longitudinal invariant $\int
p_\parallel \, ds$ \citep{2002ASSL..279.....B}. Also, as a result
of the {\it transverse} contraction of the magnetic trap,
particles are accelerated by the betatron
\index{acceleration!betatron} mechanism. This transverse
contraction can be described by the ratio $b(t) = B(t)/B_0$ which
changes from 1, when $B(0)=B_0$ to $b_m= B_m/B_0$, where $B_m$ is
the magnetic field in the magnetic mirrors.
\index{magnetic structures!mirror geometry!moving} 
Thus the particle momenta $p_\parallel $ and $p_\perp$ change as follows:
\begin{equation}
p_\parallel \,=\,  p_{0\parallel } \, \ell^{-1},
\end{equation}
where $\ell = (L/L_0)$, and
\begin{equation}
p_\perp \,=\, p_{0\perp} b^{1/2}.
\end{equation}
Both the Fermi and betatron accelerations are accompanied by a
change in particle pitch angle -- a decrease (Fermi) or increase
(betatron)\index{acceleration!and pitch-angle change}. 
When these two mechanisms act simultaneously, then the
pitch angles of accelerated particle vary according to
\begin{equation}
\tan \alpha = \frac {p_\perp }{p_\parallel } = \ell \, b^{1/2} \,
\tan \alpha_0 \, ,
\end{equation}
which can result in an increase or decrease of pitch angle,
according to the relative importance of the two effects.  The
particle kinetic energy increases in the magnetic trap until it
falls into the loss cone\index{magnetic structures!loss cone}.

Simulations show that particles with an initially Maxwellian
distribution gain energies via both acceleration mechanisms; they
retain a quasi-thermal distribution but attain much higher energies.
The change of total kinetic energy has a form of the burst of
particles with energies up to a few~MeV. However, the magnetic
energy is absorbed very rapidly by the electrons, so that a test-particle 
approach very rapidly ceases to be valid.  A fuller
treatment of this scenario therefore requires full kinetic
simulations similar to those discussed in Section~\ref{sec:8pic}.

\subsubsection{Acceleration by a termination shock} \label{sec:8termin_shock}
\index{acceleration!shock!termination}\index{termination shock!and particle acceleration}

Due to the strong curvature of the magnetic field lines in the
vicinity of the diffusion region in a helmet-type reconnection
(see Figure \ref{fig:zharkova_ma_1}), the slowly inflowing plasma
shoots away from the reconnection site as a hot jet.\index{jets!reconnection outflow}
These
oppositely-directed jets are embedded between a pair of slow
magnetosonic shocks. If the speed of this jet is super-Alfv\'enic,
a fast magnetosonic shock, also called a {\it termination shock},
is established, as shown previously in
Figure~\ref{fig:zharkova_coll_traps}. 
The appearance of such shocks was predicted in the numerical 
simulations\index{shocks!termination!numerical simulation} of
\citet{1986ApJ...305..553F}, \citet{1995ApJ...451L..83S} and 
\citet{1997ApJ...485..859S}.

Radio measurements \citep{2002A&A...384..273A,Chapter5} of SOL2003-10-28T11:10 (X17.2)
\index{flare (individual)!SOL2003-10-28T11:10 (X17.2)!termination shock}
reveal the termination shock as a standing radio source
\index{radio emission!termination-shock signature} with a
(large) half-power source area of $A_{s} = 2.9 \times
10^{20}$~cm$^2$. 
A strong enhancement of the electromagnetic
emission in the hard X- and gamma-ray range up to 10~MeV is
simultaneously observed with the appearance of the radio
signatures of the shock \citep{2004ApJ...615..526A}. These
observations indicate that the termination shock could well be the
source of the highly energetic electrons needed for the generation
of the hard X- and gamma-ray radiation.

A fast magnetosonic shock formed during the MHD simulations\index{simulations!MHD!reconnection shocks} of a
reconnection process is accompanied by a compression of both the
magnetic field and the density. Thus, it represents a moving
magnetic mirror\index{magnetic structures!mirror geometry!moving} at which charged particles can be reflected and
accelerated through a process termed {\it shock-drift
acceleration}\index{acceleration!shock-drift!electrons}. 
This process can accelerate highly energetic
electrons up to 10~MeV \citep{2006A&A...454..969M}.

Generally, shock-drift acceleration represents reflections at the
termination shocks\index{acceleration!shock!termination}\index{acceleration!shock-drift}.
Analysis of such an acceleration is most straightforwardly
performed in the {\it de Hoffmann-Teller frame}, 
\index{de Hoffmann-Teller frame} 
the frame in which the electric field
vanishes, so that the reflection takes place under conservation of
both kinetic energy and magnetic moment. From such an analysis
\citep{2006A&A...454..969M,2009A&A...494..669M}, one obtains the relationships between
the electron velocities $\beta = v/c$ parallel and perpendicular
to the upstream magnetic field before (index {\it i}) and after
(index {\it r}) the reflection:
\begin{equation}
\beta_{r,\parallel} = \frac{2\beta_{s}-\beta_{i,\parallel}(1+\beta_{s}^{2})}
    {1-2\beta_{i,\parallel}\beta_{s}+\beta_{s}^{2}}
\end{equation}
and
\begin{equation}
\beta_{r,\perp} = \frac{(1-\beta_{s}^{2})}
    {1-2\beta_{i,\parallel}\beta_{s}+\beta_{s}^{2}}
    \cdot \beta_{i,\perp} \, ,
\end{equation}
where $\beta_{s} = v_{s}\sec\theta/c$.  Here $v_{s}$ is the shock
speed, $c$ the velocity of light, and $\theta$ the angle between
the shock normal and the upstream magnetic field. In addition, the
reflection conditions:
\begin{equation}
\beta_{i,\parallel} \leq \beta_{s}
\end{equation}
and
\begin{equation}
\beta_{i,\perp} \geq \frac{\tan\alpha_{lc}}{\sqrt{1-\beta_{s}^{2}}}
    \cdot (\beta_{s}-\beta_{i,\parallel})
\end{equation}
must be satisfied in order for acceleration to occur. The
loss-cone angle\index{magnetic structures!loss cone!definition} is defined by $\alpha_{lc} = \sin^{-1}[(B_{\rm
up}/B_{\rm down})^{1/2}]$, where $B_{\rm up}$ and $B_{\rm down}$
denote the magnitude of the magnetic field in the up- and
downstream region, respectively.

\begin{figure}[t,h!]
\center
\includegraphics[width=.4\hsize]{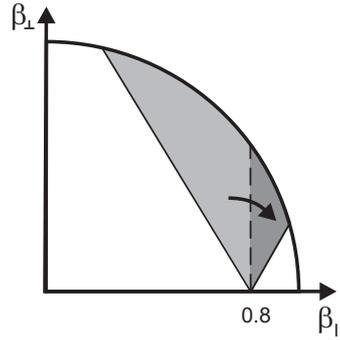}
\caption{Illustration of shock-drift acceleration in the
$\beta_{\perp}-\beta_{\parallel}$ plane for a shock with
$\beta_{s}$ = 0.8 and $\alpha_{lc}$ = 45$^{\circ}$. All particles
initially located in the light grey area are transformed into the
dark one.} \label{fig:zharkova_ma_3}
\end{figure}
\index{acceleration!shock-drift!illustration}

Shock-drift acceleration represents a transformation in the
$\beta_{\perp}-\beta_{\parallel}$ plane as illustrated in
Figure~\ref{fig:zharkova_ma_3}. As seen in
Figure~\ref{fig:zharkova_ma_3}, the accelerated distribution
function is simply a {\it shifted loss-cone distribution}. The
flux of the accelerated electrons parallel to the upstream
magnetic field is defined by
\begin{equation}
F_{acc,\parallel} = 2 \pi N_0 c^4  \int_{0}^{1}
d\beta_{\parallel} \beta_{\parallel} \int_{0}^1{y}
d\beta_{\perp} \beta_{\perp} \cdot f_{acc}(\beta_{\parallel},
\beta_{\perp}),
\end{equation}
and the differential flux is given by
\begin{equation}
j_{acc, \parallel} = \frac{dF_{acc,\parallel}}{dE}.
\end{equation}

\begin{figure}[t,h!]
\center
\includegraphics[width=.4\hsize]{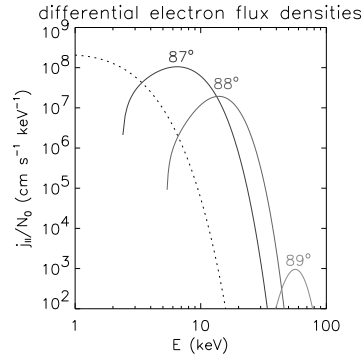}
\caption{Differential flux of electrons accelerated by shock-drift acceleration for
various shock angles. The dotted line represents the flux of a
pure Maxwellian distribution.} \label{fig:zharkova_ma_4}
\end{figure}
\index{acceleration!shock-drift!illustration}

Figure~\ref{fig:zharkova_ma_4} shows the differential fluxes
obtained from the shock-drift acceleration model for various
angles $\theta$, assuming an initial distribution in the form of a
Maxwellian with a temperature $T = 10$~MK. For comparison, the
dotted line in Figure~\ref{fig:zharkova_ma_4} represents a pure
Maxwellian distribution with the same temperature. One can see
that shock-drift acceleration indeed provides a significantly
enhanced component of energetic electrons, particularly for large
values of the angle $\theta$. These analytically-obtained results are qualitatively confirmed
by numerical simulations performed for plasma conditions
usually found at the Earth's bow shock \citep{1989JGR....9415367K,1989JGR....9415089K,1991JGR....96..143K}.

In order to compare these results with observations,
some basic parameters must be provided.
Here, SOL2003-10-28T11:10 (X17.2)
is considered as an example \citep{2009A&A...494..669M}. 
\index{flare (individual)!SOL2003-10-28T11:10 (X17.2)!termination shock}
In this event,
the termination shock appears at 300~MHz for the harmonic emission.
Matching harmonic emission to the
observed $300$~MHz frequency requires an electron number density
\index{radio emission!plasma radiation} of $2.8 \times
10^{8}$~cm$^{-3}$. According to a two-fold density model of the
corona \citep{1961ApJ...133..983N}, such a region is located $\sim$160~Mm above the photosphere, 
at which a magnetic field of $\sim$5~G is expected \citep{1978SoPh...57..279D}. These parameters
lead to an Alfv\'en speed $V_A \approx 560$~km~s$^{-1}$. For the
temperature, a value of 40~MK is adopted, resulting in a thermal
electron speed of $2.5 \times 10^4$~km~s$^{-1}$.

Assuming a density jump of a factor of two across the termination
shock, the Rankine-Hugoniot relations 
\index{Rankine-Hugoniot relations} 
give a magnetic field $B_{\rm down}/B_{\rm up}$ = 2 and
an Alfv\'en-Mach number $M_{A}$ = 3.5, corresponding to a shock
speed of $\sim$2000~km~s$^{-1}$. Due to the curvature of the
magnetic field lines convected through the termination shock,
there is a distribution of $\theta$ across the shock. Recognizing
from Figure~\ref{fig:zharkova_ma_4} that the flux is very
sensitive to $\theta$, we model this curvature by a circle; the
resulting flux of accelerated electrons can be integrated over
$\theta$, giving an accelerated electron production rate $F_{e} =
2 \times 10^{36}$~s$^{-1}$ and a related power $P_{e} = 1 \times
10^{29}$~erg~s$^{-1}$ above 20~keV. This agrees well with the
results inferred from {\em RHESSI} observations
\citep{2007CEAB...31..135W,2009A&A...494..677W}.

It must be stressed that a non-negligible part of the energy of
the outflow jet is transferred into accelerated electrons. Thus,
such a test-particle approach
\index{test-particle approach!limitations} 
has its limitations: the termination shock
acceleration process must be treated in a fully kinetic manner, in
which the accelerated particles themselves are included in the
conservation laws across the shock.

\subsection{Particle acceleration in a single 3-D RCS with complicated magnetic topology} \label{sec:8single3D}

\subsubsection{Basic equations} \label{sec:zharkova_basic_equations_rcs}

Following Section~\ref{sec:8basic}, we consider a highly-stressed
current sheet oriented in the $x$-direction, with a lateral field
\index{current sheets!and particle acceleration}
\index{acceleration!in current sheets}
\begin{equation} \label{eqn:zharkova_main}
B_z (x) = B_{0} \, \tanh \left (- \frac x d \right )
\end{equation}
and a transverse magnetic field component $B_x$ that is dependent
on $z$ according to \citep{2005MNRAS.356.1107Z,
2005SSRv..121..165Z}:
\begin{equation}\label{eqn:zharkova_transverse}
B_x (z) = B_{0} \left (\frac z a \right )^\alpha ,
\end{equation}
where $\alpha > 0$ (we consider an illustrative value $\alpha=1$).
A weak, constant longitudinal component of the magnetic field
$B_y$ is also assumed to be present:
\begin{equation}\label{eqn:zharkova_guiding}
B_y = \beta B_{0}.
\end{equation}
\index{relativistic equations of motion}
\index{equations of motion!relativistic}
Particle trajectories are calculated from the relativistic
equations of motion
\begin{eqnarray}
\frac{d {\bf r}}{d t}  &=& \frac{{\bf p}}{m_0 \gamma} \\
\frac{d {\bf p}}{d t}  &=& q \, \left( {\bf E} + \frac{1}{c} \,
\frac{{\bf p}}{m_0 \gamma} \times {\bf B} \right),
\label{eqn:zharkova_m_equa_r}
\end{eqnarray}
where $t$ is time, $\mathbf{r}$ and $\mathbf{p}$ are the particle
position and momentum vectors, $\mathbf{E}$ and $\mathbf{B}$ are
the electric and magnetic field vectors, $q$ and $m_0$ are the
charge and rest mass of the particle, $c$ is the speed of light,
and $\gamma = \sqrt{(p/m_0 c)^2 + 1}$ is the Lorentz factor.
\index{equations of motion} The integration time steps are chosen
to be much smaller than the corresponding gyro-period, i.e., $dt
\le 0.1 \, (m/q) \, B_z^{-1}$.  For protons, this typically
requires a time step $\lapprox 10^{-7}$~s, while for electrons the
required time step is substantially shorter, $\lapprox 4 \times
10^{-11}$ s.

It should be noted that in the above description, the electric and
magnetic field vectors $\mathbf{E}$ and $\mathbf{B}$ are {\it
prescribed} quantities; the self-consistent electric and magnetic
fields produced by the accelerated particles themselves are
neglected.  Such a test particle approach \index{test-particle
approach!limitations} is valid for up to only $\sim$10$^5$ test
particles \citep{nla.cat-vn71926}, and we shall return to this
point in Section~\ref{sec:8test_p}.  It should also again be noted
(see Section~\ref{sec:8basic}) that in this simulation the reconnection
electric field is assumed to extend over the entire region covering the whole diffusion region.
Thus, the caveats noted in this earlier Section also apply here.
\index{caveats!limitation of test-particle approach}

\subsubsection{Particle motion inside a 3-D RCS} \label{sec:zharkova_separ}
\index{current sheets!3-D}

\begin{figure}
\includegraphics[scale=0.8]{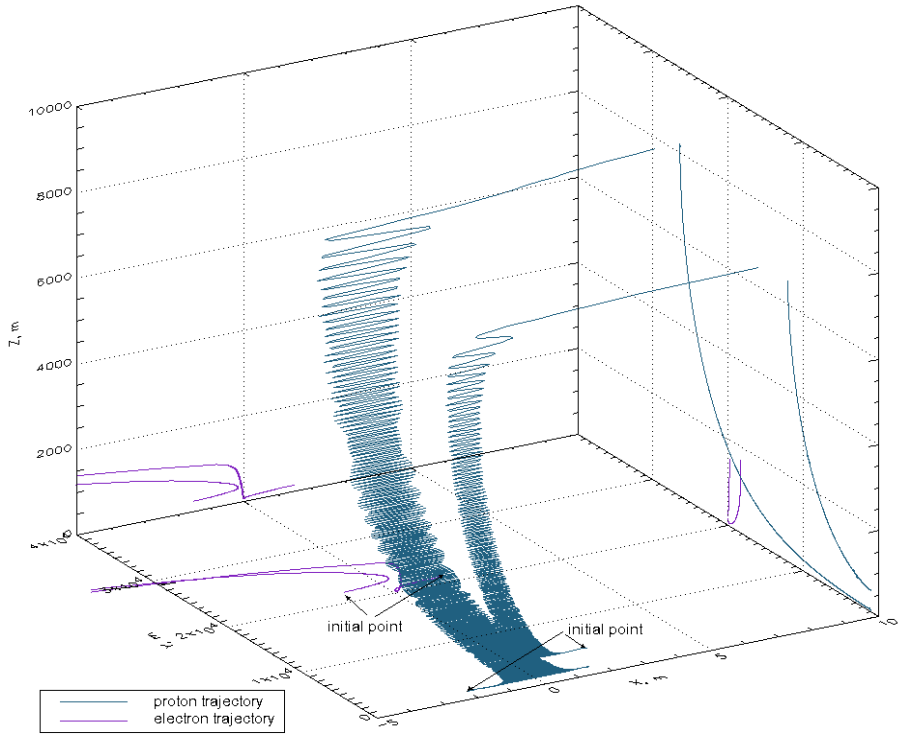}
\caption{Proton (blue lines) and electron trajectories (purple
lines) in a 3-D RCS with a lateral magnetic field $B_x = 10$~G and
an drift electric field $E_y = 100$~V~m$^{-1}$. From
\citet{2009JPlPh..75..159Z}.} \label{fig:zharkova_tr_asym}
\end{figure}

In the test-particle description used here, the electron
trajectory and the direction of ejection are determined uniquely
by its initial coordinate, velocity and the magnetic field
geometry. Particles inflowing into a 3-D RCS are accelerated
mostly near the lateral field reversal at the $x$-axis. In the
presence of a non-zero $B_y$, electrons and protons are ejected in
opposite directions from the midplane ($x=0$); whether the
particle is ejected to $x>0$ or $x < 0$ depends on the sign of the
particle charge, and the signs and magnitudes of $B_x$ and $B_y$.
If $B_{y}$ is strong enough, then all protons (regardless of the
side of the RCS they entered from) are ejected to the $x>0$
semispace, while all electrons are ejected to the opposite
semispace, as illustrated in Figure~\ref{fig:zharkova_tr_asym}
\citep{2009JPlPh..75..159Z}. For particles of the same charge,
there are two fundamental types of trajectories inside an RCS:
particles that enter from and then are ejected into the {\it same}
semispace are hereafter referred to as ``bounced'' \index{bounced
particles}particles, whereas particles that are ejected into the
{\it opposite} semispace are referred to \index{transit particles}
as ``transit'' particles.

Electron/proton asymmetry in the particle trajectories is
dependent on the magnitude of the guiding field
\citep{2004ApJ...604..884Z}. The asymmetry is {\it full} (equal to
unity) if $B_y$ is strong enough ($B_y/B_z \gapprox 1.5 \times
10^{-2}$) that electrons are ejected into one semiplane and
protons to the opposite \index{acceleration!electron-proton
asymmetry} one. If the guiding field is weak ($B_y/B_z \lapprox
10^{-6}$), electrons and protons can be ejected into the {\it
same} semiplane \citep{2004ApJ...604..884Z, 2005SSRv..121..165Z}.
For a guiding field of intermediate magnitude, electrons and
protons are ejected in a partially neutralized manner, with
protons dominating in one semiplane and electrons in the other
\citep{2005SSRv..121..165Z}. 
The result of this separation of
particle species is a polarization electric field appearing
between the protons at the edge and electrons at the midplane of
the RCS.\index{polarization!electric field}\index{electric fields!polarization}
Such a field was anticipated by \citet{1993SoPh..146..127L}
and \citet{2000AstL...26..750O}, and has been investigated in some
detail by \citet{2009JPlPh..75..159Z}.

\begin{figure}
\centering
\includegraphics[width=1.0\textwidth]{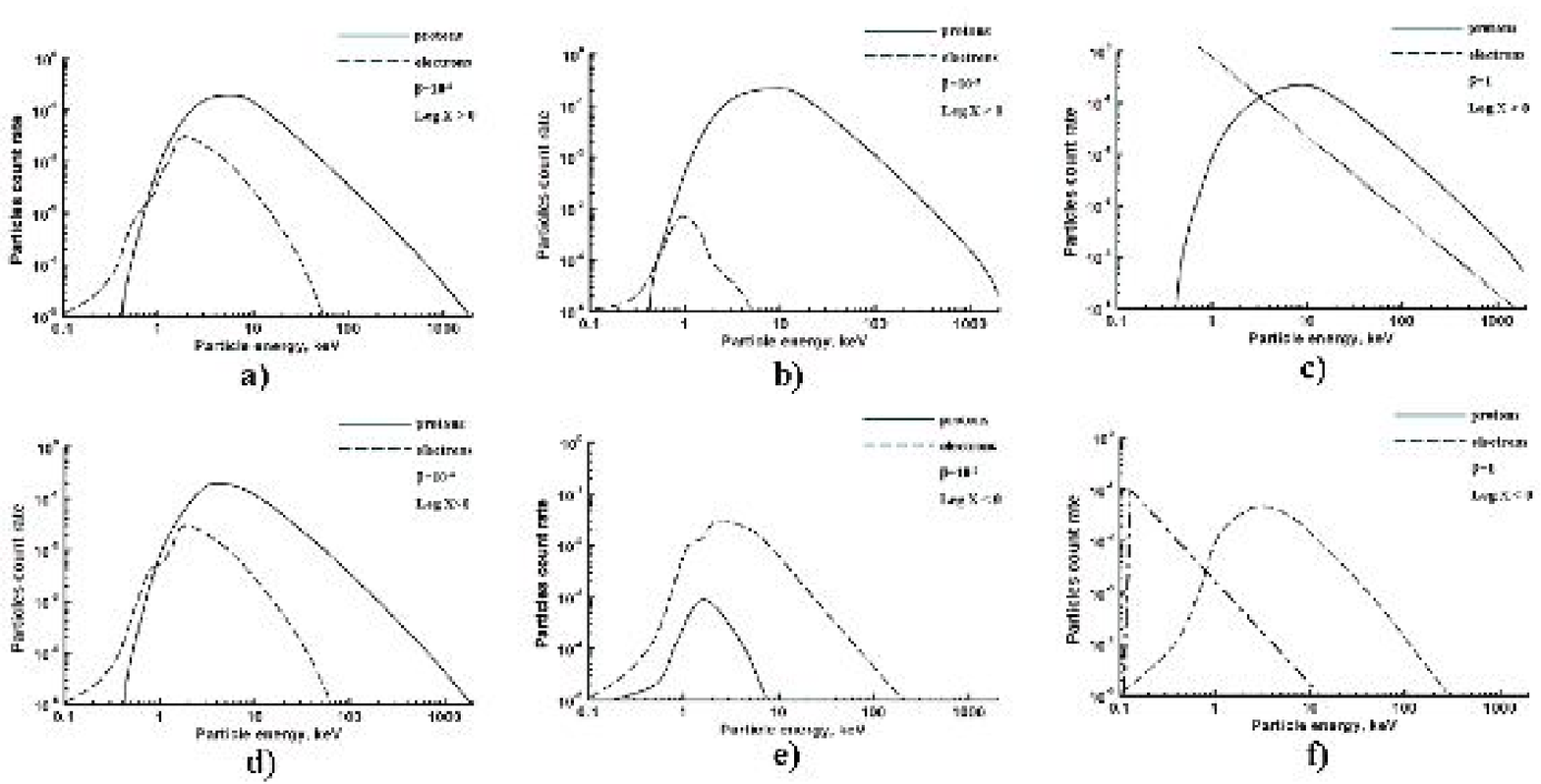}
\caption{Energy spectra of electrons (dashed lines) and protons
(solid lines) ejected from the RCS. Spectra (a) - (c) are
calculated for $\beta (= B_y/B_0)$ equal to $10^{-4}$, $10^{-2}$
and $1$, respectively, and correspond to the semispace $x > 0$.
Spectra (d) - (f) are calculated for the same values of $\beta$,
but for the semispace $x < 0$. The estimated proton spectrum is
presented as a dotted line in plot~(c) and the estimated electron
spectrum is shown as a dash-dotted line in plot~(f). From
\citet{2005SSRv..121..165Z}.} \label{fig:zharkova_sp_mhd}
\end{figure}

The model produces power-law energy distributions
\index{accelerated particles!!energy spectra} with spectral indices $\delta =
1.5$~for protons and $\delta =2$~for electrons. Such a result is
confirmed by both analytic \citep{1996SoPh..167..321L} and
numerical approaches \citep{2005SSRv..121..165Z,
2005SoPh..226...73W}. For comparison, the particle energy spectra
calculated for the model RCS of \citet{2000AstL...26..750O} and
presented in Figure~\ref{fig:zharkova_sp_mhd} show features beyond
such simple power-law forms: there is a rapid rise in flux to a
maximum at $E = E_m$; above this, the spectrum is approximately
power-law. $E_m \simeq 10-12$~keV for protons ($\simeq 2-5$~keV
for electrons); these values increase slightly with the value of
the parameter $\beta=B_y/B_z$. The energy $E_m$ can be considered
as the lower energy cutoff \index{accelerated particles!!energy
spectra!low-energy cutoff} to the electron spectrum (see
Section~\ref{sec:8photons}), and the values obtained here agree
quite well with the values deduced from {\em RHESSI} observations
\citep{Chapter3}.

The spectral index $\delta$ for the power-law part of the spectrum
($E > E_m$) is found to vary slightly with the magnitude of the
guiding field $B_y$. For a weak guiding field ($\beta = 10^{-4}$
-- Figures \ref{fig:zharkova_sp_mhd}, [a] and [d]), the
particle trajectories are nearly symmetrical
\citep{2005SSRv..121..165Z}; the particles are ejected equally
into both semispaces as approximately neutral beams. The particle
energy spectra in both semispaces ($x < 0$ and $x > 0$) are very
similar, with a sharp increase from zero to $E_m$, followed by a
power-law with spectral indices $\delta \simeq 1.8$ for protons
and $\delta \simeq 2.2$ for electrons.  The power-law shapes
extend up to about 100~keV (electrons) or 1~MeV (protons).

For a moderate guide field ($\beta = 10^{-2}$ -- Figures
\ref{fig:zharkova_sp_mhd}~[b] and [e]), the protons and electrons
are also ejected into both semispaces; however, the symmetry of
their trajectories is partially destroyed. Hence, most protons are
ejected into the semispace $x > 0$ with $\delta \simeq 1.7$, while
a few low-energy protons are ejected into the semispace $x < 0$
with very soft, thermal-like, spectra ($\delta \simeq 4.8$). The
opposite picture holds for electrons: the bulk are ejected into
the semispace $x < 0$ with an energy spectrum $\delta \simeq
2.0$, while a smaller number of the low-energy electrons are
ejected into the semispace $x > 0$ with a thermal-like spectrum
having a maximum $E_m \approx 1$~keV\index{acceleration!electron-proton asymmetry!dependence on strength of guide field}.

\begin{figure}
\centering
\includegraphics[scale=0.4]{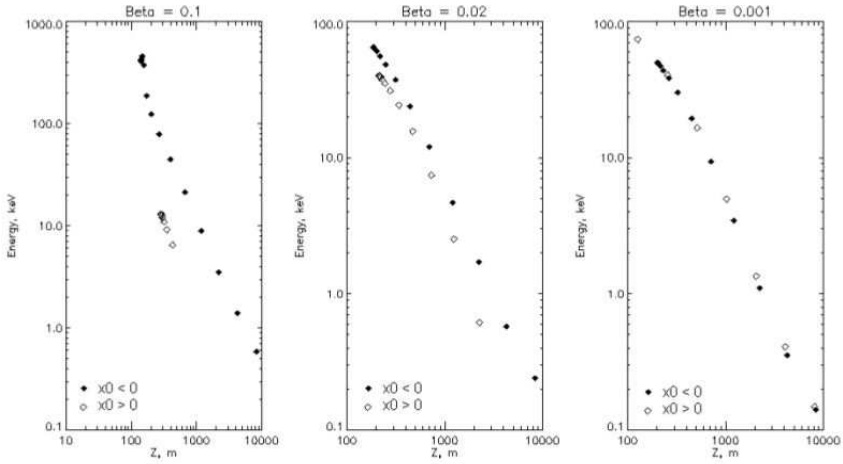}
\caption{The energy spectra resulting from the protons entering
the current sheet from opposite sides, for different ratios $\beta
= B_y/B_0$ and different magnitudes of the parameter $\alpha$.
From \citet{2009JPlPh..75..159Z}.}
\label{fig:zharkova_p_en_sp_beta}
\end{figure}

Finally, if the guiding field is very strong ($\beta \simeq 1$ --
Figures \ref{fig:zharkova_sp_mhd}, [c] and [f]), the particle
trajectories are completely asymmetric\index{current sheets!strong guide field}.
Particles are ejected
completely separately from the RCS midplane into opposite
semispaces: all protons are ejected into the semispace $x > 0$
while all electrons are ejected into the semispace $x < 0$. Their
energy distributions show a sharp increase of particle number up
to around a few keV followed by a power-law energy spectrum of
index $\delta \simeq 1.5$ for protons and $\delta \simeq 1.8$ for
electrons. (These spectral indices vary with the magnitude of a
guiding field (Fig. \ref{fig:zharkova_p_en_sp_beta}) reflecting different
reconnection scenarios and with different phases of magnetic
reconnection, as discussed in Section \ref{sec:8fan3d}.)

\index{acceleration!in current sheets!illustration}

\begin{figure}
\begin{center}
\includegraphics[scale=0.6]{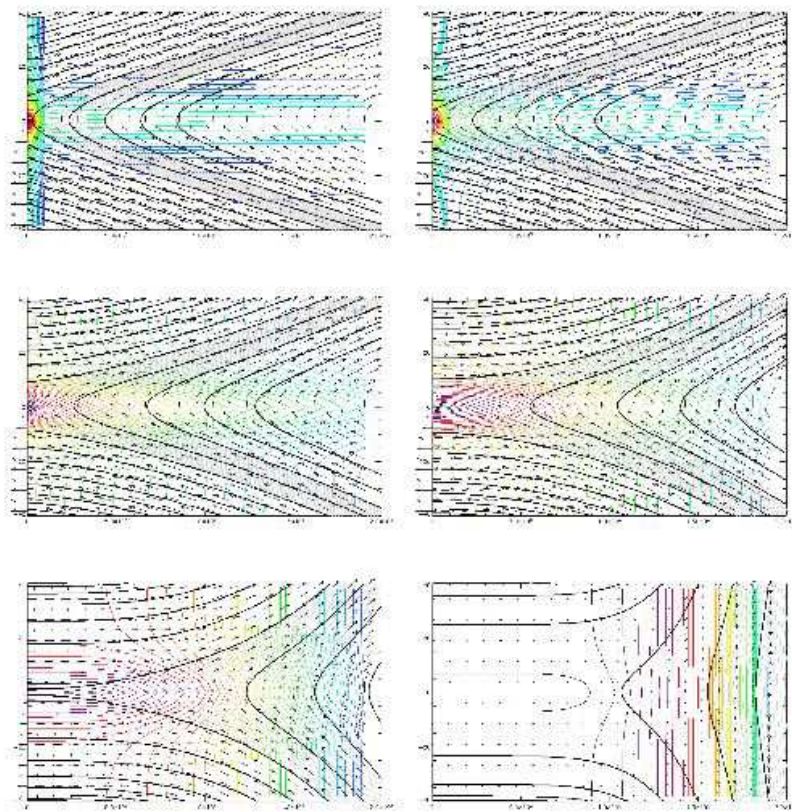}
\end{center}
\caption{Views of the current sheet for different values of
$\alpha = 0.5; 0.7; 1.0; 1.5; 2.0$~and 5.0 (plotted by rows from
the top left to the right bottom plot). From
\citet{2009JPlPh..75..159Z}.} \label{fig:zharkova_compress_alpha}
\end{figure}

As shown in Figure~\ref{fig:zharkova_compress_alpha}, during the
reconnection process, the shape of the reconnecting current sheet
changes from a non-compressed one with a symmetric X-type null
point to a strongly-compressed shape with a well-defined
elongation along the $z$-direction and a compression in the
$x$-direction.  
\index{magnetic structures!null!change in aspect ratio} 
The extent of this deformation depends on the value
of the parameter $\alpha$
\citep{2000mare.book.....P,2000ASSL..251.....S} and possibly
corresponds to different reconnection rates
\citep{2006PhPl...13k2105A}. At the reconnection onset
($\alpha=0$) the diffusion region is located very near the
null point. As the reconnection progresses ($\alpha$ increases from
0 to 1), the diffusion region becomes much wider in both $x$ and
$z$-directions.  As $\alpha$ increases further, the reconnection rate
begins to decrease \citep{2009JPlPh..75..159Z}.

\index{acceleration!in current sheets!illustration}
The spectral index of the accelerated particle energy spectra is
associated with the parameter $\alpha$ and depends on the
magnitude of the guiding field (Fig. \ref{fig:zharkova_p_en_sp_beta}). For an RCS with a density
that increases exponentially with distance $z$ and for a variation
of density $n(z) \propto (z/a)^\lambda$, the following results are
obtained \citep{2009JPlPh..75..159Z}: for electrons in a weak
(strong) guiding field $B_y$,
\begin{equation}
\delta_{e, {\rm weak}} = 1+\frac{2+\lambda}{2\alpha}; \qquad
\delta_{e, {\rm strong}} = \frac 1 2 + \frac{1+
\lambda}{2\alpha},\label{eqn:zharkova_e_weak_strong}
\end{equation}
while for protons in a guiding field of {\it any} strength,
\begin{equation}
\delta_p = 1+ \frac{1+\lambda}{2\alpha}. \label{eqn:zharkova_p_ws}
\end{equation}
In principle, therefore, by simultaneously measuring the spectral
indices of electrons and protons 
\index{accelerated particles!energy spectra!spectral index} 
in the same flare, one can deduce the
magnetic field geometry under which these particles were
accelerated \citep{2005SSRv..121..165Z,2009JPlPh..75..159Z}.

We again stress that all of these results have been arrived at
using a test-particle approach, \index{test-particle
approach!limitations} in which the feedback of the electromagnetic
fields associated with the accelerated particles is not taken into
account.  Such a limitation is addressed in the
PIC approach\index{simulations!PIC} discussed in Section~\ref{sec:8pic}.

\subsection{Particle acceleration in 3-D MHD models with fan and spine reconnection}
\label{sec:8fan3d}
\index{reconnection!fan}
\index{reconnection!spine}

A further development in modeling particle acceleration in a
reconnecting current sheet involves more advanced MHD models of
fan and spine reconnection \citep{1996RSPTA.354.2951P}.
\index{acceleration!in current sheets!fan and spine geometry}
\index{magnetic structures!fan}
\index{magnetic structures!spine} 
\index{magnetic structures!null} 
The axis of symmetry of the magnetic field, called the {\it spine} and
designated here as the $x$ axis, is a critical field line
connecting to the null point. In the plane $x=0$, the magnetic
field lines are straight lines through the null point describing a
fan; hence, the $(y,z)$-plane is termed the {\it fan plane}. The
spine and the fan are the 3-D analogs of the separatrix
\index{separatrix} planes in the classic 2-D X-point geometry. The
field configuration near the potential magnetic nulls is given by
Equations~\ref{eqn:zharkova_main}~--~\ref{eqn:zharkova_guiding}.

Two regimes of reconnection at a 3-D magnetic null are possible:
{\it spine reconnection} and {\it fan reconnection}
\citep{1990ApJ...350..672L, 1996RSPTA.354.2951P}. While the
magnetic field configuration is the same in the two regimes, they
are characterized by different plasma flow patterns and different
electric fields. Spine reconnection \index{reconnection!spine}
\index{reconnection!fan} has a current concentration along the
critical spine field line, while fan reconnection has a current
sheet in the fan plane.
\index{current sheets!fan plane}

For a potential null\index{potential null},\index{magnetic structures!null!potential} the spine reconnection regime is
characterized by plasma flows lying in planes containing the
spine, while fan reconnection has non-planar flows that carry the
magnetic field lines in a ``swirling''-like motion symmetric about
the fan plane.  It should be noted that the electric field and
flow pattern are non-axisymmetric in all cases.  The model of
\citet{1996RSPTA.354.2951P} provides a useful analytical model for
initial studies; it allows some of the key features of topology
and geometry to be identified, and it is the closest 3-D analog of
the widely-studied constant out-of-plane electric field used in
2-D studies. It describes only the outer ideal reconnection\index{reconnection!ideal region} region
and does not include acceleration due to the parallel electric
field in the inner dissipation or resistive region; indeed, it
breaks down very close to the spine/fan, as the electric fields
formally diverge. However, since few particles will enter the
small-volume dissipation regions, the results are likely to be
appropriate for the bulk of the energy spectrum.
This model has been used as a basis for  particle acceleration at 3-D nulls in
the regimes of both spine reconnection \citep{2005A&A...436.1103D,
2006ApJ...640L..99D} and fan reconnection
\citep{2008A&A...491..289D}. Particle trajectories are obtained
numerically by solving the relativistic equations of motion
(Equation~\ref{eqn:zharkova_m_equa_r}). Again, a test-particle approach is
used (for limitations, see Section \ref{sec:8test_p}).

The strength of the driving electric field is quantified by the
parameter
\begin{equation} \label{eqn:zharkova_def_mu}
\tilde{\mu} = \frac {v_\perp^2}{v^2_E \, (L)},
\end{equation}
where $v_{\perp}$ is the perpendicular speed associated with the
gyro-motion, which may be equated to the thermal speed of the
plasma. This can also be expressed as the dimensionless magnetic
moment\index{magnetic moment!dimensionless} when speeds are normalized to the
drift speed \citep{1997PhPl....4.2261V}:
\begin{equation}
\tilde{\mu} = \frac {m v_\perp^2/2B_0} {m v^2_E \, (L)/2B_0},
\end{equation}
where $B_0$ is the amplitude of the magnetic field. If $\tilde{\mu} \ll 1$, the electric drift speed
at global length scales is strong compared with the thermal
gyro-motion and we are in the {\it strong drift regime}\index{drift speed!strong drift regime}
corresponding to fast reconnection; in practice, this regime is
reached when $\tilde{\mu} \approx 1$. In this case, particles may
undergo significant acceleration \citep{2005A&A...436.1103D} even
in the outer ``ideal'' reconnection region, similar to those
discussed in Section~\ref{sec:8single3D}.

\begin{figure}
\centering
\hspace*{20pt}
\includegraphics[width=0.5\textwidth]{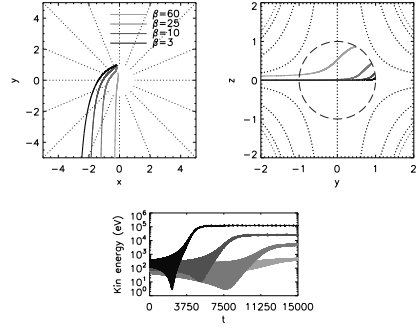}     
\caption {Trajectories and energy gains in fan reconnection, for
particles injected on the boundary at different latitudes $\beta$.
From \citet{2008A&A...491..289D}. } \label{fig:zharkova_fantraj}
\end{figure}

Typical individual trajectories in the fan reconnection are shown
in Figure~\ref{fig:zharkova_fantraj}, while similar plots for
spine reconnection can be found in \citet{2005A&A...436.1103D}.
Note the strong dependence of the energy gain on the injection
location, similar to those found for a single RCS in
Section~\ref{sec:8single3D}: particles injected closer to the fan
plane (at lower latitudes $\beta$) are accelerated the most as
they approach the strong-electric-field region near the fan plane.

\begin{figure}
\centering
\hspace*{20pt}
\includegraphics[width=0.5\textwidth]{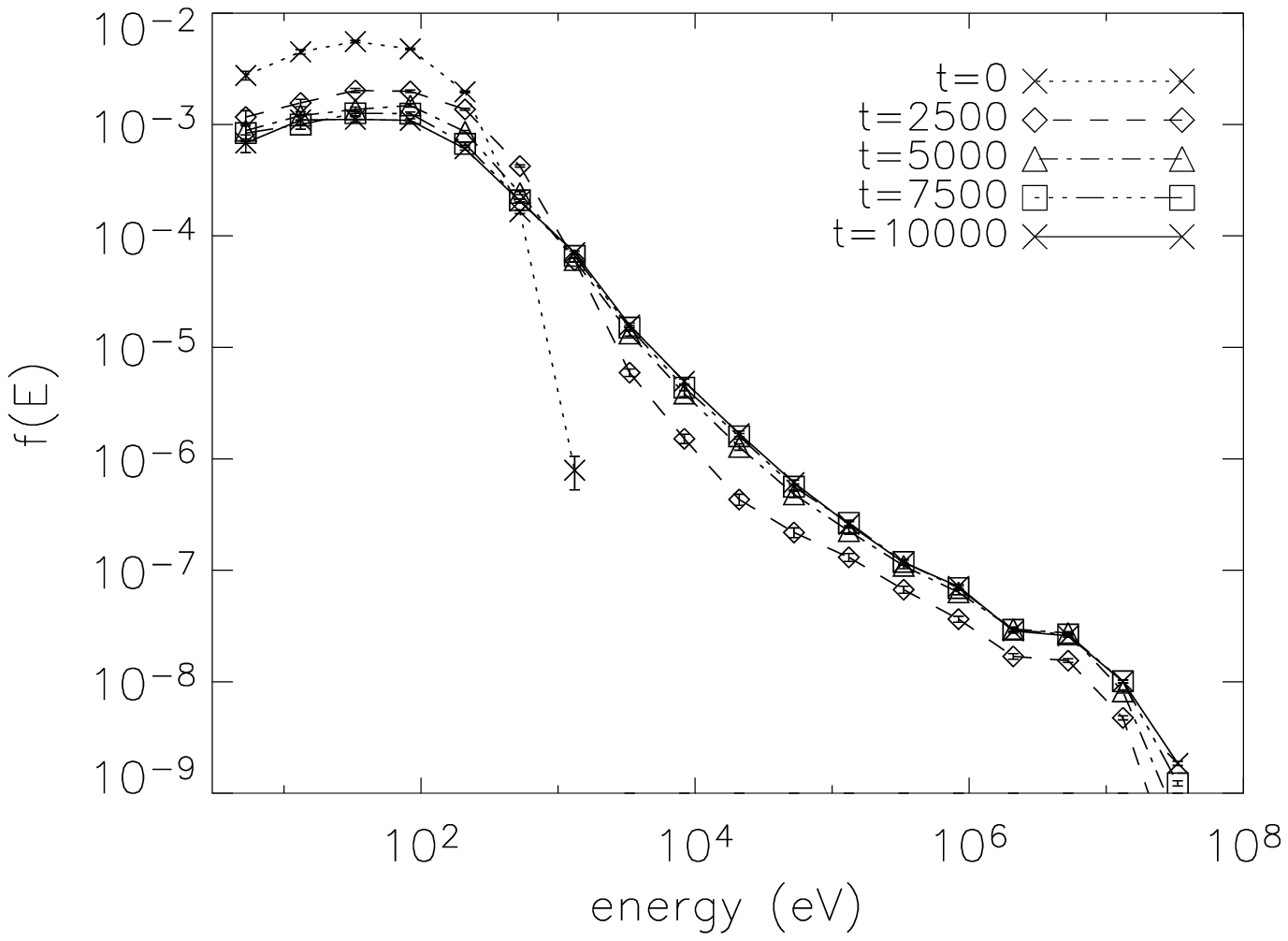}     
\caption {Energy spectra of particles in spine reconnection at
successive times, showing evolution towards steady state. From
\citet{2006ApJ...640L..99D}. } \label{fig:zharkova_spinespec}
\end{figure}

\begin{figure}
\centering
\hspace*{20pt}
\includegraphics[width=0.5\linewidth]{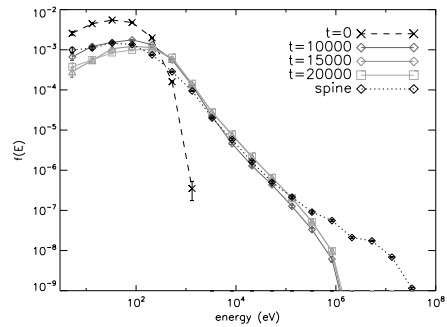}     
\caption {Energy spectra of particles in fan reconnection at
successive times, showing evolution towards steady state. From
\citet{2008A&A...491..289D}. } \label{fig:zharkova_fanspec}
\end{figure}

Energy spectra for the spine and fan reconnection
(Figures~\ref{fig:zharkova_spinespec}
and~\ref{fig:zharkova_fanspec}, respectively) show that a
significant fraction of protons can be accelerated to energies of
up to 10~MeV, comparable to the values obtained in the simple 3-D
topology presented in Section \ref{sec:8single3D}. As appears to
be natural during any reconnection process, a generally power-law
spectrum 
\index{accelerated particles!energy spectra!3-D reconnection} 
is obtained for the accelerated particles, but with a significant
``bump'' at higher energies, particularly for the spine
reconnection, for which the bump arises from particles approaching
close to the null itself.

This ``bump-on-tail'' distribution is similar to those reported
for a single 3-D current sheet in
Figure~\ref{fig:zharkova_p_en_sp_beta} in
Section~\ref{sec:8single3D} \citep[see
also][]{2005A&A...432.1033Z, 2009JPlPh..75..159Z} and is formed by
the two populations of protons: ``transit'' and ``bounced'' ones,
respectively. 
\index{distribution functions!bump-on-tail}
Note that the slopes of the spectra are similar for
the spine and fan reconnection in the intermediate energy range,
although fan reconnection is significantly less effective at
generating high-energy particles.

\begin{figure}
\centering \hspace*{20pt}
\includegraphics[width=0.8\textwidth]{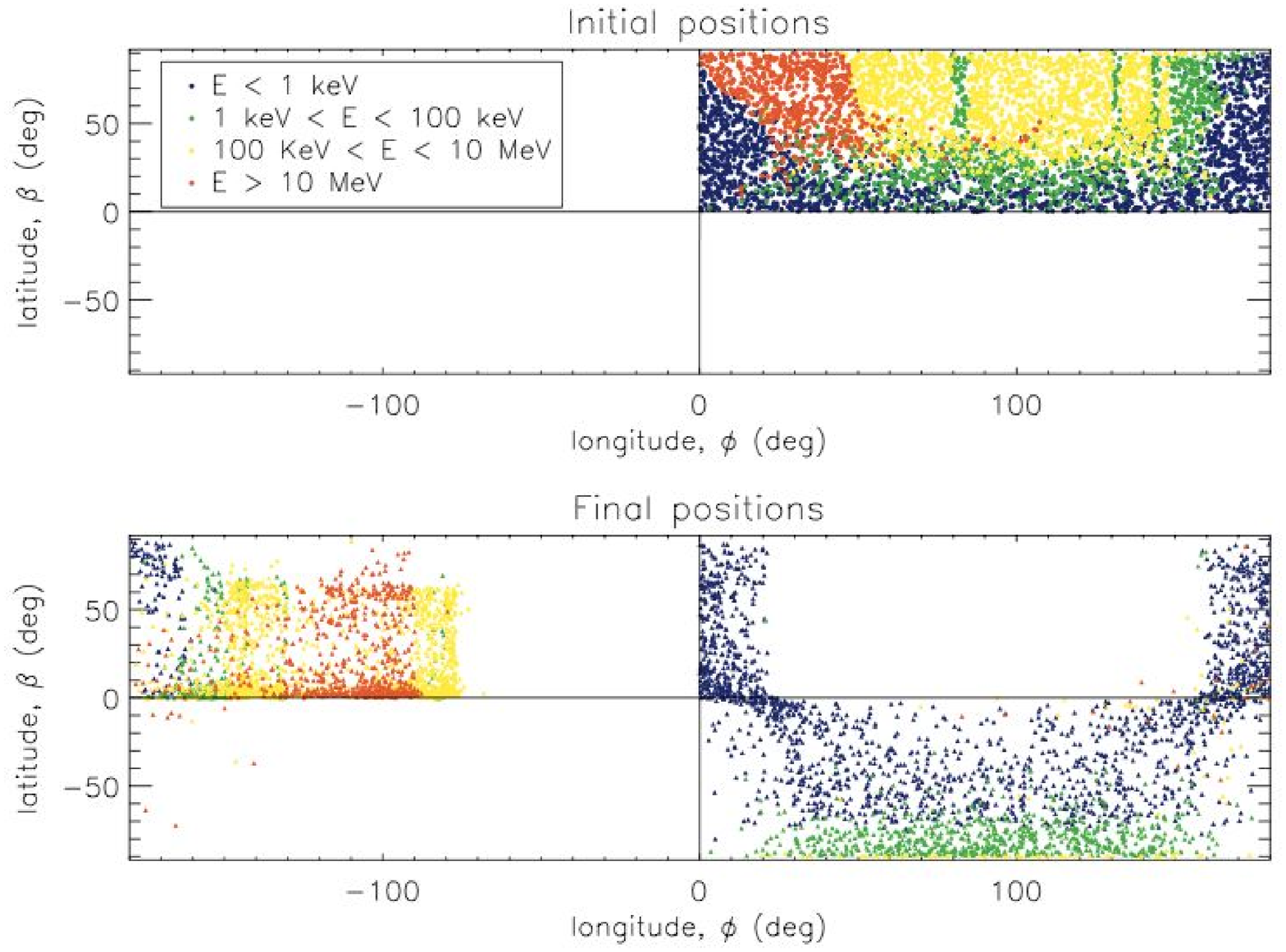}     
\caption {Positions of particles at initial and final times for
{\it spine} reconnection, color-coded according to their final
energies \citep[from][]{2006ApJ...640L..99D}. }
\label{fig:zharkova_spinepos}
\end{figure}

\begin{figure}
\centering
\hspace*{20pt}
\includegraphics[width=0.8\linewidth]{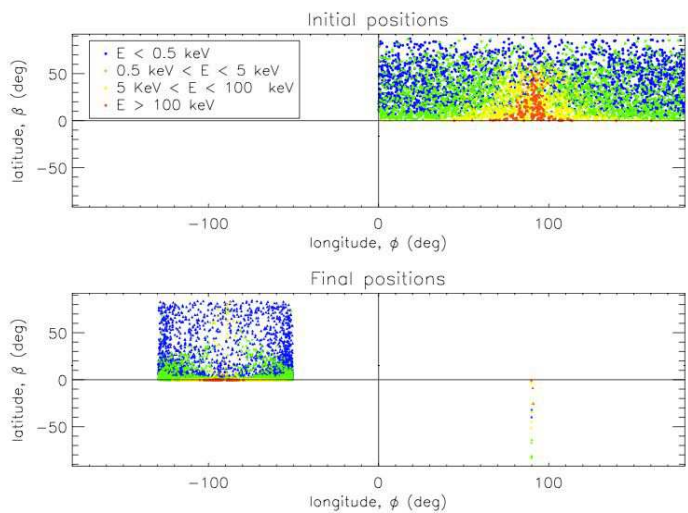}     
\caption {Positions of particles at initial and final times for
{\it fan} reconnection, color-coded according to their final
energies. From \citet{2008A&A...491..289D}. }
\label{fig:zharkova_fanpos}
\end{figure}

Detailed information on both the spatial distributions of the
accelerated of particles, and where these particles originate, are
shown in Figures~\ref{fig:zharkova_spinepos}
and~\ref{fig:zharkova_fanpos}, for spine and fan reconnection,
respectively. These show the angular location of the particles at
the initial (top plot) and the final (bottom plot) times. Each
location is identified by its latitude $\beta$, i.e., the
elevation angle from the $x$-$y$ plane, and longitude $\phi$.
Inflow regions are the top-right and bottom-left quadrants in the
$(\phi$, $\beta)$ representation, but for clarity, only particles
originating in the latter are shown (the other quadrant can be
added by symmetry). Each point represents one particle and is
color-coded according to the final particle energy.

\index{reconnection!spine and fan!test-particle acceleration}
In the case of spine reconnection, particles injected in the plane
in which the electric field is weakest (longitude $\phi = 0 $)
actually have the greatest energy gain, although their
acceleration times are longest \citep{2005A&A...436.1103D}. 
The high-energy particles which escape mostly emerge in jets along the
spine.\index{jets!reconnection outflow} 
However, most of the highest-energy particles actually
remain trapped in the magnetic field \citep{2006ApJ...640L..99D}.
The occurrence of these high-energy trapped particles may be an
explanation for coronal hard X-ray sources\index{hard
X-rays!coronal sources}\index{coronal sources} \citep{2008A&ARv..16..155K}.

In fan reconnection, particles starting near the fan plane are
those that gain the largest energy, as would be expected since the
fan plane is the location where the electric field is largest. As
one moves towards increasing latitudes, particles will reach the
fan plane with greater difficulty and so gain less energy.
Figure~\ref{fig:zharkova_fanpos} shows that, at the final time,
particles are confined to a range of longitudes
[-130$^{\circ}$,-50$^{\circ}$]. This is a result of the tendency
of the electric field drift in fan reconnection to push particles
towards the $x=0$ plane after they have passed the region near the
null. Figure~\ref{fig:zharkova_fanpos} also shows that the initial
azimuthal locations near $\phi$=90$^{\circ}$ are the most
favorable for particle acceleration since, near this plane, the
transverse component of the electric field drift, pushing
particles towards the fan plane, is largest. The fan reconnection
regime is less efficient in seeding test particles to the region
of strong electric field, compared with  spine reconnection, and
acceleration times are longer.

These results, obtained for a single-current-sheet reconnection
model with a more complicated magnetic and electric field topology
simulated by \citet{1996RSPTA.354.2951P}, advance further the
conclusions derived from the basic 3-D reconnection models
discussed in Section~\ref{sec:8single3D}. This approach confirms
that acceleration by a super-Dreicer electric field\index{electric fields!super-Dreicer} generated
during magnetic reconnection is a viable mechanism, which
naturally explains the energy input from magnetic field
dissipation and its output in the form of power-law energy spectra
of accelerated particles, with spectral indices related to the
magnetic field geometry occurring during the reconnection.
Apparently, this approach offers a good diagnostic tool for the
magnetic field topology from the observed signatures of
accelerated particles. We yet again note that the limitations of
the test-particle approach (see Section~\ref{sec:8test_p}) can
restrict this diagnostic capability, and that a full kinetic
approach to particle acceleration in such geometries is required.

\subsection{Particle acceleration in multiple current sheets formed by MHD turbulence} \label{sec:8complex}
\index{acceleration!in current sheets!multiple}

Although particle acceleration in a single current sheet can, in
principle, account for some flares or even flare ribbons, it is
highly likely that flares are due to a number of unresolved
magnetic tubes interacting with each other, so that more complex
coronal geometries with multiple acceleration sites\index{ribbons}\index{magnetic structures!complex} 
must be considered. 
One obvious benefit of such a scenario is that a particle can be accelerated
at more than one location, perhaps enhancing the energization
process.

\subsubsection{ Acceleration in Cellular Automata Models} \label{sec:zharkova_CA}
\index{flare models!cellular automaton}

The idea of multiple dissipation sites in flares is longstanding,
but the first serious effort at modeling this was by
\citet{1991ApJ...380L..89L}, who took the ideas associated with
self-organized criticality (SOC) and applied them to the solar
corona.\index{self-organized critical state}
They showed that if the right conditions were met, the
triggering of dissipation at a single point could lead to a
``spreading'' of dissipation across a large coronal volume and
that the distribution of event size as a function of energy
followed a power law (${\cal N}({\cal E}) \propto {\cal E}^{-1.8}$)
similar to that observed in flares
\citep[e.g.,][]{1993SoPh..143..275C}. However, a fundamental
question remains: can existing MHD models verify that the main
rationale behind the SOC theory is valid?

\begin{figure}
\centering
\includegraphics[width=1.0\textwidth]{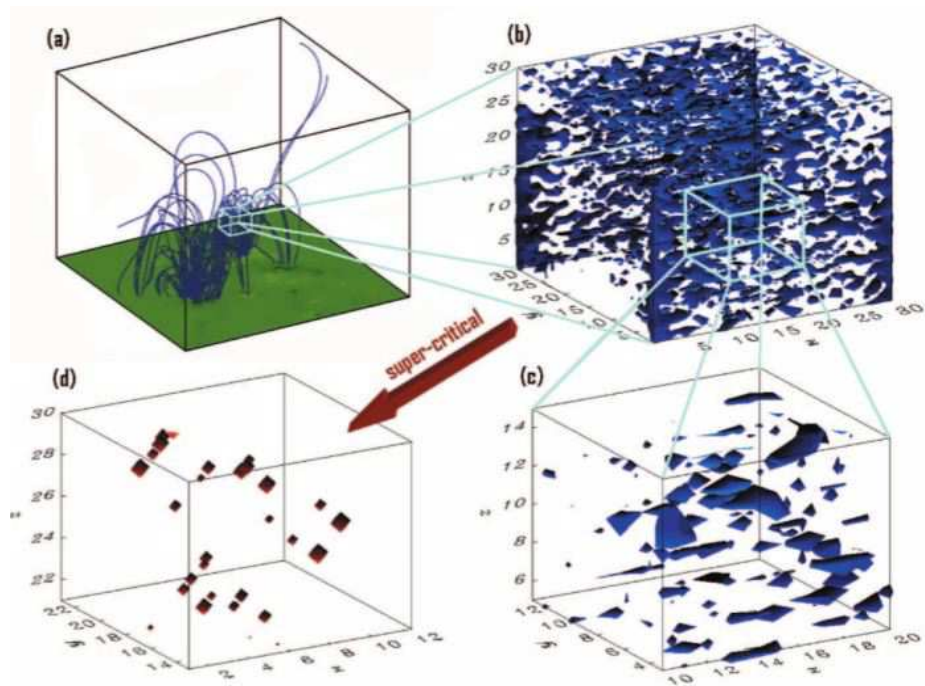}
\caption{The model of multiple current sheet formed by MHD
turbulence in the CA model. From \citet{2004ApJ...608..540V}.}
\label{fig:zharkova_vlahos1}
\end{figure}

Self-organized criticality models identify dissipation occurring
at many spatially-separated regions as illustrated in
Figure~\ref{fig:zharkova_vlahos1}, taken from
\citet{2004ApJ...608..540V}\index{flare models!cellular automaton}.
The upper left panel shows a coronal
field geometry, reconstructed from photospheric magnetograms, in a
volume with characteristic dimensions of $10^9$~cm. The current
density in a localized region is shown in the upper right panel,
and a snapshot of the locations where currents exceed the
threshold for dissipation in the cellular automaton model
\index{cellular automata} 
\index{acceleration!cellular automaton model}
(referred to as Unstable Current Sheets) is shown
in the bottom right panel. Energy release in this model occurs at
a number of locations of unstable current sheets (bottom
left panel).

\index{flare models!cellular automaton!illustration}

\begin{figure}
\centering
\includegraphics[width=0.8\textwidth]{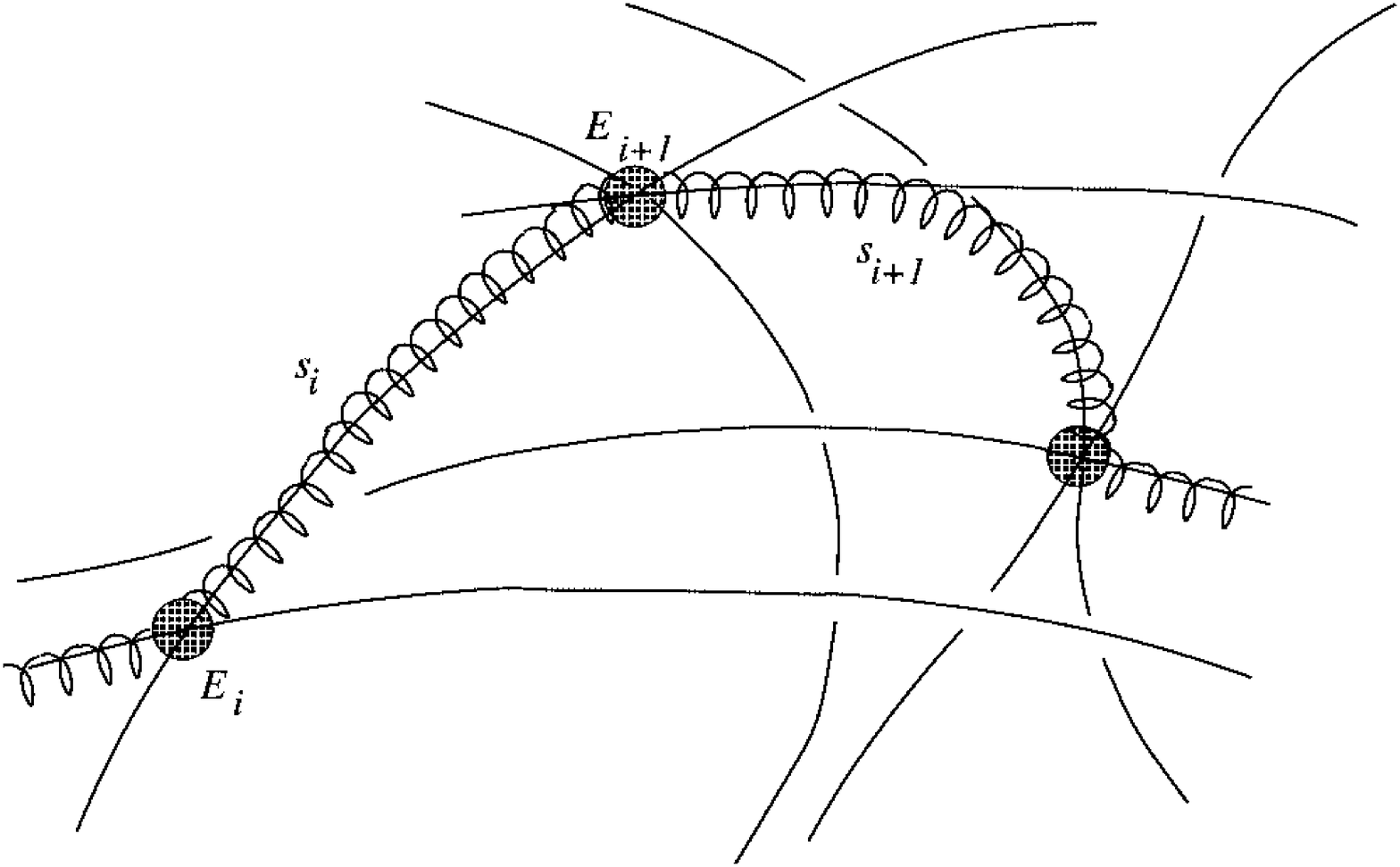}
\caption{Particle trajectories in the CA model. From
\citet{2004ApJ...608..540V}.} \label{fig:zharkova_vlahos2}
\end{figure}

To assess particle acceleration in such a coronal geometry,
\citet{2004ApJ...608..540V} have developed a model in which the
particles move from one unstable current sheet to another unstable
current sheet,
\index{acceleration!stochastic!multiple current sheets} 
\index{magnetic structures!multiple unstable current sheet}
gaining (or losing) energy at each and undergoing
ballistic motion in a background magnetic field in between. This
is illustrated in Figure~\ref{fig:zharkova_vlahos2}, where a
particle first interacts with a site $i$, escapes along the
magnetic field, encounters a site $i + 1$, escapes again, and so
on. The energy gain thus occurs in many small increments, rather
than at the single location of the current-sheet approach. By
using the self-organized criticality hypothesis, one can identify
the structure of the spatial distribution of the unstable current
sheets inside a complex active region which exhibits a specific
fractal structure \citep{2002PhRvE..65d6125M}.\index{magnetic structures!fractal dimension}\index{active regions!fractal structure}
Knowledge of
the spatial structure of the unstable current sheets can help us
to reconstruct the probability distribution of its spatial scales,
and thus,  the probability of
a particle colliding with a given element of it
\citep{2003PhRvE..67b6413I}. We do not yet understand the
statistical properties of the ``kernel'' -- the energy gain or
loss for a particle colliding with an unstable current sheet -- but
this is relatively easy to study using three-dimensional MHD
codes.\index{simulations!3-D MHD!and cellular automata}

We can therefore start from a global reconstruction of the
distribution of unstable current sheets and recover the kinetic
properties of the particles, under the assumptions of the
self-organized criticality theory. 
\index{probability densities!in cellular automata}: 
The acceleration is thus determined by three probability densities:
one, defining the distance between a pair of accelerators; the second, the electric
field strength at each accelerator; and a third, either the time
spent in, or the length of, the unstable current sheet. The first
two are assumed to be power laws, and the third is taken to be a
Gaussian. Note that this model only addresses acceleration by
direct\index{acceleration!direct electric field} electric fields,
since that can be simply parameterized at the dissipation sites.

In the example shown in Figure~\ref{fig:zharkova_vlahos1}, the
particle moves in a box of size $10^{10}$~cm and the distance
between accelerators ranges from $10^4$ to $10^{10}$~cm.  The
magnetic fields may be such that the particles become partially
trapped and the trajectory becomes rather complex. The upper limit
of the free travel distance is set to a value larger than the
coronal volume in order to allow particles to leave the coronal
region without undergoing any interaction with electric fields,
besides the initial one. Note that for the maximum time $\Delta t
= 1$~second for which the system is monitored, these escaping
particles will move at most a distance $c \Delta t = 3 \times
10^{10}$~cm, which is of the order of the length of the coronal
volume considered.

The electric field at each unstable current sheet varies between
sub-Dreicer and $10^8$ times \index{Dreicer field} Dreicer, and
the acceleration time has a mean value of $2 \times 10^{-3}$~s
\citep{2004ApJ...608..540V}. A large number of test particles are
tracked through this coronal geometry using the relativistic
equation of motion described by
Equations~\ref{eqn:zharkova_m_equa_r}, with acceleration taking
place very rapidly ($\ll 1$~s).

\begin{figure}
\centering
\includegraphics[width=1.0\textwidth]{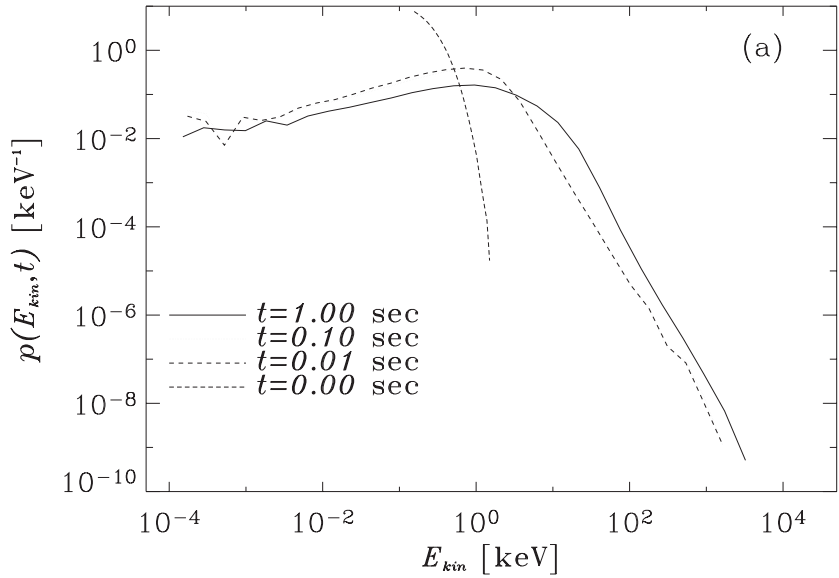}
\caption{Accelerated electron spectra in the CA model
\citep{2004ApJ...608..540V}.} \label{fig:zharkova_vlahos3}
\end{figure}
\index{flare models!cellular automaton!illustration}

Figure \ref{fig:zharkova_vlahos3} shows the distribution function
of electrons \index{accelerated particles!energy spectra} at four
different times. The initial distribution is a Maxwellian with a
temperature of 100~eV (dash-dot line). One can see that (a) prompt
acceleration to energies well in excess of 100~keV takes place,
(b) just under half of the particles end up in the high energy
tail and (c) the tail is a power law with slope $\delta \simeq 4$.
When the test particles are protons, much longer acceleration
times must  be considered to allow for similar energies to be
reached within the prescribed acceleration time.

Another approach based on the SOC models was proposed by
\citet{2004A&A...422..323A} to predict the kinetic energy
distributions of electrons. The acceleration of electrons is still
due to the presence of the randomly placed localized electric
fields produced by the energy release process and simulated by the
CA model\index{flare models!cellular automaton!PIC simulation}.
It is assumed that electrons interact successively with
reconnecting current sheets with the electric field linked to the
energy release time series. At each encounter, they randomly gain
or lose energy. Electrons can be accelerated from a Maxwellian
initial distribution up to a few~MeV, and can be represented either
by a power-law or an exponential distribution.

As the number of interactions increases, the electron energy
distribution begins to diverge from a well defined power law, and
an exponential tail develops
\citep[e.g.,][]{1998A&A...334.1099T,2003A&A...412..865V}.
Thick-target radiation from these energetic electrons was also computed,
\index{flare models!thick-target}
leading to hard X-ray spectra at low energies (10-100~keV)
which are much flatter than usually observed. At higher energies
(100-1000~keV), the X-ray spectra were consistent with the
observations at these energies.

\subsubsection{Acceleration in three-dimensional MHD turbulence}\label{sec:zharkova_MHD}
\index{acceleration!in MHD turbulence}

Numerical experiments
\index{acceleration!in MHD turbulence!test-particle experiments} 
for test-particle acceleration in the turbulent magnetic and electric
fields obtained from pseudo-spectral direct numerical solutions of
the compressible three-dimensional MHD equations, with the Hall
term included, have been carried out by
\citet{2003ApJ...597L..81D} and \citet{2006JGRA..11112110D}. In
both experiments electrons and protons were accelerated within
very short times before particles left the simulation box (with
length size of the order of a few turbulent correlation lengths).
The momentum distribution functions very quickly develop long
high-energy tails.

For electrons, there is no substantial difference found between
the results with and without the Hall term in the MHD solution,
while for protons significant differences are seen in the low-momentum part of the distributions. 
When a background uniform
magnetic field (the guiding field) is added, the particle
acceleration is anisotropic. Electrons develop large parallel
momentum, while protons develop large perpendicular momentum
similar to those reported for a single current sheet in
Section~\ref{sec:8single3D}.

\subsubsection{Acceleration in stochastic current sheets}
\index{current sheets!stochastic}

\citet{2005ApJ...620L..59T,2006A&A...449..749T} have investigated
the issue of particle acceleration in a 3-D coronal MHD model
developed by \citet{1996ApL&C..34..175G}. The initial setting for
the model is a straight stratified coronal loop with a magnetic
field which extends between the two photospheric footpoints. The
response of the loop to random photospheric motions is then
followed. A turbulent cascade develops, leading to a formation of
multiple current sheets \index{acceleration!stochastic current
sheets} in the corona. The resulting electric fields comprise both
the inductive electric field and the resistive one, the latter
appearing from the hybrid resistivity used in the model of
\citet{2002A&A...383..685G}\index{resistivity!hybrid}.
However, it is found that the impact
of the (mainly perpendicular) inductive field on acceleration is
weak. The distribution of current sheets changes in time with the
turbulence development, giving rise to the possibility of very
bursty particle acceleration events.

In order to study particle acceleration in this environment, the
magnetic field topology deduced from the MHD 
simulation\index{simulations!MHD!as framework for test-particle acceleration} is assumed
to be quasi-stationary, similar to that in
Section~\ref{sec:8basic}. Then, test particles are followed through
their motion in the electromagnetic fields associated with the
multiple current sheets, using the relativistic equations of
\index{equations of motion} motion~(Equation~\ref{eqn:zharkova_m_equa_r}).
The model also allows particles to leave the loop once they hit
the boundaries, thus limiting their energy gain. The simulation
results show different types of particle dynamics, with particles
both gaining and losing energy; however, not surprisingly, most
leave the loop with net energy gain.  About one quarter of the
particles do not interact with the current sheet at all and thus
gain no energy.

The results show that the resistive electric field, being a
parallel field, is the principal means of accelerating particles.
The acceleration is very fast and efficient, accelerating
particles to relativistic energies in a very short time. The
resulting distribution function of accelerated particles has a
clear double-power-law structure, with a thermal component at low
energies \citep{2006A&A...449..749T}.

The MHD model currently in use produces particles with higher
energies at somewhat shorter times compared to the Cellular
Automata model discussed above \citep{2004ApJ...608..540V}\index{flare models!cellular automaton}.
This
is likely because of different choices in the basic parameters,
especially the magnetic field strength and density. The same MHD
model accelerates both electrons and protons without any need to
adjust the electric fields, contrary to the CA-based model. The
double power law does not arise in the CA model because of the
possibility of particle loss in the MHD model. Both approaches
show that fragmented current sheets in the corona are a reasonable
scenario for particle acceleration.

\citet{2006ApJ...637L..61D} performed a test-particle numerical
experiment to simulate particle acceleration in the low-frequency
turbulent environment generated by footpoint motions in a
coronal loop.
\index{footpoints!motions!generation of turbulence}
Only the effect of the resistive electric field $E$
is retained, which is mainly parallel to the axial magnetic field.
In the spatial frequency\index{frequency!spatial} spectrum, the contribution of small
scales is dominant.
The spatial structure of $E$ is obtained by a
synthetic turbulence method ($p$-model)\index{turbulence!$p$-model}, which allows
intermittency with a power-law distribution to be reproduced.
Spatial intermittency plays a key role in acceleration, enhancing both
an extension of a power-law range, and an increase in the maximum energy.

By solving the relativistic equations of motion, the time
evolution of the particle distribution can be determined.
Electrons are accelerated to energies of the order of 50~keV in
less than 0.3~s, and the final energy distribution also exhibits a
power law. A correlation is found between heating events and
particle acceleration that is qualitatively similar to that in the
stochastic models discussed in Section~\ref{sec:8stochastic}.

\subsubsection{Limitations of MHD turbulence models} \label{sec:zharkova_complex_lim}
\index{acceleration!in MHD turbulence}
\index{caveats!MHD turbulent acceleration!list of problems}

With all their attractiveness in accounting for observed electron
distributions in solar flares, models invoking MHD turbulence in
an ensemble of current sheets nevertheless have a few essential
problems\index{acceleration!stochastic!limitations of models}.

First,\index{acceleration!multiple current sheets!issues} MHD
models involving multiple current sheets inevitably inherit the
same size- and time-step problems as the PIC
simulations discussed in Section~\ref{sec:8pic}, allowing us to
simulate MHD turbulence and electron acceleration in only a very
small region (about 1~m$^3$). It is then assumed, without
justification, that the results from this small volume can be
extended to the coronal scale.

Second, the basic model of a current sheet (discussed in
Section~\ref{sec:8basic}) is still used for the energy gains by
accelerated particles in these reconnecting current sheets, while
more realistic 3-D models of the reconnecting current sheets (as
discussed in Sections \ref{sec:8single3D} and \ref{sec:8fan3d})
show the energy gains to be significantly different.

Third, even if, as suggested above, multiple current sheets are
formed in the whole flare volume, the diffusion regions in each
have to be connected somewhere to reconnecting magnetic field
lines.  In a single current sheet model (see
Sections~\ref{sec:8basic} and \ref{sec:8single3D}), and as also
shown in the original magnetic field restoration from the
magnetograms taken at the photosphere \citep{2004ApJ...608..540V},
these lines are embedded in the photosphere.  The trajectory
simulations presented in Section \ref{sec:8single3D} show that the
particle drifts follow the magnetic field lines of the main
magnetic component and move towards the photosphere, not parallel
to the solar surface. Therefore, in the proposed multi-current
sheet models, the issue arises as to how the test particle can
move across the magnetic field, in order to get to the
another current sheet \citep[see, in this
context,][]{1995ApJ...446..371E}.

Last, but not least, a major problem arises from the fact that
these simulations inherit the most common limitations related to
the test-particle approach, as discussed \index{test-particle
approach!limitations} in Section~\ref{sec:8test_p}.

\subsection{Limitations of the test particle approach} \label{sec:8test_p}
\index{caveats!acceleration in test-particle approach}

As discussed above in Sections~\ref{sec:8reconnection} (basic
reconnection in a single 2-D current sheet), \ref{sec:8stochastic}
(acceleration by shocks and/or plasma turbulence),\index{plasma turbulence}
\ref{sec:8single3D} and~\ref{sec:8fan3d} (reconnection in 3-D
geometries), and~\ref{sec:8complex} (MHD models incorporating
multiple reconnection sites), a considerable amount of simulation
of the particle acceleration process involves a use of ``test
particles'' moving in a prescribed electromagnetic field
configuration. Such a test-particle approach is certainly useful
at some level -- for example, it allows us to investigate the
individual trajectories of electrons and protons inside 3-D
current sheets
\citep{2004ApJ...604..884Z,2005A&A...436.1103D,2005SoPh..226...73W},
and to obtain the aggregated energy distributions for a large
volume of particles
\citep{2005MNRAS.356.1107Z,2005SSRv..121..165Z,2005A&A...436.1103D,2007MmSAI..78..255B},
which can be compared with those inferred from observations.  The
approach can be used iteratively to extend the results of exact
PIC simulations to a much larger region (see
Section~\ref{sec:8pic}).

However, considerable caution must be exercised when applying a
test-particle approach to situations in which the number of
accelerated particles is sufficiently large (see
Section~\ref{sec:zharkova_e_numbers}). In such cases, the electric
and magnetic fields associated with the accelerated particles
themselves rapidly become comparable to the fields pre-supposed at
the outset.  Recall (Section~\ref{sec:8photons}) that a major
flare can accelerate some $10^{37}$ electrons~s$^{-1}$ and that
these electrons propagate within a magnetic flux tube of order
$10^9$~cm in radius. Attempting to model this as a single
large-scale ejection leads to the well-known \citep[e.g.,][]
{1985ApJ...293..584H,1995ApJ...446..371E} paradoxes that (1) the
magnetic fields induced by the resulting current vastly exceed the
pre-supposed field, and (2) that the electric fields resulting
from both charge separation and induction effects associated with
the accelerated particle stream (see Sections \ref{sec:8single3D}
and \ref{sec:8pic}) exceed plausible values by even more absurd
margins.
\index{acceleration!test-particle approach!limitations}
\index{caveats!acceleration in test-particle approach!specific problems}
Specifically:

\begin{itemize}

\item Amp\`ere's law, applied to a steady flow of $10^{37}$
electrons~s$^{-1}$ (current $I \approx 10^{18}$ Amps), in a cylinder
of cross-sectional radius $r \approx 10^7$~m gives a corresponding
magnetic field strength $B \approx \mu_o I/ 2\pi r \approx 3 \times
10^4$~T $\approx 3 \times 10^8$~Gauss.  Using a propagation volume $V
\approx 10^{21}$~m$^3$, the associated energy content $(B^2/2 \mu_o)
\, V \approx 10^{36}$~J $\approx 10^{43}$~ergs, at least ten orders of
magnitude greater than the energy in the electron beam itself;
\index{electron beams!inductive fields}

\item The typical rise time for a hard X-ray burst in a solar
flare is $\tau \approx 1 - 10$~s.  To initiate a current $I$ over a
timescale $\tau$ in a volume of inductance $L \approx \mu_o \ell$
(where $\ell$ is the characteristic scale) requires a voltage $V
\approx \mu_o \ell I/\tau$.  Alternatively, the largest current that
can appear when a voltage $V$ is applied is $I \approx V\tau/\mu_o
\ell$.  With the characteristic voltage associated with the
acceleration of hard-X-ray-producing electrons being $\sim$30~kV,
and with solar loop sizes $\ell \approx 10^7$~m, this gives a maximum
current $I \approx 10^4$~Amps, some {\it fourteen} orders of
magnitude less than the total current involved.

\end{itemize}
It is clear, therefore, that the self-consistent electromagnetic
fields associated with electron acceleration rapidly become
comparable to the fields pre-supposed to be present in the
acceleration region; in such an environment, a test-particle
approach breaks down severely.\index{acceleration region!test-particle approach}
Given this, it is a clear
imperative to take into account the self-consistent evolution of
the electric and magnetic fields, {\it including the fields
associated with the accelerated particles themselves}. Failure to
do so can result in unrealistic numbers of accelerated particles
predicted by acceleration models and even paradoxical results,
such as the energy in accelerated particles exceeding the free
energy in the pre-flare magnetic field configuration.

\subsubsection{The polarization electric field}
\label{sec:zharkova_p_elec_f}
\index{electric fields!polarization}
\index{electric fields!Hall}

\begin{figure}
\includegraphics[width=0.8\textwidth]{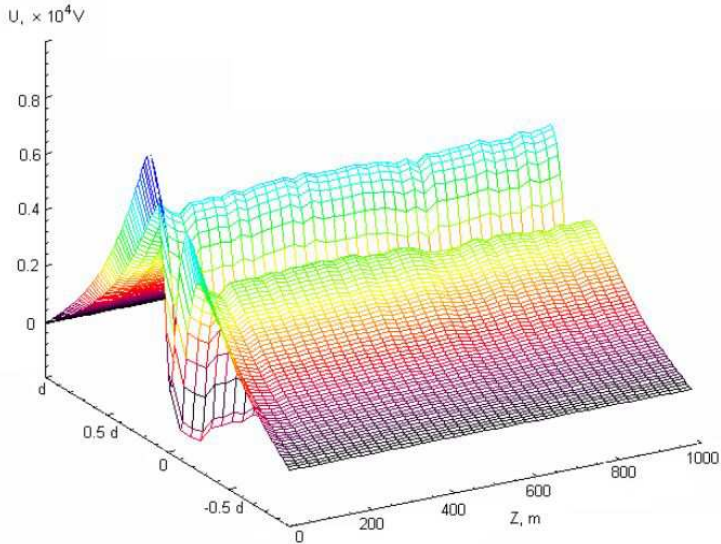}
\caption{Polarization electric field induced by the separation of
protons and electrons accelerated in a 3-D current sheet with
$B_0$= 10~G, at various distances $z$ from the null point 
\citep[from][]{2009JPlPh..75..159Z}.} \label{fig:zharkova_polar_field}
\end{figure}

As discussed in Section~\ref{sec:8single3D} (see
Figure~\ref{fig:zharkova_tr_asym}), reconnection in a 3-D current
sheet geometry leads to a separation of protons and electrons
across the midplane. As shown in Figure
\ref{fig:zharkova_polar_field}, such a charge separation results
in a substantial (up to $100$~V~cm$^{-1}$) Hall-type polarization
electric field across the midplane. As pointed out in
Section~\ref{sec:8reconnection}, inclusion of such Hall-type
electric fields can significantly  {\it increase} the
effectiveness of the whole magnetic reconnection process
\citep[][]{2001JGR...106.3715B,Huba2004,2005PhRvL..94i5001D};
see also Section~\ref{sec:8island}.

Electrons, because of their much smaller gyroradii, typically do
not show much change in behavior in the presence of the
polarization field.\index{polarization!electric field}\index{electric fields!polarization} 
On the other hand, protons, with their larger
gyroradii, gain slightly more energy when they move in the
positive $x$-direction than in the negative $x$-direction.  The
net (nonlinear) result is that protons gain energy as a result of
the polarization field, thus tending to make the proton energy
spectrum flatter. A full investigation of the effect of the
polarization electric field requires use of a
PIC approach, 
as described in Section~\ref{sec:8pic}.
\index{simulations!PIC!effect of polarization electric field}

\subsubsection{Turbulent electric fields}
\index{electric fields!turbulent}
Various turbulent electric fields typically form inside a
reconnecting current sheet -- these are caused, for example, by
the two-stream instability \citep{Bret2009} of the two electron
beams that enter the current sheet from opposite sides
\citep{2009JPlPh..75..159Z}.
\index{electron beams!and two-stream instability}
\index{plasma instabilities!two-stream}
\index{plasma instabilities!cold-beam}
Such electric fields, and their
associated currents, typically oscillate near the plasma frequency\index{frequency!plasma}
and have an exponential growth rate $\Gamma$ and an energy
saturation level $PE$ given by the analytical relations derived
for the cold-beam plasma instability \citep{1988SSRvE...6..425S,
Bret2009}:
\begin{equation}
\Gamma= \sqrt{1/3} \, \, 2^{-4/3} \, \Delta^{1/3} \omega _{pe},
\qquad \qquad PE = (\Delta/2)^{1/3} \, E_{\rm beam} \, ,
\end{equation}
where $\Delta$ is the ratio of the beam to plasma densities, and
$E_{\rm beam}$ is the beam kinetic energy. The effect of the
electric fields associated with such waves on the accelerated
particle distributions was estimated by
\citet{2009JPlPh..75..159Z}; this work confirmed the estimates of
\citet{Bret2009}.  Further progress requires full kinetic
simulations.

\subsection{Particle-in-cell simulation of acceleration in a 3-D RCS} \label{sec:8pic}
\index{current sheets!PIC simulations}
\index{simulations!PIC!3-D current sheet}

As has been pointed out repeatedly above, a full description of
the particle acceleration process in solar flares requires that
the electromagnetic fields and currents associated with the
accelerated particles be self-consistently included. In this
section, we describe early efforts to accommodate this crucial
element of the modeling.

\subsubsection{Description of the PIC approach}

In order to self-consistently include the electric and magnetic
fields induced by the accelerated particles, one can use the 2D3V
(PIC) simulation code \citep{Verboncoeur95}. This
PIC method\index{acceleration!particle-in-cell (PIC)} solves the
equations of motion for the plasma particles
(see Equations~\ref{eqn:zharkova_m_equa_r}), using values for the
electric and magnetic fields which include both the pre-specified
background fields $\mathbf{E}$ and $\mathbf{B}$ and the fields
$\mathbf{\tilde{E}}$ and $\mathbf{\tilde{B}}$ associated with the
accelerated particles, as calculated from Maxwell's equations.

In the 3-D simulations by \citet{Siversky2009}, the $y$-direction
is chosen to be invariant. The system is periodic in the
$z$-direction, so that a particle leaving the system through a
boundary far from the reconnecting current sheet (see bottom of
Figure~\ref{fig:zharkova_twoloops}) reappears at the opposite
boundary. The current sheet half-thickness is taken to be to
100~cm and the period along the $x$-direction is 2000~cm.  In the PIC
simulations discussed here, the computation region is extended well
beyond the null point in order to explore the effect on the
acceleration of different magnitudes of the transverse magnetic
field. Plasma is continuously injected from the $x = \pm1000$~cm
boundaries of the simulation region.

In order to avoid numerical instabilities in the PIC method,
constraints on the steps in time and space must be satisfied
\citep[for details, see][]{Siversky2009}.  To satisfy these
conditions while keeping the code running time reasonable, an
artificially low plasma number density $n=10^{10}-10^{12}$~m$^{-3}$ is
used, and the proton-to-electron mass ratio is artificially reduced
to $m_p/m_e=100$\index{simulations!PIC!small mass-ratio assumption}. 
The spatial simulation grid has from $10$ to
$100$ cells in the $z$ direction and $100$ cells in the $x$
direction with $\Delta z = \lambda_D$ (the Debye length) and
$\Delta x = \lambda_D /5$.  There are $\sim$100 particles per
cell on average and the time step is $6\times 10^{-10}$~s.

\subsubsection{PIC simulation results}\label{sec:zharkova_pic_density}

\begin{figure}
\centering
\includegraphics[width=0.45\textwidth]{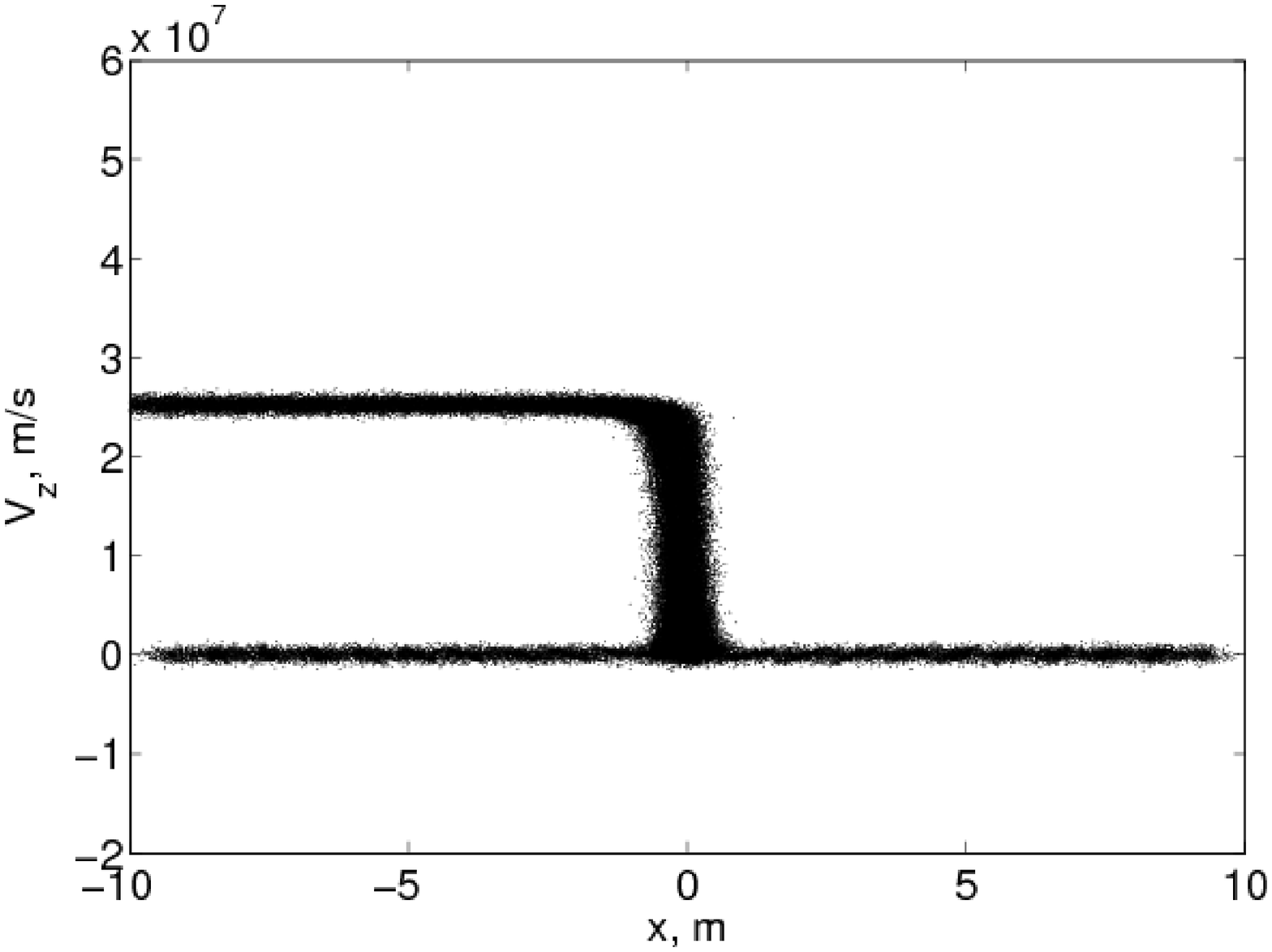}\qquad
\includegraphics[width=0.45\textwidth]{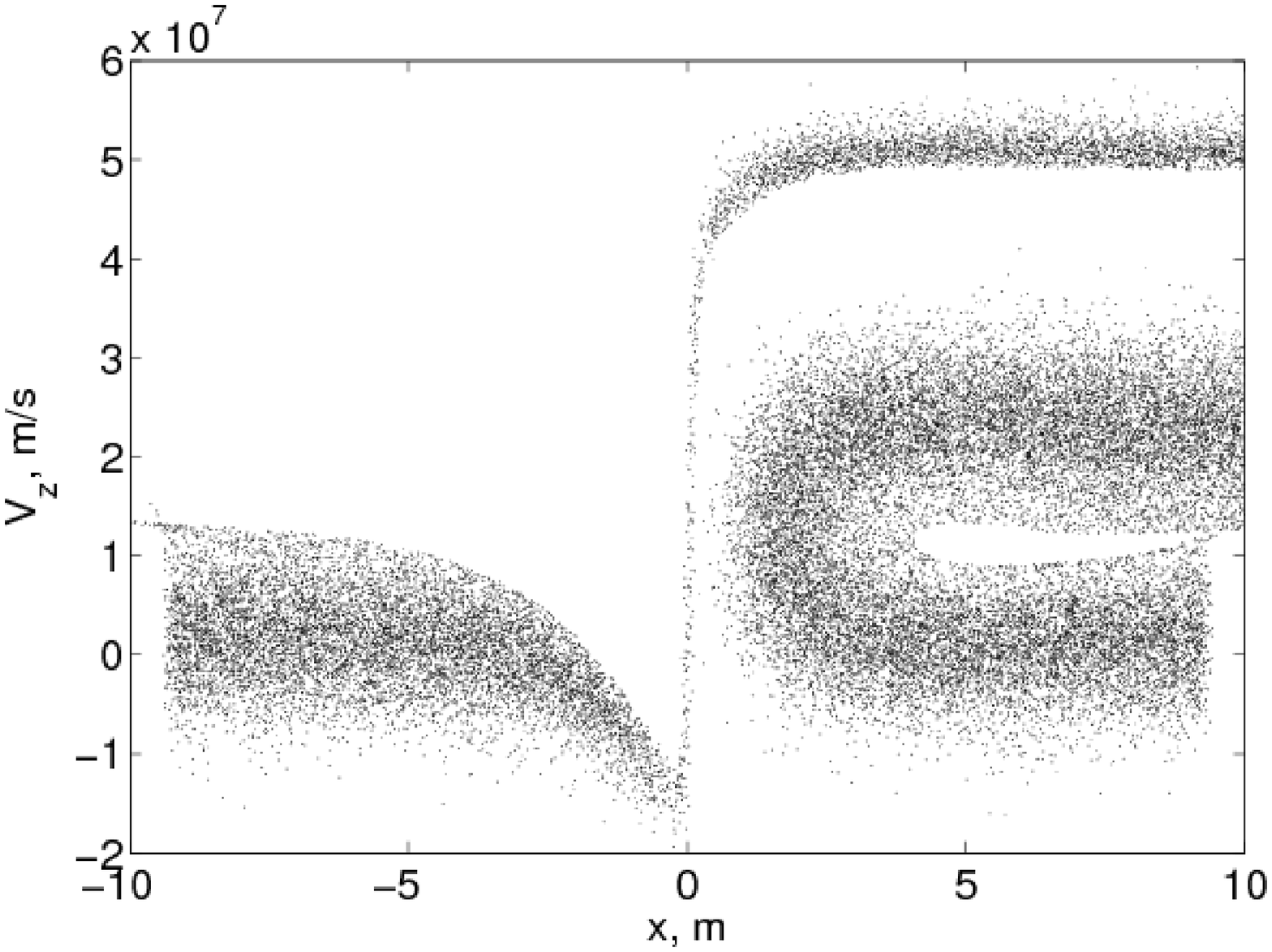}
\caption{PIC simulation snapshots of low-density ($n =
10^{10}$~m$^{-3}$) plasma particles on the $x$-$V_z$ phase plane for
protons (left plot) and electrons (right plot).  The protons and
the electrons enter the RCS from both sides and are ejected to
opposite semiplanes either as a beam (transit particles) or as a
flow (bounced particles); see the text for details.  The current-sheet
parameters are the same as in Section~\ref{sec:8single3D}
\citep[from][]{Siversky2009}.} \label{fig:zharkova_lownpne}
\end{figure}

The results of a simulation carried out with an extremely low
number density\index{simulations!PIC!low-density assumption} 
$n = 10^{4}$~cm$^{-3}$ are shown in
Figure~\ref{fig:zharkova_lownpne} for both protons (left panel)
and electrons (right panel).
This low-density simulation was
performed mainly to verify that the PIC code can reproduce the
results obtained in test-particle simulations (see
Section~\ref{sec:8single3D}, Figure \ref{fig:zharkova_tr_asym}).

Consistent with the test-particle results, the formation of
separate beams of accelerated protons and electrons is clearly
observed, and electrons and protons are ejected
\index{beams!electron-proton asymmetry}
\index{acceleration!electron-proton asymmetry} 
into the opposite
semispaces with respect to the $x=0$ midplane.  The protons are
largely unmagnetized inside the current sheet, so that the
ejection velocity of the ``bounced'' and
``transit'' protons are almost the same.
\index{transit particles}
\index{bounced particles}
On the other hand, the electrons are much more magnetized, so that
bounced electrons cannot reach the midplane and thus gain much
less energy than transit electrons. In this particular case,
however, the polarization electric field $\tilde{E}_x$ caused by
the charge separation is rather weak ($\sim$20~V~m$^{-1}$).
It must be noted, however, that the electron skin depth for this simulation is about 5000~cm, which
exceeds by a large factor the size (100~cm $\times$ 1000~cm) of the simulation region in both directions.
The ion inertial length, which is also instrumental in controlling the electron/proton
decoupling \citep{1994GeoRL..21...73M}, is about 1000~cm, comparable to the size of the simulation region.
Thus artificial decoupling of electrons and ion motions does result in this simulation.

\begin{figure}
\centering
\includegraphics[width=0.45\textwidth]{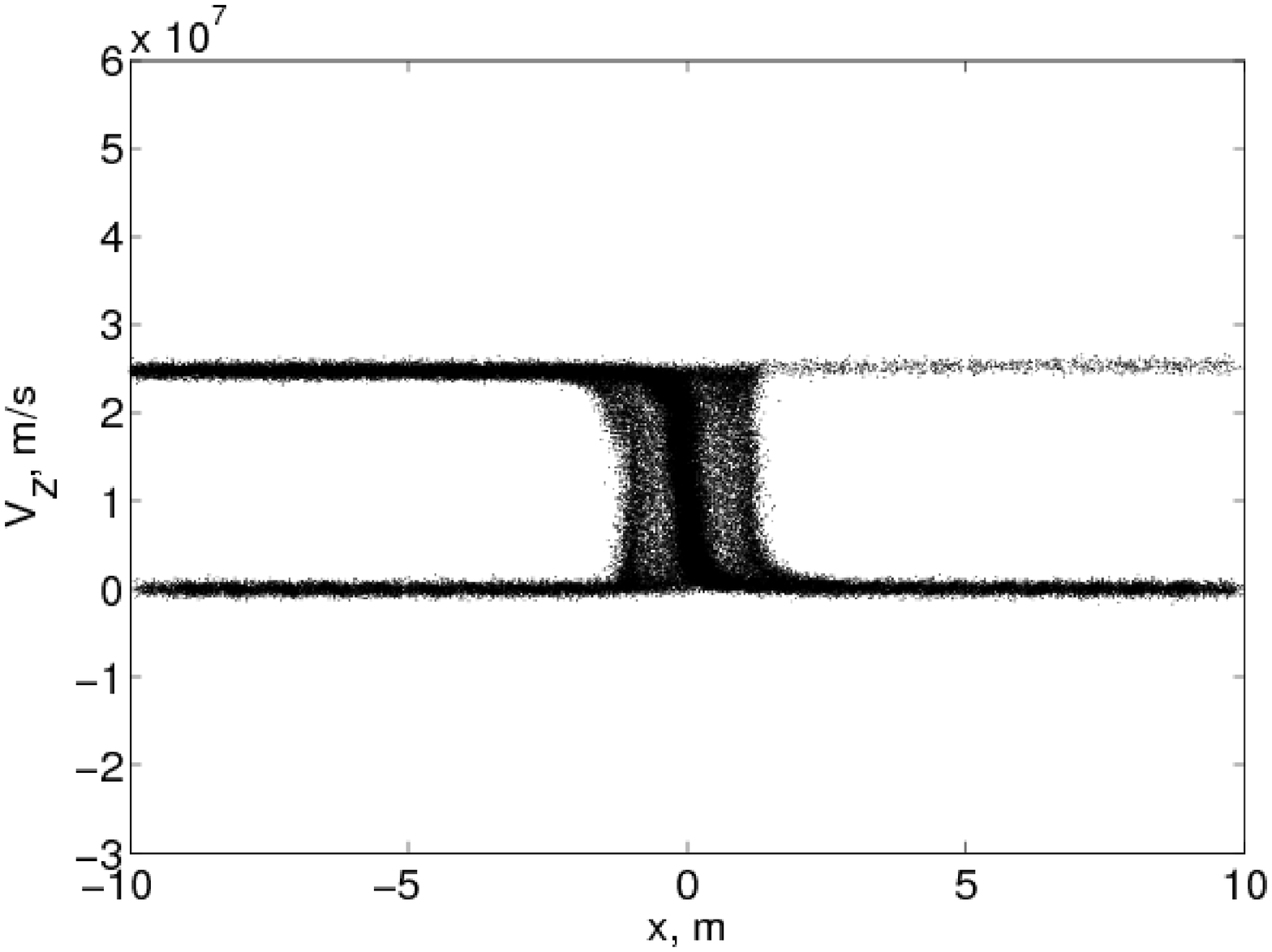}\qquad
\includegraphics[width=0.45\textwidth]{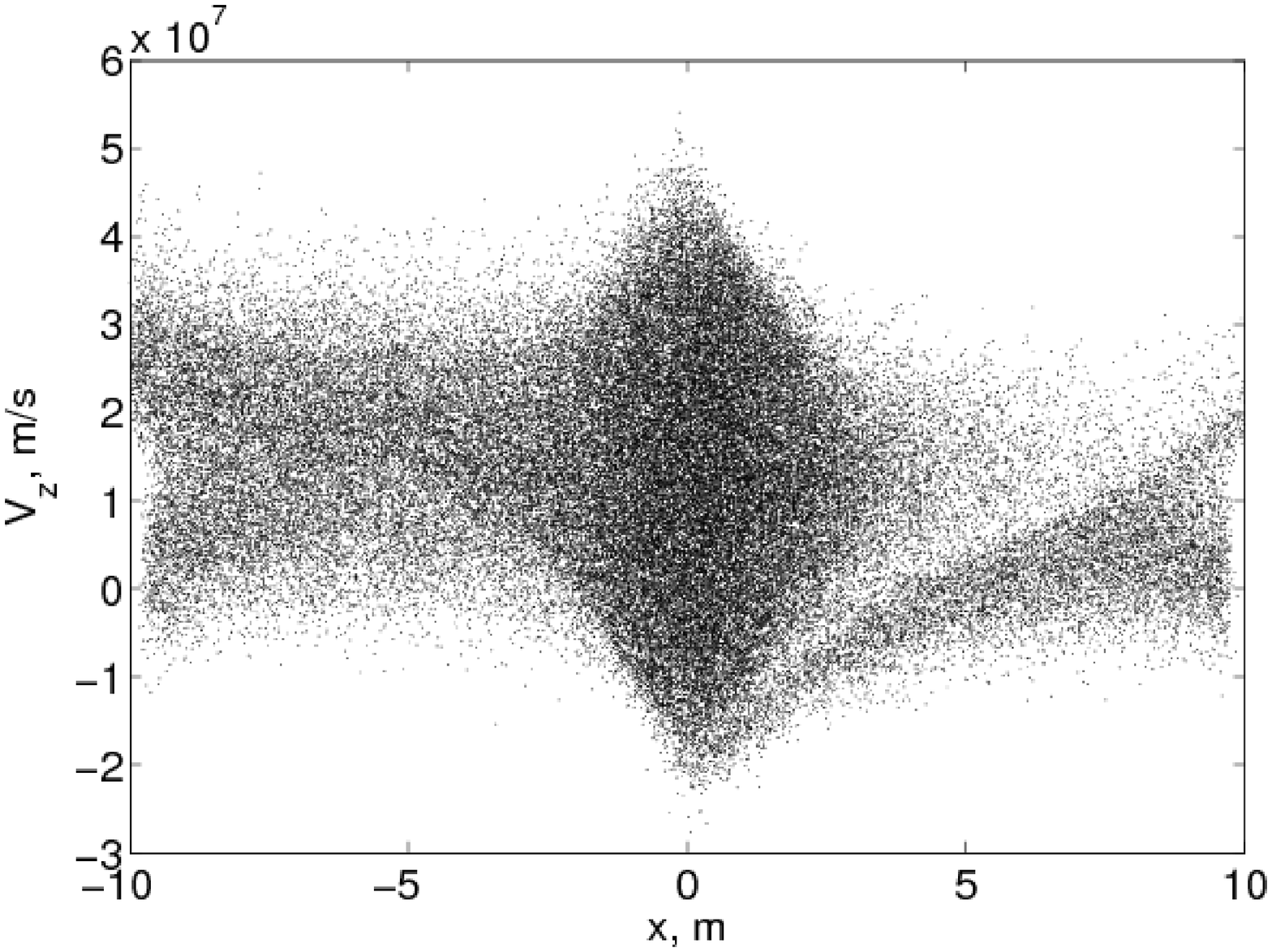}
\caption{Snapshots of higher-density plasma particles
($10^{12}$~m$^{-3}$) on the $x$-$V_z$ phase plane (PIC simulations)
for protons (left plot) and electrons (right plot) entering the
RCS from both sides. Protons are still ejected to the same
semiplane, while the electrons form a cloud which experiences
multiple ejections and returns to the acceleration region.  As a
result, the majority of the electrons ends up being ejected to the
same semiplane as the protons, with only a small fraction of
electrons being ejected into the opposite semiplane (see the text
for details). 
The current sheet parameters are the same as in
Section~\ref{sec:8single3D}  \citep[from][]{Siversky2009}.}
\label{fig:zharkova_hinpne}
\end{figure}

\begin{figure}
\centering
\includegraphics[width=0.8\textwidth]{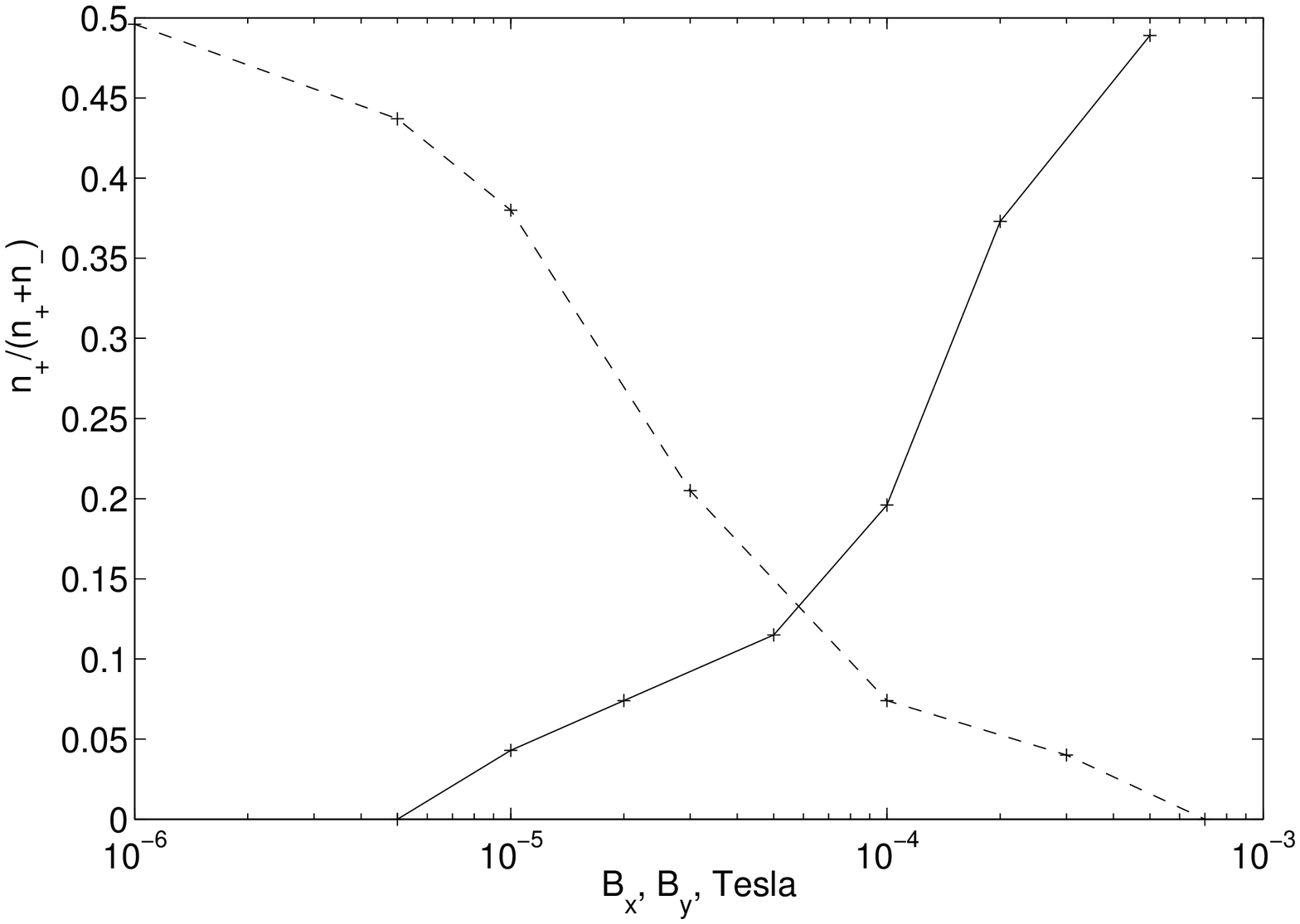}
\caption{The fraction of protons that are ejected into the $x>0$
semispace,  {\it Dashed line}: fraction vs. $B_{y0}$ ($B_{x0} =
0.2$~G).  {\it Solid line}: fraction vs. $B_{x0}$ ($B_{y0} =
1$~G). All other parameters of the current sheet are the same as
in Figure~\ref{fig:zharkova_hinpne}.} \label{fig:zharkova_separ}
\end{figure}

Simulations for a higher density, $n = 10^{6}$~cm$^{-3}$, are shown in
Figure~\ref{fig:zharkova_hinpne}, for protons (left panel) and
electrons (right panel). 
\index{skin depth} 
For this density the electron skin depth is reduced to $\sim$500~cm,
which is within the size of the simulation region in either direction.  Thus the decoupling of
electron and ion motions is not an artefact of the size of the region compared to the skin depth.
Although the assumed density is several orders of magnitude lower than in the solar corona (often taken to be $10^{10}$~cm$^{-3}$),
already several differences between the test-particle and PIC approaches appear. 
The dynamics of the relatively unmagnetized protons
are not substantially different: their trajectories are close to
those obtained from the test-particle simulations
(Figure~\ref{fig:zharkova_tr_asym}).
They are still ejected mainly
into one semispace ($x<0$) with respect to the midplane, and their
acceleration rate coincides with the theoretical magnitude given
in Section~\ref{sec:8basic}.

In addition, however, a small number of protons is ejected to the $x>0$ semispace, where
electrons are normally ejected in test-particle simulations
(Figure~\ref{fig:zharkova_hinpne}, left panel).
Figure~\ref{fig:zharkova_separ} shows the fraction of the protons
ejected into the $x>0$ semispace as a function of $B_{x0}$ and
$B_{y0}$; the dependence on $B_{y0}$ qualitatively coincides with
that obtained by \citet{2004ApJ...604..884Z}. For small $B_{y0}$,
both PIC and test-particle simulations show that the protons are
ejected symmetrically with respect to the midplane. However, in
the PIC simulation, all the protons are ejected into the one
semispace ($x<0$) when $B_{y0} > 7$~G ($\approx B_{z0}$), while in
the test-particle simulation \citep{2004ApJ...604..884Z}, the
accelerated particles are fully separated if $B_{y0}
> 1.5 \times 10^{-2} B_{z0}$. 
Thus, in PIC simulations, the particle trajectories have
less asymmetry than those in the test-particle approach.

\subsubsection{Polarization electric field induced by accelerated particles}
\index{electric fields!PIC simulations!induced polarization field}

\begin{figure}
\centering
\includegraphics[width=0.8\textwidth]{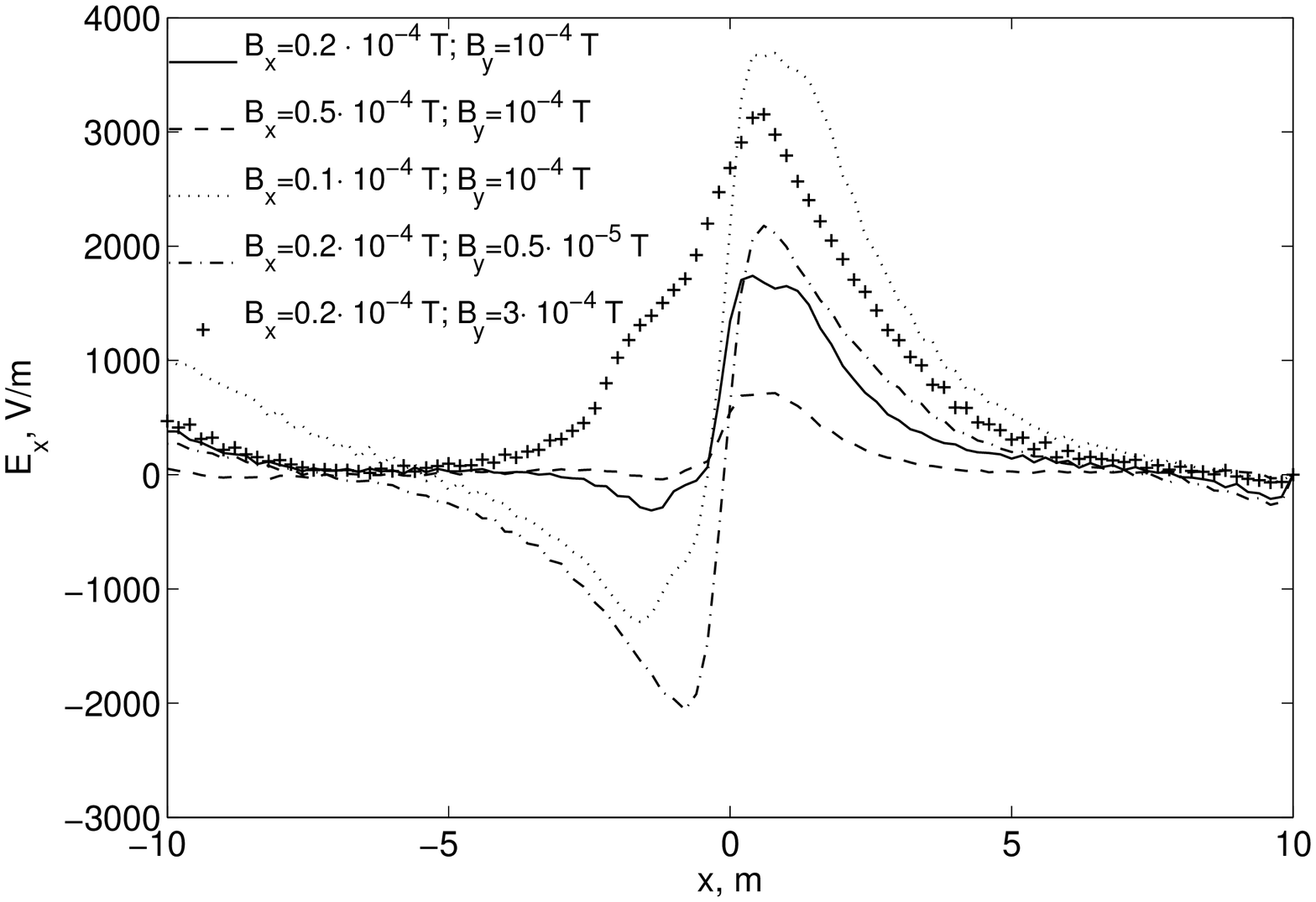}
\caption{Electric field $\tilde{E}_x$ induced by particles in the
PIC simulations for different values of $B_{x0}$ and $B_{y0}$. All
other parameters of the current sheet are the same as in
Figure~\ref{fig:zharkova_hinpne}.} \label{fig:zharkova_E_x}
\end{figure}

The PIC simulations show that the induced magnetic field
$\mathbf{\tilde{B}}$ is typically much smaller than the background
$\mathbf{B}$. On the other hand, inclusion of the induced {\it
electric} field $\mathbf{\tilde{E}}$ is essential. Indeed, its
absolute value is much larger (by an order of magnitude) than the
field $E_y$ induced by the reconnection itself (see
Sections~\ref{sec:8single3D} and~\ref{sec:zharkova_p_elec_f}).

Figure~\ref{fig:zharkova_E_x} shows the value of the charge
separation electric field
\index{electric fields!polarization!induced by asymmetric acceleration} 
$\tilde{E}_x$, perpendicular
to the current sheet, averaged over the $z$ coordinate, for
different values of $B_{x0}$ and $B_{y0}$. The field becomes
stronger when either $B_{x0}$ decreases or $B_{y0}$ increases.
Also, when $B_{y0} \rightarrow 0$, the system becomes symmetric, and
accordingly so does $\tilde{E}_x$. The distribution of the
associated charge density $\rho(x) = (1/4\pi) \, \partial
\tilde{E}_x/\partial x$ has been presented by
\citet{Siversky2009}.

\subsubsection{Particle trajectories} \label{pic_traj}
\index{simulations!PIC!particle trajectories}

As discussed above, proton trajectories do not change much in the
presence of the polarization field, the only difference being that
the ``bounced'' protons have somewhat larger orbits
\citep{Siversky2009}, leading to two smaller peaks at about $\pm
1$~m in the charge density plot \citep[Figure~7
in][]{Siversky2009}.

The trajectories of electrons are much more complicated. First,
electrons which enter from the $x<0$ semispace have dynamics
similar to those of the ``transit'' electrons: they drift towards
the midplane, become accelerated and are ejected to the $x>0$
semispace. However, the polarization field $\tilde{E}_x(x)$, which
extends well beyond the current sheet thickness and has a
component parallel to the magnetic field, acts to decelerate the
ejected electrons. For the chosen magnitudes of $B_x$ and $B_y$,
the majority of electrons cannot escape to the $x>0$ semispace;
instead, they are dragged back multiple times to the current sheet
and after some time become indistinguishable from the electrons
that entered from the $x>0$ semispace.

The electrons that come from the $x>0$ semispace
demonstrate rather different dynamics compared with the case when
$\tilde{E}_x = 0$. These electrons, which are ``bounced'' from the RCS
in the absence of an $\tilde{E}_x$, can now reach its midplane. In
the vicinity of the midplane, electrons become unmagnetized and
oscillate with a gyro-frequency\index{frequency!Larmor} determined by the magnitude of
$B_y$ \citep[Figure 8 in][]{Siversky2009}. After a large number of
oscillations, the electron is ejected. If the electron's initial
velocity is small, it can become quasi-trapped inside the RCS,
similar to the situation in the vicinity of an X-type null point
\citep{1997SoPh..172..279P}. Such electrons are accelerated on the
midplane, ejected from it, then decelerated outside the RCS and
return back to the midplane. This cycle is repeated until the
electron finally gains enough energy to escape the RCS. Since the
magnitude of the polarization field $\tilde{E}_x(x)$ is smaller at
$x<0$ than at $x>0$ (see Figure~\ref{fig:zharkova_E_x}), it is
easier for the electron to escape to the $x<0$ semispace. Thus, eventually,
{\it most of the electrons are ejected to the same semispace as
protons}, contrary to the results of the test-particle simulations,
while creating an electron cloud around an RCS before their joint ejection with protons.
It seems that the asymmetry condition is shifted to much higher
magnitudes of the guiding field than those found in the
test-particle approach
\citep[$>$0.01B$_{z0}$,][]{2004ApJ...604..884Z}.

\subsubsection{Energy spectra}
\index{simulations!PIC!energy spectra}
As shown by \citet{Siversky2009}, the ejected electrons, in general,
form a wide, single-peaked energy distribution with the width
being of the order of the mean energy. This is different from the
test-particle simulations, in which two narrow-energy electron
beams are formed. Consistent with Section~\ref{sec:8single3D}, the
mean electron energy still strongly depends on $B_x$, with lower
$B_x$ corresponding to a higher mean energy and a wider energy
distribution.

\subsubsection{Turbulent electric field}
\index{simulations!PIC!turbulent electric field}

\begin{figure}
\centering
\includegraphics[width=\textwidth]{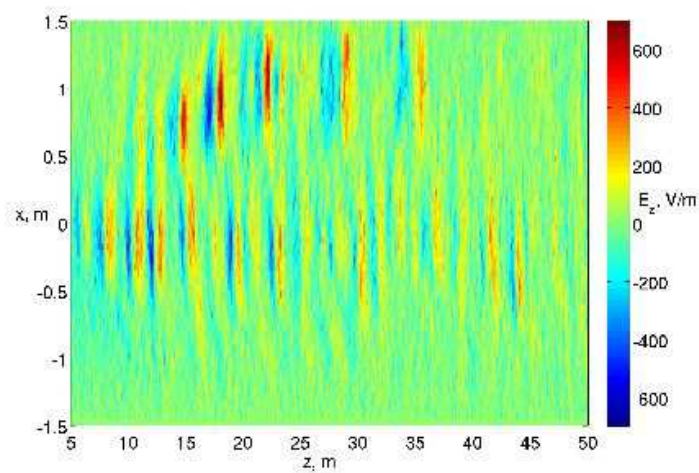}
\caption{Electric field $\tilde{E}_z$ induced by particles in PIC
simulation ($B_{z0} = 10^{-3} T $, $B_{y0} = 10^{-4} T$, $B_{x0} =
4 \times 10^{-5} T$, $E_{y0} = 250$~V~m$^{-1}$, $m_p/m_e=10$, $n =
10^{6}$~cm$^{-3}$).} \label{fig:zharkova_E_z}
\end{figure}

\begin{figure}
\centering
\includegraphics[width=0.8\textwidth]{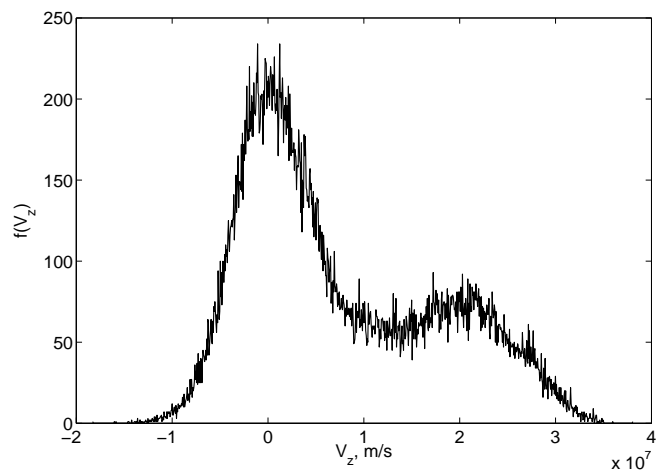}
\caption{Distribution function of electrons calculated for the
region with turbulence plotted in Figure~\ref{fig:zharkova_E_z}.}
\label{fig:zharkova_distr}
\end{figure}

Another interesting effect observed in the PIC simulations for the selected physical conditions and
proton-to-ion mass ratio is the excitation 
\index{electric fields!PIC simulations!turbulent} 
of plasma waves.

As can be seen from Figure~\ref{fig:zharkova_E_z}, the $\tilde{E}_z$
component of the induced electric field is structured with a
characteristic length scale of $\lambda_{\rm wave} \approx
2$~m. This structure propagates in time in the positive
$z$-direction at a speed $V_{\rm wave} \approx 1.3 \times
10^{7}$~m~s$^{-1}$. This corresponds to an oscillation period
$T_{\rm wave} \approx 1.5 \times 10^{-7}$~s, consistent with the
plasma frequency for a density $n = 10^{6}$~cm$^{-3}$. 
Note, also, that the oscillating component of the excited wave $\tilde{E}_z$
is parallel to the direction of propagation which corresponds to
the polarization of the\index{waves!Langmuir} 
\index{Langmuir waves} 
Langmuir wave.  
The instability mechanism is evident from study of the electron
velocity $V_z$\index{accelerated particles!energy spectra}. As
shown in Figure~\ref{fig:zharkova_distr}, the electrons have an
unstable ``bump-on-tail'' distribution\index{plasma instabilities!bump-on-tail}\index{electrons!distribution function!bump-on-tail}.
The range of velocities
$V_z$ for which the derivative $\partial f/\partial V_z$ is
positive is from $1.3 \times 10^{9}$ to $2 \times
10^{9}$~cm~s$^{-1}$, corresponding to the phase velocity of the
associated Langmuir wave $V_{\rm wave}$.

The use of plasma instabilities to produce anomalously high values of resistivity 
\index{resistivity!anomalous}
in the reconnection region have been studied by many authors.
\index{plasma instabilities!lower hybrid drift} 
\index{plasma instabilities!Kelvin-Helmholtz}
\index{plasma instabilities!kink}
\index{plasma instabilities!sausage}
\index{plasma instabilities!in PIC simulations!list}
A comprehensive review of these results
is beyond the scope of this article; however, a brief commentary is appropriate.
Both 2-D~and 2.5-D~PIC simulations have revealed various types of instabilities to be present,
from the lower hybrid drift instability (LHDI) leading to the Kelvin-Helmholtz instability
\citep{1960ecm..book.....L,Chen1997, Lapenta2003} to kink \citep{Pritchett1996,Zhu1996,Daughton1999} and
sausage \citep{Buchner1999} instabilities. 
However, full 3-D PIC simulations \citep[with an artificially
reduced speed of light $c=15 V_{Ae}$ and proton-to-electron mass ratio of 277.8;][]{Zeiler2002},
while retaining the lower hybrid drift instability, reveal no evidence of kink or sausage instabilities.
This indicates that the appearance of various instability modes may be dependent on the details of the simulation used, and hence argues for an abundance of caution at this time in applying these results to actual flares.

\subsection{Particle acceleration in collapsing magnetic islands} \label{sec:8island}

\subsubsection{Tearing-mode instability in current sheets} \label{sec:zharkova_tearing}
\index{current sheets!instability}
\index{simulations!PIC!collapsing islands}

As mentioned in Section~\ref{sec:8reconnection}, in reconnecting
current sheets, diffusion converts magnetic energy into Ohmic heat,
with a concomitant slow expansion of the current sheet at a rate
proportional to the resistivity.  
Consequently, one way of
enhancing the reconnection rate is to consider enhancement of the
resistivity through resistive instabilities such as the tearing
instability \citep{1963PhFl....6..459F}.
\index{plasma instabilities!tearing mode}

In a current sheet wide enough to meet the condition
$\tau_{dif}\gg \tau_A$, where $\tau_A= d/V_A$ is the Alfv\'en
traverse time, the tearing instability occurs on a timescale of
$\tau_{dif}(\tau_A /\tau_{dif})^{\lambda}$, with the parameter
$\lambda$ in the range $0 < \lambda < 1$. The tearing instability
occurs on wavelengths greater than the width of the current
sheet, ($ka < 1$), with a growth rate \citep{2000mare.book.....P}:
\begin{equation}
\omega\,=\,\left [\tau_{dif}^3\tau_{A}^2 (ka)^2 \right ]^{-1/5} \,
, \label{eq:8tearing_growth}
\end{equation}
where the wave number $k$ satisfies the condition
$(\tau_{dif}/\tau_{A})^{1/4}< \,ka \,< 1$. The fastest growth is
achieved for the longest wavelength (i.e., at $ka=1$), while the
smallest wavelength of the instability grows more slowly, on a
timescale $\tau_{dif}^{3/5}\tau_{A}^{2/5}$. As first demonstrated
by \citet{1963PhFl....6..459F} \citep[see
also][]{1977SoPh...53..305S}, tearing-mode instabilities generated
during reconnection in a current sheet can produce closed magnetic
``islands''\index{magnetic structures!islands} 
with an elliptical O-type topology. 
Accordingly, large current sheets\index{current sheets!observed in flares} often observed in solar flares \citep[e.g., see][and references
therein]{2003ApJ...596L.251S} can, depending on their sizes and
conditions, be subject to the formation of magnetic islands. For
example, for a current sheet of size $\sim$10,000~km \citep[as
reported by][]{2003ApJ...596L.251S}, embedded in a (coronal)
medium with magnetic field $B \approx 100$~G and density $n =
10^{10}$~cm$^{-3}$, then the Alfv\'en velocity is about
2000~km~s$^{-1}$ and $\tau_{A} \approx 5$ sec.  With a rather
conservative collision~frequency estimate $10^4$~s$^{-1}$,
$\tau_{dif} \approx10^{15}$ sec. Substituting these values into
Equation~\ref{eq:8tearing_growth} gives an island formation time
in the range $10 - 1000$~s.

\subsubsection{Studying particle acceleration in magnetic islands using a PIC approach} 
\label{sec:zharkova_pic_island}
\index{simulations!PIC!acceleration in magnetic islands}

In the past decade, the tearing-mode instability has been
extensively investigated, both in 2.5-D current sheets
\citep{Drake1997,1998GeoRL..25.3759S,2005PhRvL..94i5001D,2006GeoRL..3313105D,2006Natur.443..553D}
and in a full 3-D kinetic PIC approach
\citep{Zeiler2002,2004JGRA..10901220P,Karimabadi2007,Shay2007,
Daughton2009,Drake2010}\index{acceleration!in PIC approach}.

\begin{figure}
\includegraphics[width=0.8\textwidth]{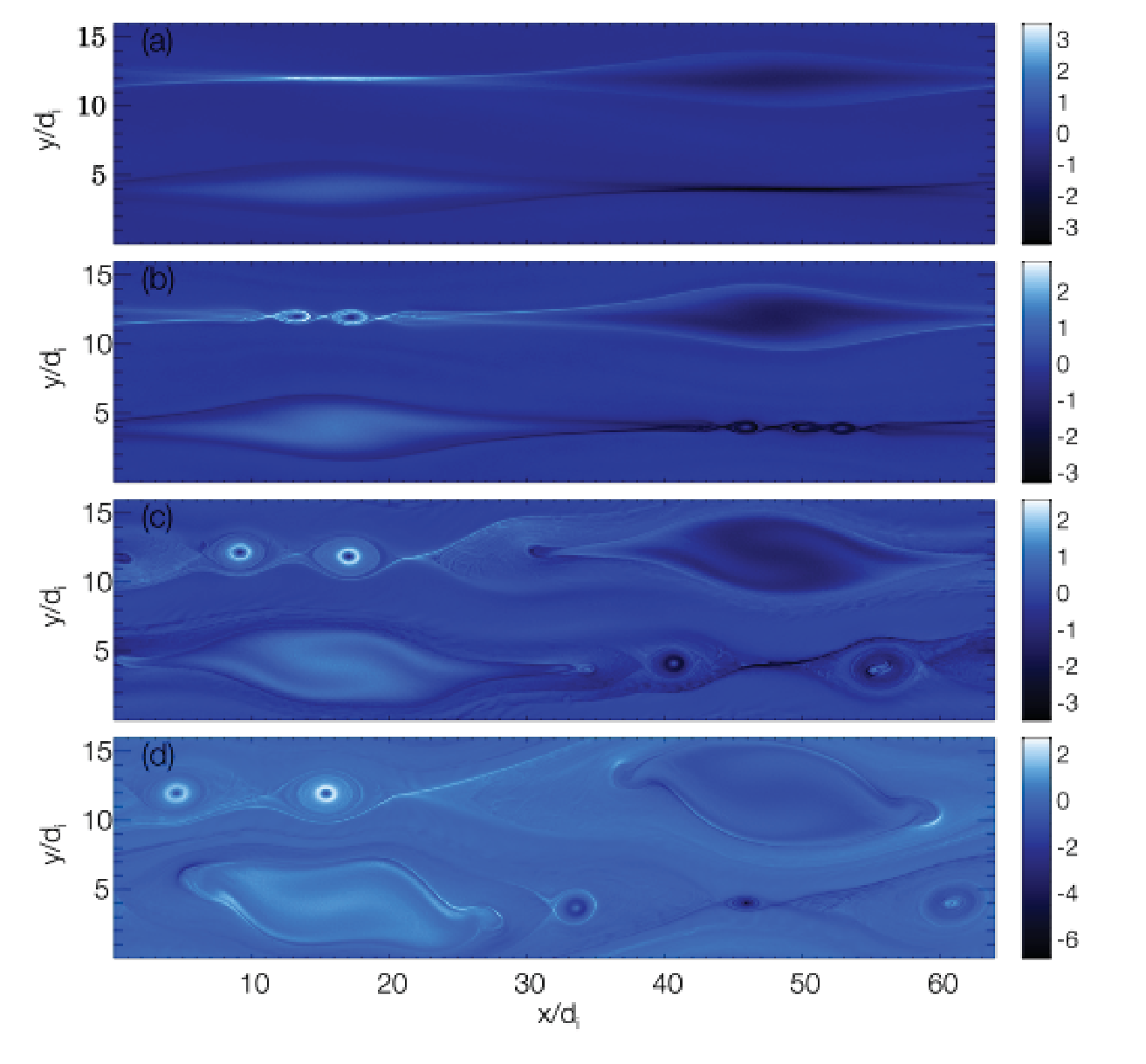}
\caption{Out-of-plane current density in 2.5-D PIC simulations of
reconnection with a guiding field, for four consecutive time
intervals, from top to bottom. The environment contains two current
sheets and is characterized by the appearance of islands formed by
the tearing instability. At the start, there are only (a) large
islands, followed at later times (b-d) by the formation of smaller
magnetic islands. Note that such islands do {\it not} appear in 2-D PIC
simulations of the same environment without a guiding field. From
\citet{2006GeoRL..3313105D}.} \label{fig:island_islands}
\end{figure}

\citet{1998GeoRL..25.3759S} studied the dynamics of whistler waves\index{waves!whistlers}
generated inside the magnetic islands, using a code in which the
electric and magnetic fields were calculated at each time step
using the Poisson condition $\nabla \cdot E = 4\pi\rho$.
The time step was normalized to the electron cyclotron time
$\Omega_{e}^{-1}$, calculated for a maximum initial magnetic field
$B_0$, the spatial step normalized to the skin depth $c/\omega
_{pe}$, and the velocity normalized to the Alfv\'en velocity for
electrons $V_{Ae} = B_0/\sqrt{4\pi n_0m_e}$.  The speed of light
was artificially reduced to $5 V_{Ae}$ in order to marginally
resolve Debye radii.  The simulation grid was $512\times512$ with
7 million particles involved.  Small magnetic perturbations and
associated currents were added as initial conditions to form seed
magnetic islands centered around the two current layers, as shown in Figure \ref{fig:island_islands}.
Using such a formalism, \citet{1998GeoRL..25.3759S} were able to resolve
the electron dissipation region, despite the use of an
artificially low ion-to-electron mass ratio of 200.

\index{magnetic structures!PIC simulation of magnetic islands!illustration}

\begin{figure}
\centering
\includegraphics[width=1.0\textwidth]{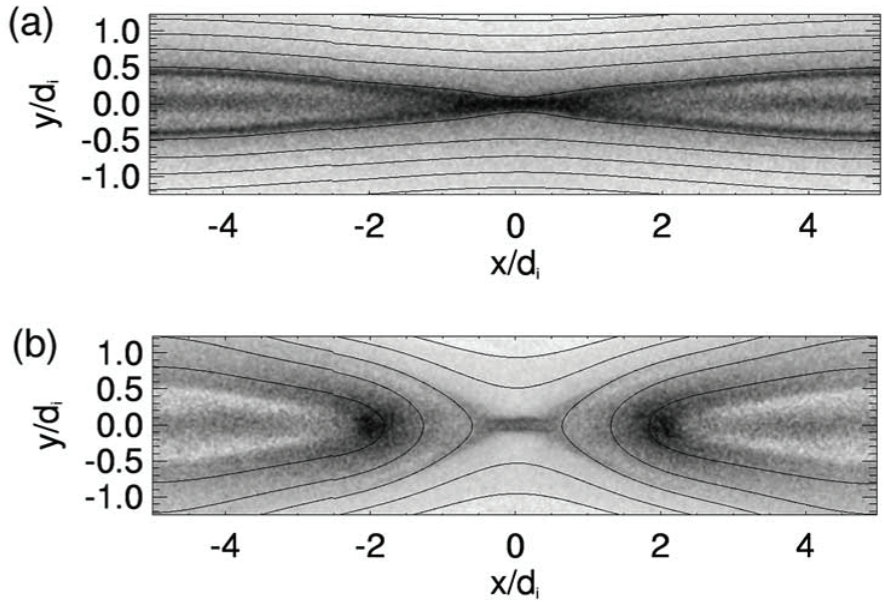}
\caption{Current density in PIC simulations for times (a) $t\,=\,
7.7/\omega_{ci}$ and (b) $t\,=\, 12.9/\omega_{ci}$. The solid
lines correspond to magnetic flux lines. Adopted from
\cite{Zeiler2002}.} \label{fig:island_e_drift}
\end{figure}

A further modification of the PIC code was used to extend it to a
3-D model with 64 grid cells in the third direction, while
increasing the proton-to-electron mass ratio to 277.8 and the
speed of light to $c = 15 V_{Ae}$ \citep{Zeiler2002}. Unlike a conventional Harris equilibrium, in which the
variation of the density produces the equilibrium current, the density was taken to be nearly constant,
with the current being produced by the ${\bf E} \times {\bf B}$ drift of the electrons.  This resulted
(see Figure \ref{fig:island_e_drift}) in a ``parallel'' electric field occurring in the plane of the island current sheet
\citep{Drake1997,1998GeoRL..25.3759S,Zeiler2002}. \citet{2004JGRA..10901220P}
have further shown that in 3-D current sheets with a guiding
magnetic field, the ``parallel'' electric field is formed only in
the two quadrants $(+x, +y)$ and $(-x, -y)$.  Electrons entering these sectors
are accelerated by this ``parallel'' electric field; the other quadrants
are characterized by an excess of ions, strongly reminiscent of
the drift separation of electrons and protons in a 3-D single
current sheet (Sections~\ref{sec:8single3D} and~\ref{sec:8pic}).

Once formed, magnetic islands tend to shrink due to magnetic
tension forces\index{magnetic structures!islands}\index{magnetic tension}.
Conservation of the longitudinal invariant leads
to efficient first-order Fermi acceleration of particles tied to
the island's field lines.  As shown by
\index{acceleration!Fermi!in shrinking islands}
\citet{2006Natur.443..553D}, the process of island shrinking halts
when the kinetic energy of the accelerated particles is sufficient
to halt the further collapse due to magnetic tension forces; the
electron energy gain is naturally a large fraction of the released
magnetic energy. A sufficient number of electrons can be produced
to account for a moderately-sized solar flare.  The resultant
energy spectra of electrons take the form of power laws, with
harder spectra corresponding to lower values of the plasma
$\beta$. For the low-$\beta$ conditions appropriate to the solar
corona, the model predicts accelerated spectra that are much
harder ($\delta \approx 1.5$) than observed.

\begin{figure}
\centering
\includegraphics[width=0.6\textwidth]{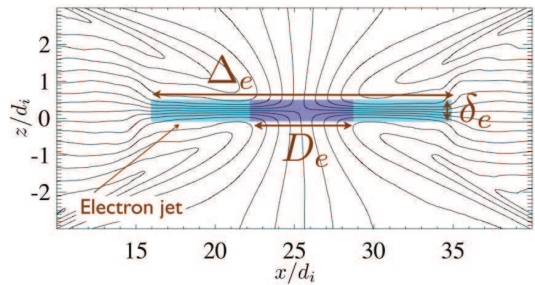}
\caption{The multi-scale structure of the electron diffusion
region around an X-type null point of total length of $\Delta _e$
and thickness $\delta _e$, as derived from 3-D PIC simulations.
The solid lines show the electron trajectories for the time
$t\omega_{ci} = 80$ ($\omega_{ci}$ is the ion cyclotron
frequency). In the inner region of length $D_e$, there is a steady
inflow of electrons and a strong out-of-plane current; the outer
region is characterized by electron outflow jets. From
\cite{Karimabadi2007}.} \label{fig:island_edr}
\end{figure}
\index{magnetic structures!null!illustration}
\index{jets!reconnection outflow!illustration}

Further progress with 3-D PIC simulations of magnetic islands,
using a larger 2560$\times$ 2560 grid, revealed that the electron
diffusion region formed by the electron flows occurs over a width
of $5d_e$ \citep[where $d_e$ is the electron inertial length
--][]{2001JGR...106.3715B}, while other simulations show an extent
up to $20d_i$ \citep[where $d_i$ is the ion inertial length
--][]{Karimabadi2007,Shay2007}; see Figure \ref{fig:island_edr}. 
The electron diffusion region becomes elongated with a two-scale
structure: an inner region (size $D_e$ in
Figure~\ref{fig:island_edr}) with a strong
electron-drift-associated electric field (see
Equation~\ref{eqn:zharkova_edrift}) and a wider diffusion region
characterized by electron jets \citep{Karimabadi2007,Shay2007}.
\citet{Shay2007} found that the electrons form super-Alfv\'enic
outflow jets\index{simulations!PIC!super-Alfv\'enic
outflow jets}, which become decoupled from the magnetic field and
extend to a large distance from the X-type null point, similar to
the manner in which electron jets decouple from the protons in a
test-particle approach to reconnection in a single 3-D current
sheet (Section~\ref{sec:8single3D}).

\begin{figure}
\centering
\includegraphics[width=0.8\textwidth]{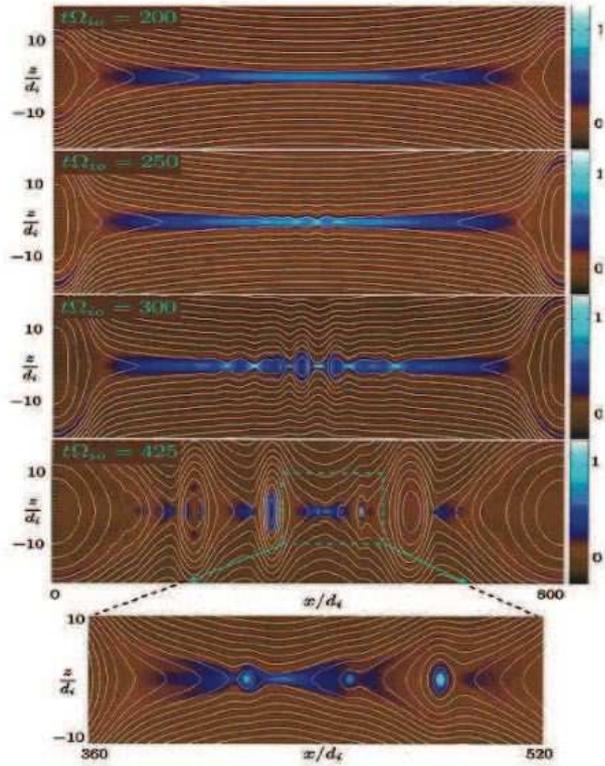}
\caption{Evolution of the current density over a region of extent
$800d_i$ (where $d_i$ is the ion inertial length), for different
times $t$. White lines are the magnetic flux surfaces.  The bottom
plot presents a close-up of a region in which formation of new
islands is occurring. From \cite{Daughton2009}.}
\label{fig:island_daughton}
\end{figure}
\index{magnetic structures!islands!illustration}

Recently, \citet{Daughton2009} showed that for collisional
Sweet-Parker reconnection\index{reconnection!Sweet-Parker!PIC} 
with Lundquist numbers $S$ higher than 1000, the
reconnection region becomes susceptible to the tearing
instability, leading to the formation of many magnetic islands with
half-thickness approaching the ion inertial length $d_i$ (see
Figure~\ref{fig:island_daughton}). This result shows that the
current sheet size is a key factor in defining timescales and the
role of different physical processes at different stages of the
reconnection process.  For example, the shrinking of a large
current sheet during MHD reconnection is halted when it approaches
a size conducive to the formation of magnetic islands\index{plasma instabilities!tearing mode}.

\subsection{Limitations of the PIC approach} \label{sec:8pic_limit}
\index{simulations!PIC!limitations}

Many PIC simulations \index{acceleration!particle-in-cell
(PIC)!limitations} \citep{2001JGR...106.3715B,2006GeoRL..3313105D,
2007PhPl...14k2905T, Shay2007,Karimabadi2007} have been concerned
with the study of reconnection rates in the vicinity of a magnetic
X-type null\index{magnetic structures!PIC simulation!magnetic null}.
Even for the very low densities used in these
simulations, it is apparent that the feedback of the fields
induced by accelerated particles can be substantial (see Section~\ref{sec:8test_p}): 
it can significantly modify the particle
trajectories and the energies gained during their acceleration.

However, because of computer power limitations, the size of the
region \index{magnetic structures!PIC simulation!model size limits} in which the
induced electric and magnetic field are simulated and coupled to
the particle motion is necessarily rather small: $\sim$100~$ \times
50 \times 1000$~cm for 2-D simulations \citep{2001JGR...106.3715B,
2007PhPl...14k2905T}, or $4000 \times 40000 \times 86$~cm for the
most recent 3-D simulations \citep{Shay2007, Drake2010}.  These
are orders of magnitude smaller than the $10^8$ - $10^{10}$~cm
sizes associated with flares \citep{1994Natur.371..495M,
2003ApJ...596L.251S, 2006ApJ...638.1140J}.  Only in the PIC
simulations discussed in Section~\ref{sec:8pic} is the computation
region extended well beyond the null point in order to explore the
variations of acceleration process, at different magnitudes of the
transverse magnetic field. Thus, {\it the narrowness of the
simulation region} still imposes strong limitations on the number
of particles being accelerated in PIC simulations.

Other puzzling features of the simulations relate to the origin of
the electric fields in the models\index{electric fields!problem of origin in simulations}.  
One example is the
``parallel'' electric field reported in many 3-D simulations
\citep{Zeiler2002,2004JGRA..10901220P,Drake2010} and produced
either as a result of an advective electric field (see the first
term in Equation~\ref{eqn:zharkova_edrift}) and/or by a gradient
of the magnetic field component across the current sheet (see the
second term in Equation~\ref{eqn:zharkova_edrift}.
\index{return current!electric field}
\index{electric fields!return current}
\index{electric fields!advective}
This 'parallel'
electric field is the electric field induced by accelerated electrons
known as a self-induced (return-current) electric field  associated with
their propagation into a flaring atmosphere, which can exceed the
local Dreicer field by a factor of up to $\sim$30 \citep{2006ApJ...651..553Z}. 
Although both the ``parallel'' electric field, or the ``self-induced (return current')' electric field
associated with the propagation of accelerated electrons can cause
various interesting effects (e.g., the charge separation electric
field can decelerate electrons and even return them back to the
acceleration region), causality arguments would strongly suggest
that neither field can be considered as the main driver of
particle acceleration since the field is {\it formed by the
accelerated particles themselves}.
\index{electric fields!return current!inadequacy as a driver}
\index{electric fields!advective!inadequacy as a driver}

A related problem is the difference in electron and ion gyration
radii/periods \citep{2001JGR...106.3715B}. Because of this
difference, within a timescale of $\lapprox 10^{-7}$~s, electrons
can move towards the midplane, be accelerated and ejected, while
the protons have only just started their gyrations in the RCS
environment. In the simulations to date, the proton-to-electron
mass ratio \index{mass ratio!artificial value of}
has been arbitrarily lowered to 100-200, an order of magnitude
smaller than the actual value. 
Even the speed of light is\index{speed of light!artificial reduction in simulations}
conveniently reduced by some one or two orders of magnitude to
allow a reduction in simulation times
\citep{2001JGR...106.3715B,2006Natur.443..553D}. Such simulations,
then, serve to highlight that by artificially adjusting
fundamental physical constants (such as the electron mass or the
speed of light) in order to achieve results with a reasonable
amount of computational effort, certain features of the results can be
altered significantly. Although \citet{1998GeoRL..25.3759S} have shown that the
{\it reconnection rate} may be indeed insensitive to the adopted values of both the electron
mass and speed of light.

Issues such as the ratio of macro- to micro-scales
can be resolved only by increasing the reconnection region
size in the PIC simulations up to sizes comparable with
observations. However,
even with exponential increases in computing power, there are
substantial doubts that one can achieve this goal within the next
few decades.  Possibly a hybrid approach, combining PIC
simulations convoluted with test-particle extensions to greater
source volumes, will allow us to make substantial progress in this
area.

\section{How well do Recent Theories Account for Observed Flare Characteristics?}\label{sec:8explain}

\subsection{Interrelation between acceleration and transport}\label{sec:8transport}
\index{acceleration!and particle transport, overview}

Before we address the relative merits of the various particle
acceleration models, it is important to realize that many flare
observations consist of radiation integrated over the source
volume; the quantity observed involves integration over not only
space but also quantities such as particle pitch angles.
Therefore, in order to relate to observed features
(Section~\ref{sec:8obs_review}), one must incorporate the effects
of particle acceleration with the {\it transport} of the
accelerated particles throughout the remainder of the flaring
source.  Although such an inclusion will in all probability
obscure information on the acceleration process, it can also, more
profitably, be used as a diagnostic of particle transport
processes 
\index{accelerated particles!transport}\index{transport!biases acceleration theory}
themselves.

As we have seen (Sections~\ref{sec:8single3D}, \ref{sec:8fan3d}
and~\ref{sec:8complex}), accelerated electrons can attain
quasi-relativistic energies within an extremely small timescale
and can thus propagate over the extent of a solar flare of size $\sim$10$^9$~cm within $\sim$0.01-0.1~s. 
The proton acceleration
time is longer than that for electrons by up to two orders of
magnitude (Section~\ref{sec:8single3D}), so that protons still
remain within the acceleration region while the electrons are
ejected from it.\index{acceleration region!charge separation}
Thus, particle transport mechanisms are essential
for completing the electric circuit\index{acceleration!and particle transport, overview!electric circuit} for particles escaping the
acceleration region \citep{1995ApJ...446..371E} and/or in
establishing co-spatial return currents\index{return current}
\citep{1977ApJ...218..306K,1995ApJ...446..371E,2006ApJ...651..553Z}\index{acceleration!and particle transport, overview!return current}.
Such return currents can be formed both by ambient electrons
\citep{1977ApJ...218..306K,1985SoPh...97...81N} and by relatively
high-energy beam electrons \citep{2006ApJ...651..553Z} that return
to the corona with nearly the same velocities as the injected
electrons\index{electron beams!high-energy return current}.
The self-induced electric fields\index{return current}\index{acceleration!and particle transport, overview!turbulence}
created by the beam electrons and/or the return current have a
structure very favorable for the generation of plasma turbulence\index{plasma turbulence}
\citep{2008SoPh..247..335K}, which naturally creates the conditions
required for stochastic acceleration of electrons to
sub-relativistic energies as discussed in Section
\ref{sec:8stochastic} \citep[see
also][]{1996ApJ...461..445M,1997ApJ...482..774P}.

Another issue that can arise is consideration of {\it radiative}
transport effects, such as radiative transfer of optically-thick
lines and continua, and reflection (or albedo
effects) of emission of hard X-rays from the
photosphere.
\index{acceleration!and particle transport, overview!radiative transport}
\index{hard X-rays!albedo} 

\subsection{Testing acceleration models against observational constraints} \label{sec:8probing}

Here we briefly summarize the extent to which the various models
discussed in Section~\ref{sec:8developments} have increased our
ability to account for the characteristics (e.g., number,
spectrum) of accelerated particles in a solar flare.  In other
words, can {\it any} hitherto-identified acceleration process
produce the rate, number and energy spectrum of accelerated
particles inferred from observations?

\subsubsection{Stochastic acceleration} \label{sec:zharkova_prob_stochastic}

Stochastic acceleration 
\index{acceleration!stochastic!MHD turbulence}\index{acceleration!and particle transport, overview!stochastic acceleration}
remains one of the most promising candidates to
account for most of the characteristics deduced from hard X-ray,
radio and gamma-ray observations. This mechanism does, however,
require a pre-supposed level of either plasma or MHD turbulence,
the basis for which has not yet been fully established.  Further,
high-frequency waves near the plasma frequency can be excluded as
drivers for
\index{acceleration!stochastic} 
stochastic acceleration, since they would couple into a high level of
decimetric radio emission in every flare, contrary to observations
\citep{2005SoPh..226..121B}.

On the other hand, acceleration by
\index{acceleration!transit-time damping}\index{acceleration!and particle transport, overview!transit-time damping}
transit-time damping
{\it is} compatible with a frequent lack of radio emission because
the frequency\index{frequency!turbulence} of the postulated MHD turbulence is far below the
plasma frequency. 
The non-thermal electron distribution
\index{accelerated particles!energy spectra} ``grows'' out of the
thermal population, resulting in a ``soft-hard-soft''
\index{hard X-rays!soft-hard-soft}\index{soft-hard-soft}
evolution of the hard X-ray spectrum that is consistent with observations. 
\index{hard X-rays!coronal sources!soft-hard-soft}
\index{hard X-rays!coronal sources!pivot point}
Further, such ``soft-hard-soft'' behavior has been found in coronal sources,
\citep{2007A&A...466..713B} and is therefore inherent to the
acceleration process and not to collisional transport effects.
{\em RHESSI} observations showing the existence of a ``pivot
point'' energy at which the differential electron flux does not
vary with time, leading to the observed ``soft-hard-soft''
behavior, have been used \citep{2004A&A...426.1093G} as evidence
for a stochastic acceleration model.

Recent models of stochastic acceleration that involve the
interaction of turbulence with high energy particle beams injected
from elsewhere \citep{Bykov2009} can also account for some basic
features of accelerated particles and their evolution with time.
During the initial (linear) phase of the acceleration, effective
particle acceleration by the longitudinal large scale turbulent
motions, leading to visible spectral hardening. Then, because the
accelerated particles eventually accumulate a considerable
fraction of the turbulent energy, they deplete the energy in the
turbulence, leading to a spectrum softening.  Again, this is
consistent with observations \citep{2007A&A...466..713B} of
``soft-hard-soft'' spectral evolution in the main stage and even
some hardening (soft-hard-harder behavior) during later stages
\citep{2007ApJ...663L.109K}.

\subsubsection{Collapsing current sheets} \label{sec:zharkova_prob_traps}

Estimates of the electron energy gained in collapsing magnetic
traps
\index{acceleration!collapsing magnetic trap}\index{acceleration!and particle transport, overview!collapsing traps} 
associated
with reconnection in a helmet-type geometry, with and without a
termination shock, reveal that there is a sufficient amount of
energy released in accelerated particles during such a process.
In addition, betatron acceleration, possibly coupled with first-order
Fermi acceleration in collapsing current sheets, can account for
the electron number and energies gained, although the energy
spectrum of the accelerated particles resembles more the higher
energy tail of a Maxwellian distribution\index{collapsing magnetic trap}.
It remains to be
established how the accelerated particles can move across the
magnetic field of the trap in order to precipitate into the flare
footpoints.
\index{footpoints!and collapsing traps}

\subsubsection{Acceleration in 3D reconnecting current sheets} \label{sec:zharkova_prob_rcs}

New MHD simulations of particle acceleration rates and spectra
from collapsing current sheets\index{acceleration!and particle transport, overview!current sheets}
\index{acceleration!stochastic!multiple current sheets} have
revised our views on the efficiency of particle acceleration
inside a diffusion region, although the number of particles (both
electrons and protons) that can be accelerated in a current sheet
is limited by the volume encompassed by the current sheet
\citep[][see
Section~\ref{sec:8basic}]{1985ApJ...293..584H,1996SoPh..167..321L}.
\index{current sheets!collapsing trap}
However, more complex 3-D current sheets and/or multiple current
sheets may be reconcilable with this model (see
Sections~\ref{sec:8single3D} and~\ref{sec:8fan3d}). Acceleration
of electrons in a 3-D RCS occurs mostly around the RCS midplane,
over a very short timescale comparable with the growth rate of
hard X-ray emission in solar flares. Acceleration of protons has a
much longer timescale and results in a spread of protons in a
location outside the midplane. The particle orbits in the RCS lead
to a separation of electrons and protons into the opposite
semiplanes, corresponding to \index{acceleration!electron-proton
asymmetry} ejection in opposite directions and into opposite loop legs.
The separation of electrons and protons during acceleration in an RCS
leads to the formation of a polarization electric field (induced
by Hall currents),
which sets the frame for the generation of
strong plasma turbulence that can lead to further particle
acceleration by stochastic processes.
\index{electric fields!polarization}\index{plasma turbulence} 
Single 3-D current sheets
with guiding fields have been found, through both test-particle
and PIC approaches, to be able to supply sufficient energy into
both electrons and protons in order to account for {\em RHESSI}
hard X-ray and gamma-ray observations.

Spine reconnection in 3-D MHD reconnection simulations can
naturally produce two symmetric jets of energetic particles
escaping along the spine, while the fan reconnection produces
azimuthally localized ribbons of high energy particles in the fan
plane.\index{ribbons!and fans}\index{acceleration!and particle transport, overview!3-D reconnection}
Both features, in principle, are often observed in flares
(e.g., magnetic jets in SOL2000-07-14T10:24, X5.7, and  
\index{flare (individual)!SOL2000-07-14T10:24 (X5.7)!ribbons in} and 
\index{flare (individual)!SOL2002-07-23T00:35 (X4.8)!ribbons in} SOL2002-07-23T00:35, X4.8) and
two-ribbon flares as discussed in \cite{Chapter2}.\index{jets}\index{flare types!two-ribbon}

Another possibility involves tearing instabilities in current
sheets, which leads to the formation of large numbers of magnetic
islands, on which a large number of particles can be rapidly
accelerated; however, the simulated electron energy spectra are
much harder than inferred from hard X-ray observations\index{plasma instabilities!tearing mode}\index{magnetic structures!islands}\index{acceleration!and particle transport, overview!magnetic islands}.

Despite these qualified successes, there are two outstanding
problems with this kind of acceleration model: (1) incorporation
of the self-consistent electrodynamic fields associated with the
accelerated particles; and (2) production of sufficient numbers of
accelerated particles to match the observations. The latter can be
marginally accounted for if one notes that strongly stressed 3-D
current sheets, discussed in Sections~\ref{sec:8single3D}
and~\ref{sec:8fan3d}, are characterized by much larger diffusion
regions than in simple reconnection models
(Section~\ref{sec:8reconnection}).

\subsubsection{Complex reconnection models} \label{sec:zharkova_prob_complex}
\index{acceleration!and particle transport, overview!self-organized criticality}\index{self-organized critical state}

The other way to tackle the particle number problem involves
multiple dissipation sites associated with self-organized
criticality (SOC) that could lead to ``spreading'' of the
dissipation across a large coronal volume (see
Section~\ref{sec:8complex}). Particles move between the current
sheets, either gaining or losing energy in each one, resulting in
a systematic second-order gain in energy. Particle energy
distributions from the simulations fit the observed
spectra for electrons very well, although proton spectra have yet to be
simulated.\index{accelerated particles!number problem}

However, a few essential issues
\index{acceleration!stochastic!limitations of models} do remain:

\begin{enumerate}

\item The models invoke an {\it ad hoc} distribution of RCS sizes
and physical conditions in order to account for the number and
spectra of the accelerated particles;

\item The basic model of a current sheet (discussed in
Section~\ref{sec:8basic}) is used to derive the energy gains by
accelerated particles in a component current sheet, while the
actual gains are significantly different in more realistic 3-D
models of reconnecting current sheets;

\item Even if multiple current sheets are formed in the whole
volume of the corona, there remains the issue of where the
reconnecting magnetic field lines are embedded and how the test
particles can move across the strong magnetic field threading the
reconnection volume in order to reach another current sheet;

\item Some of these models also suffer from problems associated
with the neglect of the feedback of the electric and magnetic
fields associated with the accelerated particles
themselves\index{acceleration!test-particle approach!limitations};

\item Multiple RCS models simulated with a more realistic PIC
approach are applicable to only a very small region (about
$10^6$~cm$^3$), and it is assumed without justification that the
results can be scaled to flare volumes.
\end{enumerate}

\section{Summary} \label{sec:8summary}

Over the past decade, and especially in the {\em RHESSI} era,
significant progress has been made in advancing various particle
acceleration models and assessing their ability to account for the
observed features in solar flares. {\em RHESSI} observations have
been instrumental in encouraging the further development and
exploration of both ``traditional'' and ``non-traditional''
mechanisms, including more complicated magnetic field geometries
and topologies, to enhanced MHD and PIC models of magnetic
reconnection in a single current sheet, to multiple current sheets
and stochastic acceleration by various plasma waves. Substantial
progress has been made through the use of phenomenological
estimations made in collapsing traps, termination shocks of
Alfv\'en waves, and direct simulations using both test-particle and
self-consistent PIC approaches.

Consideration of the full three-dimensionality of the magnetic
reconnection process has been found to significantly enhance the
acceleration efficiency in current-sheet diffusion regions.
However, the question still remains whether such a mechanism can
account for the very large number of electrons responsible for hard
X-ray emission and whether consideration of the whole electric
circuit formed by precipitating and returning beams in
reconnecting loops can resolve the particle number problem.  PIC
simulations have demonstrated that the acceleration of substantial
numbers of charged particles causes a number of important
secondary effects: for example, polarization electric fields
induced by separation of electrons and protons towards the
midplane, return currents formed from precipitating and ambient
electrons, and plasma turbulence related to beam instabilities.\index{plasma turbulence}

Interactions of large-scale loops with strong magnetic fields can
lead either to acceleration in collapsing traps by betatron and/or
first-order Fermi mechanisms, or to acceleration in the
termination shock formed by such the collapsing sheets (or by
large-scale Alfv\'en waves produced during the reconnection).
Interaction of smaller-scale magnetic fields can produce magnetic
islands or multiple current sheets in which particles can be also
accelerated in a cascade processes if the particle travels and gains
energy from one feature or another.

The exact acceleration scenario will depend on the magnetic field
topology and on the absolute magnitudes of physical parameters.
Further, it is possible, or even likely, that all the acceleration
processes discussed above play some role in solar flares and
elsewhere in astrophysics.  The strict demands imposed by the {\em
RHESSI} observations have necessitated a thorough and critical
review of all these models, leading to substantial new
understanding along the way.

\begin{acknowledgements} \label{sec:8acknowledgements}

AGE acknowledges support from the NASA Heliophysics Division and from UC Berkeley. EPK acknowledge the support of a
PPARC/STFC Rolling Grant, PPARC/STFC UK Advanced Fellowship and
Royal Society Conference Grant and by the European
Commission through the SOLAIRE Network (MTRN-CT-2006-035484). AGE and EPK were also supported in
part by a grant from the International Space Science Institute
(ISSI) in Bern, Switzerland. NV acknowledges support from Centre National
d'Etudes Spatiales (CNES) and from the French program on Solar-Terrestrial Physics
(PNST) of INSU/CNRS for the participation to the RHESSI project.
VZ acknowledges the support of the Science Technology and Facility
Council (STFC) project PP/E001246/1. The computational work for the test-particle and PIC simulations was
carried out on the joint STFC and SFC (SRIF) funded cluster at the
University of St Andrews, UK.
\end{acknowledgements}

\bibliographystyle{ssrv}

\bibliography{ch8}

\printindex

\end{document}